\documentclass[manuscript,screen]{acmart}

\setcopyright{acmlicensed}
\copyrightyear{2018}
\acmYear{2018}
\acmDOI{XXXXXXX.XXXXXXX}

\usepackage{amsmath} 

\usepackage{calculi}
\usepackage{microtype}
\usepackage{natbib}
\usepackage{amsthm}
\usepackage{thmtools}
\usepackage{listings}
\usepackage{hyperref}
\usepackage[colorinlistoftodos,prependcaption,textsize=tiny]{todonotes}

\definecolor{strings}{rgb}{0,0.5,0}
\definecolor{emphs}{rgb}{0.64,0.08,0.08}
\definecolor{comments}{rgb}{0.17,0.57,0.68}
\colorlet{keywords}{blue!50!cyan}

\declaretheorem[style=plain,name=Theorem]{theorem}

\declaretheorem[style=plain,name=Corollary]{corollary}
\declaretheorem[style=plain,name=Lemma]{lemma}

\declaretheorem[style=definition,qed=$\blacksquare$,name=Definition]{definition}

\declaretheorem[style=definition,qed=$\blacksquare$,name=Remark]{remark}

\lstdefinestyle{tinysol}{
  language=C,
  captionpos=b,
  numbers=left,
  numberstyle=\tiny,
  frame=lines,
  showspaces=false,
  showtabs=false,
  breaklines=true,
  showstringspaces=false,
  breakatwhitespace=true,
  emph={contract,field,func,call,if,then,else,while,do},
  emphstyle={\rmfamily\bfseries\color{emphs}},
  commentstyle=\color{comments},
  morekeywords={contract, interface, account, this, var, field, func, method, send, value, balance, sender, then, fallback, call, dcall, args, id},
  keywordstyle={\bfseries\color{keywords}},
  stringstyle=\color{strings},
  basicstyle=\ttfamily\small,
  escapechar=@
}

\usetikzlibrary{
  patterns
}

\newbool{showedits}
\boolfalse{showedits}

\makeatletter
\def\edited{%
  \def\edited@temp{edited}%
  \ifx\@currenvir\edited@temp
    \@xp\edited@environmentcase 
  \else
    \@xp\edited@commandcase 
  \fi
}

\ifbool{showedits}{
  \newcommand{\edited@commandcase}[2][red]{\textcolor{#1}{#2}}
  \newcommand{\edited@environmentcase}[1][red]{\color{#1}}
}{
  \newcommand{\edited@commandcase}[2][\relax]{#2}
  \newcommand{\edited@environmentcase}[1][\relax]{}
}
\makeatother

\def\TINT{\text{\normalfont\sffamily int}}

\def\TBOOL{\text{\normalfont\sffamily bool}}

\def\SUBS{\ensuremath{\mathrel{\text{\code{<:}}}}}

\def\TYPES{\SETNAME{T}}
\def\BASETYPES{\SETNAME{B}}

\def\TYPEENVS{\SETNAME{E}}

\def\SUCHTHAT{\ensuremath{\mathrel{.}}}
\newcommand{\TFAMILY}[1][\Gamma]{\SETNAME{R}_{#1}}
\newcommand{\WRONG}[2][\Gamma]{\ensuremath{\text{\normalfont\sffamily Wrong}_{#1}\kern-2pt\ARGS{#2}}}
\newcommand{\SAFE}[2][\Gamma]{\ensuremath{\text{\normalfont\sffamily Safe}_{#1}\kern-2pt\ARGS{#2}}}
\newcommand{\NSAFE}[2][\Gamma]{\ensuremath{\text{\normalfont\sffamily NSafe}_{#1}\kern-2pt\ARGS{#2}}}

\def\EPI{\ensuremath{\prescript{e}{}{\kern-1pt\PI}}}
\def\TINYSOL{\textsc{TinySol}}

\def\SWAP#1#2#3{\ensuremath{\ARGS{#1,#2}\kern-2pt\boldsymbol{\cdot}\kern-2pt #3}}

\makeatletter
\newcommand*\bigcdot{\mathpalette\bigcdot@{.5}}
\newcommand*\bigcdot@[2]{\mathbin{\raisebox{-.7ex}{\hbox{\scalebox{#2}{$\m@th#1\bullet$}}}}}
\makeatother

\def\CHANEQ{\ensuremath{\xleftrightarrow{\kern-1.5pt\bigcdot}}}

\def\CSOR{\ensuremath{~{\syntaxfont{[\kern-2pt]}}~}}

\newcommand{\makeenv}[1]{\{#1\}}
\newcommand{\joine}[2]{\{#1\} \cup #2}

\def\TYPES{\ensuremath{\textbf{\textit{Types}}}}

\def\INCRSYM{\ensuremath{\text{\syntaxfont{+}}}}

\def\NCOMP#1#2{\ensuremath{#1\cdot#2}}

\def\NAMESPACE#1{\SETNAME{N}_{\kern-1pt\HOLE[#1]}}
\def\LNS#1{\prescript{\INCRSYM\kern-2pt}{}{\NAMESPACE{#1}}}

\def\LNC{\prescript{\INCRSYM\kern-2pt}{}{N}}

\newcommand{\LNSSUBST}[1][s_1,s_2]{\ensuremath{\prescript{\INCRSYM\kern-2pt}{}{\sigma_{#1}}}}

\def\TYPES{\ensuremath{\LANG{T}}}

\def\SUCHTHAT{\ensuremath{\;.\;}}

\newlength\QEDSymbolSpace
\newenvironment{fixqed}
{\begin{minipage}[b]{\dimexpr\linewidth-\QEDSymbolSpace}\centering}
{\end{minipage}\vspace{-1em}}

\newcommand{\code}[1]{\texttt{#1}}
\newcommand{\CONF}[1]{\ensuremath{\left<#1\right>}}
\newcommand{\UPDATE}[2]{\ensuremath{\kern-2pt\left[#1 \mapsto #2\right]}} 

\def\VALUES{\ensuremath{\normalfont\text{\sffamily Val}}}
\def\EXPR{\ensuremath{\normalfont\text{\sffamily Exp}}}
\def\STM{\ensuremath{\normalfont\text{\sffamily Stm}}}

\def\ANAMES{\ensuremath{\normalfont\text{\sffamily ANames}}} 
\def\MNAMES{\ensuremath{\normalfont\text{\sffamily MNames}}} 
\def\VNAMES{\ensuremath{\normalfont\text{\sffamily VNames}}} 
\def\FNAMES{\ensuremath{\normalfont\text{\sffamily FNames}}} 
\def\TNAMES{\ensuremath{\normalfont\text{\sffamily TNames}}} 

\newcommand{\ENV}[1]{\ensuremath{\normalfont\text{\sffamily env}_{#1}}}
\newcommand{\DEC}[1]{\ensuremath{\normalfont\text{\sffamily Dec}_{#1}}}
\newcommand{\SETENV}[1]{\ensuremath{\normalfont\text{\sffamily Env}_{#1}}}

\def\TRUE{\ensuremath{\text{\sffamily T}}}
\def\FALSE{\ensuremath{\text{\sffamily F}}}

\def\BOOLEANS{\ensuremath{\mathbb{B}}}
\def\INTEGERS{\ensuremath{\mathbb{Z}}}

\def\LVAL{\ensuremath{\text{\sffamily LVal}}}
\def\MVAR{\ensuremath{\text{\sffamily MVar}}}
\def\TRANSACTIONS{\ensuremath{\text{\sffamily Tr}}}
\def\BLOCKCHAINS{\ensuremath{\SETNAME{B}}}

\def\ANAMES{\ensuremath{\text{\sffamily ANames}}} 

\newcommand{\CALL}[4]{\ensuremath{\code{$#1$.$#2$($#3$)\$$#4$}}}
\newcommand{\DCALL}[3]{\ensuremath{\code{dcall $#1$.$#2$($#3$)}}}

\def\SECLEVELS{\SETNAME{S}}
\def\ordleq{\sqsubseteq}

\def\subs{\mathrel{<:}} 
\def\SUBS{\mathrel{<:}} 

\newcommand{\TVAR}[1]{\ensuremath{\normalfont\text{\sffamily var}(#1)}}
\newcommand{\TCMD}[1]{\ensuremath{\normalfont\text{\sffamily cmd}(#1)}}
\newcommand{\TSPROC}[2]{\ensuremath{\normalfont\text{\sffamily proc}(#1)\code{:}#2}}

\def\BSVEC{\VEC{B_s}}  

\def\TENV{\SETNAME{E}}
\def\STOP{\ensuremath{s_\top}}
\def\SBOT{\ensuremath{s_\bot}}
\def\ITOP{\ensuremath{I^{\top}}}

\newcommand{\TRANSACT}[5]{\ensuremath{#1 \stackrel{#5}\leadsto #2\code{.}#3\code{(}#4\code{)}}}

\newcommand{\EXTEND}[2]{\ensuremath{\kern-2pt\left[#1 \mapsto #2\right]}} 

\def\TYPEENVS{\SETNAME{E}} 

\NewDocumentCommand{\ETYPI}  { O{B_s} O{\Delta} }{\ensuremath{\SETNAME{R}^{#1}_{\Sigma;\Gamma;#2} }}
\NewDocumentCommand{\ETYPING}{ O{B_s} O{\Delta} }{\ensuremath{\nu\SETNAME{R}^{#1}_{\Sigma;\Gamma;#2} }}

\NewDocumentCommand{\STYPI}  { O{s} O{\Delta} }{\ensuremath{\SETNAME{R}^{#1}_{\Sigma;\Gamma;#2} }}
\NewDocumentCommand{\STYPING}{ O{s} O{\Delta} }{\ensuremath{\nu\SETNAME{R}^{#1}_{\Sigma;\Gamma;#2} }}

\newcommand{\TYPEOF}[2][\Gamma]{\ensuremath{\text{\textsc{TypeOf}}_{#1}(#2)}}

\begin{document}

\title{A Sound Type System for Secure Currency Flow}

\author{Luca Aceto}
\email{luca@ru.is}
\orcid{https://orchid.org/0000-0002-2197-3018}
\affiliation{%
  \institution{Reykjavík University, Menntavegur 1, 102}
  \city{Reykjavík}
  \country{Iceland}
}
\affiliation{%
  \institution{Gran Sasso Science Institute}
  \city{L'Aquila}
  \country{Italy}
}

\author{Daniele Gorla}
\email{gorla@di.uniroma1.it}
\orcid{https://orchid.org/0000-0001-8859-9844}
\affiliation{
  \institution{Sapienza, Università di Roma}
  \city{Rome}
  \country{Italy}
}

\author{Stian Lybech}
\email{stian21@ru.is}
\orcid{https://orcid.org/0000-0001-8219-2285}
\affiliation{%
  \institution{Reykjavík University, Menntavegur 1, 102}
  \city{Reykjavík}
  \country{Iceland}
}

\begin{abstract}
In this paper, we focus on \TINYSOL{}, a minimal calculus for Solidity smart contracts, introduced by Bartoletti et al. We start by rephrasing its syntax (to emphasise its object-oriented flavour) and give a new big-step operational semantics for that language. We then use it to define two security properties, namely call integrity and noninterference. 
These two properties have some similarities in their definition, in that they both require that some part of a program is not influenced by the other part. 
However, we show that the two properties are actually incomparable. 
Nevertheless, we provide a type system for noninterference and show that well-typed programs satisfy call integrity as well; hence, programs that are accepted by our type system satisfy both properties.
We finally discuss the practical usability of the type system and its limitations by means of some simple examples.
\end{abstract}

\maketitle

\section{Introduction}
The classic notion of noninterference \cite{goguen_mesegier1982security_policies} is a well-known concept that has been applied in a variety of settings to characterise both integrity and secrecy in programming.
In particular, this property has been defined in \cite{volpano2000secure_flow_typesystem} by Volpano, Smith and Irvine in terms of a lattice model of security levels (e.g.\@ `High' and `Low', or `Trusted' and `Untrusted'); the key point is that information must {\em not} flow from a higher to a lower security level.
Thus, the lower levels are unaffected by the higher ones, and, conversely, the higher levels are `noninterfering' with the lower ones.

Ensuring noninterference seems particularly relevant in a setting where not only information, but also \emph{currency}, flows between programs.
This is a core feature of \emph{smart contracts}, which are programs that run on top of a blockchain and are used to manage financial assets of users, codify transactions, and implement custom tokens; see e.g.\@ \cite{seijas2016scripting_smart_contracts} for an overview of the architecture.
The code of a smart contract resides on the blockchain itself and is therefore both immutable and publicly visible.
This is one of the important ways in which the `smart-contract programming paradigm' differs from conventional programming languages.

Public visibility means that \emph{vulnerabilities} in the code can be found and exploited by a malicious user.
Moreover, if a vulnerability is discovered, immutability prevents the contract creator from correcting the error.
Thus, it is obviously desirable to ensure that a smart contract is safe and correct \emph{before} it is deployed onto the blockchain.

The combination of immutability and visibility has led to huge financial losses in the past (see, e.g., \cite{ABC17,dao2016,wallet17a,wallet17b,tolmach2020survey_smart_contracts}).
A particularly spectacular example was the infamous DAO-attack on the Ethereum platform in 2016, which led to a loss of 60 million dollars \cite{dao2016}.
This was made possible because a certain contract (the DAO contract, storing assets of users) was \emph{reentrant}, that is, it allowed itself to be called back by the recipient of a transfer \emph{before} recording that the transfer had been completed.

\begin{figure}[t]
\begin{lstlisting}[style=tinysol]
contract X {                             contract Y {
  ...                                      ...
  field called := @\FALSE@;                       deposit(x) {
  transfer(z) {                              x.transfer(this)$0
    if @$\neg$@called @$\land$@ this.balance @$\geq$@ 1         }
      then z.deposit(this)$1;            } 
           this.called := @\TRUE@
    else skip
  }
}
\end{lstlisting}
\caption{Illustration of reentrancy written in the language \TINYSOL{}.}
\label{fig:reentrancy}
\end{figure}

\emph{Reentrancy} is a pattern based on mutual recursion, where one method $f$ calls another method $g$ while also transferring an amount of currency along with the call.
If $g$ then immediately calls $f$ back, it may yield a recursion where $f$ continues to transfer funds to $g$.
We can illustrate the problem as in Figure~\ref{fig:reentrancy}, using a simple, imperative and class-based model language called \TINYSOL{}~\cite{bartoletti2019tinysol}. 
This model language, which we shall formally describe in Section~\ref{sec:tinysol_new}, captures some of the core features of the smart-contract language Solidity \cite{solidity2022}, which is the standard high-level language used to write smart contracts for the Ethereum platform. 
A key feature of this language is that contracts have an associated \code{balance}, representing the amount of currency stored in each contract, which cannot be modified \emph{except} through method calls to other contracts.
Each method call has an extra parameter, representing the amount of currency to be transferred along with the call, and a method call thus represents a (potential) outgoing currency flow.

In Figure~\ref{fig:reentrancy}, \code{X.transfer(z)} first performs a sanity check to ensure that it has not already been called and that the contract contains sufficient funds, which are stored in the field \code{balance}.
Then it calls \code{z.deposit(this)} and transfers \code{1} unit of currency along with the call, where \code{z} is the address received as parameter.
However, suppose that the received address is \code{Y}.
Then \code{Y.deposit(x)} immediately calls \code{X.transfer(z)} back, with \code{this} as the actual parameter, thus yielding a mutual recursion, because the field \code{called} will never be set to $\TRUE$.
A transaction that invokes \code{X.transfer(Y)} with any number of currency units will trigger the recursion.

The problem is that currency cannot be transferred without also transferring control to the recipient, and execution of \code{X.transfer(z)} comes to depend on unknown and untrusted code in the contract residing at the address received as the actual parameter.
Simply swapping the order of lines 6 and 7 in \code{X} solves the problem in this particular case, but it might not always be possible to move external calls to the last position in a sequence of statements.
Furthermore, the execution of a function $f$ can also depend on external fields and not only on external calls.
Thus, reentrancy is not just a purely syntactic property. 

The property of reentrancy in Ethereum smart contracts has been formally characterised by Grishchenko, Maffei and Schneidewind in \cite{grishchenko2018}.
Specifically, they define another property, named \emph{call integrity}, which implies the absence of reentrancy (see~\cite[Theorem~1]{grishchenko2018}) and has been identified in the literature as one of the safety properties that smart contracts should have.
Informally, this property requires any call to a method in a `trusted' contract (say, $X$) to yield the very same sequence of currency flows (i.e.\@ method calls) even if some of the other `untrusted' contracts (or their stored values) are changed.
In a sense, the code and values of the other contracts, which could be controlled by an attacker, must not be able to affect the currency flow from $X$.

A disadvantage of the definition of call integrity given in \cite{grishchenko2018} is that it relies on a universal quantification over all possible execution contexts, making it difficult to check in practice.
However, call integrity seems intuitively to be related to noninterference, in the sense that both stipulate that changes in one part of a program should not have an effect upon another part.
Even though we discover that the two properties are incomparable, one might hope to be able to apply techniques for ensuring noninterference to also capture call integrity.
Specifically, in \cite{volpano2000secure_flow_typesystem}  Volpano et al. show that noninterference can be soundly approximated using a type system.
In the present paper, we shall therefore create an adaptation of that type system for secure-flow analysis to the setting of smart contracts and show that the resulting type system \emph{also} captures call integrity.

To recap, our main contributions in this paper are:
(1) a thorough study of the connections between call integrity and noninterference for smart contracts written in the language \TINYSOL{}, and (2) a sound type system guaranteeing both noninterference and call integrity for programs written in that language. 
We choose \TINYSOL{} because it provides a minimal calculus for Solidity contracts and thus allows us to focus on the gist of our main contributions in a simple setting. 
In doing this, we also provide a more standard operational semantics for this language than the one given in \cite{bartoletti2019tinysol}; this can be considered a third contribution of our work.

The paper is organised as follows:
In Section~\ref{sec:tinysol_new}, we describe a revised version of the smart-contract language \TINYSOL{}~\cite{bartoletti2019tinysol}.
In Section~\ref{sec:call_integrity_tinysol}, we adapt the definition of call integrity from \cite{grishchenko2018} and of noninterference from \cite{SV98} to this language; we then show that these two desirable properties are actually incomparable.
Nevertheless, there is an overlap between them.
In Section~\ref{sec:typesystem_noninterference}, we create a type system for ensuring noninterference in \TINYSOL{}, along the lines of Volpano et al.\@ \cite{volpano2000secure_flow_typesystem}, and prove the corresponding type soundness results (Theorems~\ref{thm:subject_reduction}--\ref{thm:extended_soundness}).
Our main results are Corollary~\ref{cor:noninterference} and Theorem~\ref{thm:welltypedness_callintegrity}, which show that well-typedness provides a sound approximation to \emph{both} noninterference \emph{and} call integrity.
This is used on a few examples in Section~\ref{sec:examples}, where we also discuss the limitations of the type system.
We survey some related work in Section~\ref{sec:relwork} and conclude the paper with some directions for future research in Section~\ref{sec:concl}.
Some of the proofs are relegated to the appendix, to streamline reading.

With respect to the conference version \cite{AGL/2024/ecoop/flowtypes}, this paper extends our syntax of \TINYSOL{} with the construct of {\em delegate call}, which allows us to write more sophisticated programming examples (see, e.g., the {\em pointer to implementation pattern} provided in Fig.\ref{fig:pimpl}). 
Moreover, here we introduce a static inheritance between contract types; this is done through a different (w.r.t.\@ the conference version) notion of {\em interfaces} (see Definition~\ref{def:interfaces}) and yields a different, and in a sense simpler, notion of subtyping (see Figures~\ref{fig:subtyping_interface_members} and~\ref{fig:subtyping_expressions}). 
Also, the type system itself, albeit equivalent to the one of the conference version, is presented here in a more elegant and readable way.

\section{The \TINYSOL{} language}\label{sec:tinysol_new}
In \cite{bartoletti2019tinysol}, Bartoletti, Galletti and Murgia present the \TINYSOL{} language, a standard imperative language (similar to Dijkstra's \texttt{While} language \cite{nielson_nielson2007semantics_with_applications}), extended with classes (contracts) and two constructs: (1) a \code{throw} command, representing a fatal error, and (2) a procedure call, with an extra parameter $n$, denoting an amount of some digital asset, which is transferred along with the call from the caller to the callee. 
\TINYSOL{} captures (some of) the core features of Solidity, and, in particular, it is sufficient to represent reentrancy phenomena.

In this section, we present a version of \TINYSOL{} that has been adapted to facilitate our later developments of the type system.
Compared to the presentation in \cite{bartoletti2019tinysol}, we have, in particular, added explicit declarations of variables (local to the scope of a method) and fields (corresponding to the \emph{keys} in the original presentation) to have a place for type annotations in the syntax. 
Furthermore, differently from \cite{AGL/2024/ecoop/flowtypes,bartoletti2019tinysol}, here we also consider a second kind of procedure call taken from Solidity (named {\em delegate}), which allows a method to be executed in the environment of the caller; i.e.\@ \emph{as if} the method was declared in the caller contract, rather than in the callee contract.

\subsection{Syntax}

\begin{figure}[t]
\begin{center}
\begin{syntax}[h]
  DF \in \DEC{F} \IS \epsilon \OR \code{field p := $v$;} DF         \tabularnewline
  DM \in \DEC{M} \IS \epsilon \OR \code{$f(\VEC{x})$ \{ $S$ \} } DM \tabularnewline
  DC \in \DEC{C} \IS \epsilon \OR \code{contract $X$ \{ }           \tabularnewline
                  & & \qquad\ \ \ \code{field balance := $n$; $DF$} \tabularnewline
                  & & \qquad\ \ \ \code{send() \{ skip \} $DM$}     \tabularnewline
                  & & \qquad\ \code{\} $DC$}                        \tabularnewline
  m \in \MVAR \IS \code{this} \OR \code{sender} \OR \code{value}    \tabularnewline
  L \in \LVAL \IS x \OR \code{this.$p$}                             \tabularnewline%
  e \in \EXPR \IS v \OR x \OR m \quad |\quad \code{$e$.balance} \OR \code{ $e$.$p$ } \OR \op(\VEC{e}) \tabularnewline
  S \in \STM  \IS \code{skip} \OR \code{throw} \OR \code{var $x$ := $e$ in $S$} \OR \code{$L$ := $e$} \OR \code{$S_1$;$S_2$} 
            \OR \code{if $e$ then $S_{\TRUE}$ else $S_{\FALSE}$ } 
            \ISOR \code{while $e$ do $S$} \OR \CALL{e_1}{f}{\VEC{e}}{e_2} \OR \DCALL{e}{f}{\VEC{e}}
\tabularnewline%
v \in \VALUES \IS \mathbb{N} \UNION \mathbb{B} \UNION \ANAMES 
\tabularnewline%
\tabularnewline%
\multicolumn{3}{l}{
\text{where } x, y \in \VNAMES \text{ (variable names)},
p, q \in \FNAMES \text{ (field names)},
}
\tabularnewline%
\multicolumn{3}{l}{
\qquad\quad X, Y \in \ANAMES \text{ (address names)},
f, g \in \MNAMES \text{ (method names)}
}
\end{syntax}
\end{center}
\caption{The syntax of \TINYSOL.}
\label{fig:syntax_tinysol}
\end{figure}

The syntax of \TINYSOL{} is given in Figure~\ref{fig:syntax_tinysol}, where we use the notation $\mathop{\ \widetilde\cdot\ }$ to denote (possibly empty) sequences of items.
The set of \emph{values}, ranged over by $v$, consists of the sets of integers $\mathbb{Z}$, ranged over by $n$, booleans $\mathbb{B} = \SET{\TRUE, \FALSE}$, ranged over by $b$, and address names $\ANAMES$, ranged over by $X, Y$.
 
We introduce explicit declarations for fields $DF$, methods $DM$, and contracts $DC$. 
The latter also includes account declarations: an \emph{account} is a contract that contains only the declarations of a special field \code{balance} and of a single special method \code{send()}, which does nothing and is used only to transfer funds to the account. 
In contrast, a contract usually contains other declarations of fields and methods.
For simplicity, we make no syntactic distinction between an account and a contract, but, for the purpose of distinguishing, we can assume that the set $\ANAMES$ is divided into contract addresses and account addresses.

We have four `magic' keywords in our syntax:
\begin{itemize}
  \item \code{balance}, a special integer field that records the current balance of the contract (or account).
    It can be read from, but not directly assigned to, except through method calls.
    This ensures that the total amount of currency `on-chain' remains constant during execution.
  \item \code{value}, a special integer variable that is bound to the currency amount transferred with a method call.
  \item \code{sender}, a special address variable that is always bound to the address of the caller of a method. 
  \item \code{this}, a special address variable that is always bound to the address of the contract that contains the method that is currently executed.  
\end{itemize}

\noindent The last three of these are local variables, and we collectively refer to them as `magic variables' $m \in \MVAR$.
The declaration of variables and fields are very similar: the main difference is that variable bindings will be created at runtime (and with scoped visibility). Hence we can let the initial assignment of a local variable be an \emph{expression} $e$; in contrast, the initial assignment to a field must be a \emph{value} $v$ (see the grammar for $\DEC F$ in Fig. \ref{fig:syntax_tinysol}).

The core of the language is the declaration of the expressions $e$ and of the statements $S$, which are almost the same as in \cite{bartoletti2019tinysol}.
The main differences are: (1) we introduce fields $p$ in expressions, instead of keys; (2) we explicitly distinguish between (global) fields and (local) variables, where the latter are declared with a scope limited to a statement $S$; and (3) we introduce explicit \emph{lvalues} $L$, to restrict what can appear on the left-hand side of an assignment (in particular, this ensures that the special field \code{balance} can never be assigned directly). 

Besides the ordinary method call, Solidity also features another kind of method call with a different calling style, known as the \emph{delegate call}.
The purpose of this construct is to facilitate code reuse and the creation of contract libraries. 
This is desirable, since the cost of deploying a smart contract on the Ethereum chain depends on the code size,\footnote{See the description of the \code{CREATE} and \code{CREATE2} opcodes in \cite{evmcodes}.} which therefore encourages developers to use libraries for common tasks.
The special feature of this call style is that the method body is executed in the context of the \emph{caller} contract, rather than in the context of the contract in which it is declared, which allows the calling contract to use the called method \emph{as if} it were part of the calling contract itself.
A natural way to use delegate calls would be as a body of a stub method declaration like
\begin{center}
\code{f(}$\VEC{x}$\code{) \{ dcall X.f(}$\VEC{x}$\code{) \}}
\end{center}

\noindent which simply forwards the call to an implementation that resides in $X$, rather than locally.

\begin{figure}
\begin{lstlisting}[style = tinysol]
contract C {
  field owner := Bob;
  field impl := X;
  ...
  Update(addr) { if sender = owner then impl := addr else skip }
  @$f_1$@(@$\VEC{x}$@) { dcall impl.@$f_1$@(@$\VEC{x}$@) }
  @$\vdots$@
  @$f_n$@(@$\VEC{x}$@) { dcall impl.@$f_n$@(@$\VEC{x}$@) }
}  
\end{lstlisting}
\caption{The \emph{pointer to implementation} pattern.}
\label{fig:pimpl}
\end{figure}

A related usage is the so-called \emph{pointer to implementation} pattern described in \cite{dickerson2018proof_carrying_smart_contracts}, which is illustrated in Figure~\ref{fig:pimpl}.
This pattern can be used to allow a contract to update its implementation, despite the immutable nature of a blockchain.
It consists of a contract \code{C}, which acts as a proxy for the functionality provided by a different contract, which resides at the address \code{X}.
This address is stored in the field \code{impl}, and \code{C} simply forwards all calls to methods $f_1$, \ldots{} $f_n$ to their actual implementations in \code{X}.
Finally, \code{C} provides an \code{Update} method, which allows the contract owner \code{Bob} to overwrite the address stored in \code{impl}.
Thus, if the actual implementation needs to be updated, \code{Bob} can deploy a new version of \code{X} and then simply update the pointer to the implementation in \code{C} without affecting any users of the contract.

Finally, as in the original presentation of \TINYSOL{}, we can also use our new formulation of the language to describe transactions and blockchains.
A \emph{transaction} is simply a call, where the caller is an account $A$, rather than a contract.
We denote this by writing $\TRANSACT{A}{X}{f}{\VEC{v}}{n}$, which expresses that the account $A$ calls the method $f$ on the contract (residing at address) $X$, with actual parameters $\VEC{v}$, and transferring $n$ units of currency with the call.
We can then model blockchains as follows:

\begin{definition}[Syntax of blockchains]
A \emph{blockchain} $B \in \BLOCKCHAINS$ is a list of initial contract declarations $DC$, followed by a sequence of transactions $T \in \TRANSACTIONS$:
\begin{center}
\begin{math}
  B \DCLSYM \code{$DC$ $T$} \qquad\qquad
  T \DCLSYM \epsilon \ORSYM \TRANSACT{A}{X}{f}{\VEC{v}}{n}, T
\end{math}
\end{center}

\noindent Notationally, a blockchain with an empty $DC$ will be simply written as the sequence of transactions.
\end{definition}

\subsection{Big-step semantics}\label{sec:tinysol_bigstep}
To define the semantics, we need some environments to record the bindings of 
variables (including the three magic variable names \code{this}, \code{sender} and \code{value}),
fields, methods, and contracts.
We define them as sets of partial functions as follows:

\begin{definition}[Binding model]
We define the following sets of partial functions:
\normalfont
\begin{align*}
  \ENV{V} \in \SETENV{V} & : \VNAMES \UNION \MVAR \PARTIAL \VALUES                  &
  \ENV{S} \in \SETENV{S} & : \ANAMES \PARTIAL \SETENV{F}            \\
 \ENV{F} \in \SETENV{F} & : \FNAMES \UNION \SET{\code{balance}} \PARTIAL \VALUES   
 &
  \ENV{T} \in \SETENV{T} & : \ANAMES \PARTIAL \SETENV{M}                  \\
\ENV{M} \in \SETENV{M} & : \MNAMES \PARTIAL \VNAMES^* \times \STM   
\end{align*}
We regard each environment $\ENV{J}$, for any $J \in \SET{V, F, M, S, T}$, as a (finite) set of pairs $(d, c)$ where $d \in \DOM{\ENV{J}}$ and $c \in \CODOM{\ENV{J}}$.
The notation $\ENV{J}\EXTEND{d}{c}$ denotes the update of $\ENV{J}$ mapping $d$ to $c$.
We write $\ENV{J}^{\EMPTYSET}$ for the empty $J$-environment.
To simplify the notation, when two or more environments appear together, we shall use the convention of writing the subscripts together (e.g.\@ $\ENV{MF}$ instead of $\ENV{M}, \ENV{F}$).
\end{definition}

Our binding model consists of two environments:
a \emph{method table} $\ENV{T}$, which maps addresses to method environments,
and a \emph{state} $\ENV{S}$, which maps addresses to lists of fields and their values.
Thus, for each contract, we have the list of methods it declares and its current state;
of course, the method table is constant, once all declarations are performed, whereas the state will change during the evaluation of a program.

\subsubsection{Declarations}
The semantics of the declarations builds the field and method environments, $\ENV{F}$ and $\ENV{M}$, and the state and method table $\ENV{S}$ and $\ENV{T}$.
We give the semantics in a classic big-step style 
and their defining rules are given in Figure~\ref{fig:semantics_declarations}.
We assume that the field and method names are distinct within each contract;
therefore, the rules in Figure~\ref{fig:semantics_declarations} define finite partial functions.

\subsubsection{Expressions}
Figure~\ref{fig:semantics_expressions} gives the semantics of expressions $e$.
Expressions have no side effects, so they cannot contain method calls, but can access both local variables and fields of any contract.
Thus, expression evaluations are of the form $\ENV{SV} \vdash e \trans_e v$, i.e.\@ they are relative to the state and variable environments.
We use $k$ to range over \code{this}, \code{sender}, \code{value} and variables $x$ (i.e. $k \in \DOM{\ENV{V}})$, and $q$ to range over \code{balance} and fields $p$ (i.e.\@ $q \in \DOM{\ENV{F}}$).

We do not give explicit rules for the boolean and integer operators subsumed under $\op$, but simply assume that they can be evaluated to a {\em unique} value by some semantics $\op(\VEC{v}) \trans_{\op} v$.\footnote{To simplify the definitions, we assume that all operations are total. Were this not the case, we would have needed some exception handling for partial operations (e.g., division by zero).} 
It follows that each expression evaluates to a unique value relative to some given state and variable environments.
Note that we assume that no operation is defined for addresses $X$, so we disallow any form of pointer arithmetic.

\begin{figure}
\begin{tabular}{cc}
\begin{minipage}{0.29\textwidth}
\center
\begin{semantics}
  \RULE[decf$_1$]
    { }
    { \CONF{\epsilon, \ENV{F}} \trans_{DF} \ENV{F} }

  \RULE[decm$_1$]
    { }
    { \CONF{\epsilon, \ENV{M}} \trans_{DM} \ENV{M} }

  \RULE[decc$_1$][ts_dec_c1]
    { }
    { \CONF{\epsilon, \ENV{ST}} \trans_{DC} \ENV{ST} }

  \RULE[decc$_2$][ts_dec_c2]
    {
        \CONF{DF, \ENV{F}^{\EMPTYSET}} \trans_{DF} \ENV{F} \AND
        \CONF{DM, \ENV{M}^{\EMPTYSET}} \trans_{DM} \ENV{M} \AND
        \CONF{DC, \ENV{ST}} \trans_{DC} \ENV{ST}'  
    }
    { \CONF{\code{contract $X$ \{ $DF$ $DM$ \} $DC$}, \ENV{ST}} \trans_{DC} \joine{(X, \ENV{F})}{\ENV{S}'}\ \ ,\, \joine{(X, \ENV{M})}{\ENV{T}' }}
\end{semantics}
\end{minipage}
~
&
~
\begin{minipage}{0.66\textwidth}
\center
\begin{semantics}
  \RULE[decf$_2$]
    { \CONF{DF, \ENV{F}} \trans_{DF} \ENV{F}' }
    { \CONF{\code{field $p$ := $v$;$DF$}, \ENV{F}} \trans_{DF} \joine{(p, v)}{\ENV{F}' }}

  \RULE[decm$_2$]
    { \CONF{DM, \ENV{M}} \trans_{DM} \ENV{M}' }
    { \CONF{\code{$f$($\VEC{x}$) \{ $S$ \} $DM$}, \ENV{M}} \trans_{DM} \joine{(f, (\VEC{x}, S))}{\ENV{M}' }}

\vspace*{2.2cm}

\end{semantics}
\end{minipage}
\end{tabular}
\caption{Semantics of declarations.}
\label{fig:semantics_declarations}
\end{figure}

\begin{figure}
\begin{tabular}{cc}
\!\!\!\!\!\!\begin{minipage}{0.35\textwidth}
\center
\begin{semantics}
  \RULE[var][Exp-Var]
    { k \in \DOM{\ENV{V}} \AND
      \ENV{V}(k) = v}
    { \ENV{SV} \vdash k \trans_e v }

  \RULE[op][Exp-Op]
    { \ENV{SV} \vdash \VEC{e} \trans_e \VEC{v} \AND \op(\VEC{v}) \trans_{\op} v}
    { \ENV{SV} \vdash \op(\VEC{e}) \trans_e v }

\end{semantics}
\end{minipage}
~
&
~
\begin{minipage}{0.64\textwidth}
\center
\begin{semantics}
  \RULE[val][Exp-Val]
    { }
    { \ENV{SV} \vdash v \trans_e v }
    
  \RULE[field][Exp-Field]
    { \ENV{SV} \vdash e \trans_e X \AND q \in \DOM{\ENV{S}(X)} \AND
      \ENV{S}(X)(q) = v}
    { \ENV{SV} \vdash e.q \trans_e v }
\end{semantics}
\end{minipage}
\end{tabular}
\caption{Semantics of expressions.}
\label{fig:semantics_expressions}
\end{figure}

\subsubsection{Statements}
The semantics of statements describes the actual execution steps of a program.
In Figure~\ref{fig:tinysol_semantics_statements_bss} we give the semantics in big-step style, where a step describes the execution of a statement in its entirety.
Statements can read from the method table and they can modify the state (i.e., the variable and field bindings).
The result of executing a statement is a new state, so transitions must here be of the form $\ENV{T} \vdash \CONF{S, \ENV{SV}} \trans_S \ENV{SV}'$ (recall that $\ENV{SV}'$ stands for $\ENV{S}', \ENV{V}'$), since both the field values in $\ENV{S}$ and the values of the local variables in $\ENV{V}$ may have been modified by the execution of $S$.

\begin{figure}[t]
\begin{center}
\begin{semantics}
  \RULE[skip][ts_bss_skip]
    {}
    { \ENV{T} \vdash \CONF{\code{skip}, \ENV{SV}} \trans_S \ENV{SV} }

  \RULE[seq][ts_bss_seq]
    { \ENV{T} \vdash \CONF{S_1, \ENV{SV}} \trans_S \ENV{SV}'' \AND \ENV{T} \vdash \CONF{S_2, \ENV{SV}''} \trans_S \ENV{SV}' }
    { \ENV{T} \vdash \CONF{\code{$S_1$;$S_2$}, \ENV{SV}} \trans_S \ENV{SV}' }

  \RULE[if][ts_bss_if]({b \in \SET{\TRUE, \FALSE}})
    { \ENV{SV} \vdash e \trans_e b \AND \ENV{T} \vdash \CONF{S_b, \ENV{SV}} \trans_S \ENV{SV}' }
    { \ENV{T} \vdash \CONF{\code{if $e$ then $S_{\TRUE}$ else $S_{\FALSE}$ }, \ENV{SV}} \trans_S \ENV{SV}' } 

  \RULE[while$_{\TRUE}$][ts_bss_whiletrue]
    {\ENV{SV} \vdash e \trans_e \TRUE \qquad
      \ENV{T}  \vdash \CONF{S, \ENV{SV}} \trans_S \ENV{SV}''  \qquad
      \ENV{T}  \vdash \CONF{\code{while $e$ do $S$}, \ENV{SV}''} \trans_S \ENV{SV}'
    }
    { \ENV{T} \vdash \CONF{\code{while $e$ do $S$}, \ENV{SV}} \trans_S \ENV{SV}' }

  \RULE[while$_{\FALSE}$][ts_bss_whilefalse]
    { \ENV{SV} \vdash e \trans_e \FALSE }
    { \ENV{T} \vdash \CONF{\code{while $e$ do $S$}, \ENV{SV}} \trans_S \ENV{SV} }

  \RULE[decv][ts_bss_decv]
    {x \notin \DOM{\ENV{V}} \qquad \ENV{SV} \vdash e \trans_e v \qquad
    \ENV{T} \vdash \CONF{S, \ENV{S}, \joine{(x, v)}{\ENV{V}}} \trans_S \ENV{S}'\ , \joine{(x, v')}{\ENV{V}'}
    }
    { \ENV{T} \vdash \CONF{\code{var $x$ := $e$ in $S$, \ENV{SV}}} \trans_S \ENV{SV}' }
   
  \RULE[assv][ts_bss_assv]
    { x \in \DOM{\ENV{V}} \qquad \ENV{SV} \vdash e \trans_e v }
    { \ENV{T} \vdash \CONF{\code{$x$ := $e$}, \ENV{SV}} \trans_S \ENV{S}, \ENV{V}\EXTEND{x}{v} }

  \RULE[assf][ts_bss_assp]
    { \ENV{V}(\code{this}) = X \AND \ENV{S}(X) = \ENV{F} \AND p \in \DOM{\ENV{F}} \AND \ENV{SV} \vdash e \trans_e v}
    { \ENV{T} \vdash \CONF{\code{this.$p$ := $e$}, \ENV{SV}} \trans_S \ENV{S}\EXTEND{X}{\ENV{F}\EXTEND{p}{v}}, \ENV{V} }

  \RULE[call][ts_bss_call]
    { 
  \begin{array}{l}
      \ENV{SV} \vdash e_1 \trans_e Y           
      \qquad
     \ENV{S}(Y) ~= \ENV{F}^Y 
     \qquad
     (\ENV{T}(Y))(f) ~= (\VEC{x}, S)
     \\
     |\VEC{x}| ~= |\VEC{e}| = h
    \qquad\qquad\!
    \ENV{SV} \vdash \VEC{e} \trans_e \VEC{v}
     \qquad\ \ \ 
      \ENV{SV} \vdash e_2 \trans_e n            
      \\
     \ENV{V}(\code{this}) ~= X 
     \qquad
     \ENV{S}(X) ~= \ENV{F}^X
     \qquad
     n ~\leq \ENV{F}^X(\code{balance})
    \\ 
    \ENV{S}'' ~= \ENV{S}%
    \EXTEND{X}{\ENV{F}^X [\code{balance -= } n]}%
    \EXTEND{Y}{\ENV{F}^Y [\code{balance += } n]}
     \\
    \ENV{V}'' ~= \{(\code{this}, Y) , (\code{sender}, X) , (\code{value}, n) , (x_1, v_1) , \ldots , (x_h, v_h) \}
     \\
      \ENV{T} \vdash \CONF{S, \ENV{SV}''} \trans_S \ENV{SV}'
    \end{array}
    }
    { \ENV{T} \vdash \CONF{\CALL{e_1}{f}{\VEC{e}}{e_2}, \ENV{SV}} \trans_S \ENV{S}', \ENV{V} }

\RULE[dcall][ts_bss_dcall]
    { 
    \begin{array}{l}
      \ENV{SV} \vdash e \trans_e Y  \qquad
       \ENV{T}(Y)(f) = (\VEC{x}, S)  \qquad
       |\VEC{x}|  = |\VEC{e}|  = h \qquad
      \ENV{SV} \vdash \VEC{e} \trans_e \VEC{v}                     
       \tabularnewline
       \ENV{V}''  = \{(\code{this}, \ENV{V}(\code{this})), (\code{sender}, \ENV{V}(\code{sender})),  (\code{value}, \ENV{V}(\code{value})), (x_1, v_1), \ldots, (x_h, v_h)\}
       \tabularnewline    
      \ENV{T}  \vdash \CONF{S, \ENV{S}, \ENV{V}''} \trans_S \ENV{SV}'
    \end{array}
    }
    { \ENV{T} \vdash \CONF{\DCALL{e}{f}{\VEC{e}}, \ENV{SV}} \trans_S \ENV{S}', \ENV{V} }
\vspace*{-.4cm}
\end{semantics}
\end{center}
\caption{Big-step semantics of statements in \TINYSOL.}
\label{fig:tinysol_semantics_statements_bss}
\end{figure}

Most of the rules are straightforward.
The rule \nameref{ts_bss_decv} is used when we declare a new variable $x$, with scope limited to the statement $S$; we implicitly assume alpha-conversion to handle the shadowing of an existing name.
In the premise, we evaluate the expression $e$ to a value $v$, and then execute the statement $S$ with a variable environment 
where we have added the pair $(x,v)$.
During the execution of $S$, this variable environment may of course be updated (by applications of the rule \nameref{ts_bss_assv}), which may alter any value in the environment, including $v$.
However, outside of the scope of the declaration, $x$ is not visible and so the pair $(x,v')$ is removed from the environment once $S$ finishes. 
In contrast, any other change that lead to $\ENV{V}'$ (as well as any change that lead to the global state $\ENV{S}'$) is retained.

The \nameref{ts_bss_call} rule is the most complicated, because we need to perform a number of actions: 
\begin{itemize}
\item First of all, we obviously have to evaluate the address and the parameters (viz., $e_1$, $\VEC{e}$ and $e_2$), relatively to the current execution environment $\ENV{SV}$; we use the obtained address $Y$ of the callee to retrieve the field environment $\ENV{F}^Y$ for this contract and, through the method table, to extract the list of formal parameters $\VEC{x}$ and the body of the method $S$. Also, we have to check that the number of actual parameters is the same as the number of formal parameters. \item We have to check that the \code{balance} of the caller is at least $n$, and, in that case, update the state environment by subtracting $n$ from the balance of $X$ and adding $n$ to the balance of $Y$, in their respective field environments; this yields a new state $\ENV{S}''$, where we write $\ENV{F}[\code{balance -= } n]$ and $\ENV{F}'[\code{balance += } n]$ for these two operations.
%
\item We create the new execution environment $\ENV{V}''$ by creating new bindings for the special variables \code{this}, \code{sender} and \code{value}, and by binding the formal parameters $\VEC{x}$ to the values of the actual parameters $\VEC{v}$. 
\item Finally, we execute the statement $S$ in this new environment. 
This yields the new state $\ENV{S}'$, and also an updated variable environment $\ENV{V}'$, since $S$ may have modified the bindings in $\ENV{V}''$. 
However, these bindings are local to the method and therefore we discard them once the call finishes.
So, the result of this transition is the updated state ($\ENV{S}'$) and the original variable environment of the caller ($\ENV{V}$).
\end{itemize}
It should be noted that a \emph{local} method call, i.e.\@ a call to a method within the same (calling) contract, is merely a special case of the rule \nameref{ts_bss_call}.
Local calls would have the form $\CALL{\code{this}}{f}{\VEC{e}}{0}$, as transferring any amount of currency will not alter the balance of the contract.

Finally, rule \nameref{ts_bss_dcall} accounts for delegate calls; there are a few things to note about the semantics of this construct.
First, the method body $S$ is executed in the context of the caller; i.e.\@ the magic variables \code{this}, \code{sender} and \code{value} are \emph{not} rebound. In particular, this means that any access to fields (through \code{this}) are to fields in the \emph{caller} contract, rather than in the contract $Y$ (where $f$ is defined).
Second, there is no \code{value} parameter, so funds cannot be transferred to the callee with this type of call. This is consistent with the idea that the code is executed in the context of the caller; i.e.\@ \emph{as if} the method had been declared in the caller contract. With the conventional \code{call} construct, calls to local methods \emph{can} in principle transfer funds, but this has no effect, since the caller and callee are the same contract.
Finally, since delegate calls are similar to local calls, except for the fact that the actual code resides in another contract, it therefore makes sense that there is no \code{value} parameter in the call syntax.

\begin{figure}[t]
\begin{tabular}{cc}
\begin{minipage}{0.4\textwidth}
\center
\begin{semantics}
  \RULE[Genesis][ts_genesis]
    { \CONF{DC, \ENV{ST}^{\EMPTYSET}} \trans_{DC} \ENV{ST} }
    { \CONF{\code{$DC$ $T$}, \ENV{ST}^{\EMPTYSET}} \trans_B \CONF{T, \ENV{ST}} }

  \RULE[Trans][ts_trans]
    { \ENV{T} \vdash \CONF{\CALL{X}{f}{\VEC{v}}{n}, \ENV{S}, \{(\code{this}, A)\} } \trans_S \ENV{S}', \ENV{V} }
    { \CONF{\TRANSACT{A}{X}{f}{\VEC{v}}{n}, T, \ENV{ST}} \trans_B \CONF{T, \ENV{S}', \ENV{T}} }
\end{semantics}
\end{minipage}
~
&
~
\begin{minipage}{0.5\textwidth}
\center
\begin{semantics}
  \RULE[Revelation][ts_revel]
    { }
    { \CONF{\epsilon, \ENV{ST}} \trans_B \ENV{ST} }
\vspace*{1.2cm}
\end{semantics}
\end{minipage}
\end{tabular}
\caption{Semantics of blockchains.}
\label{fig:semantics_blockchains}
\end{figure}

\subsubsection{Transactions and blockchains}
The semantics for blockchains is given as a transition system defined by the rules given in Figure~\ref{fig:semantics_blockchains}.
Here, the rule \nameref{ts_genesis} describes the `genesis event' where contracts are declared, whilst \nameref{ts_trans} describes a single transaction.
This is thus a \emph{small-step} semantics, invoking the big-step semantics for declarations and statements in the premises of its rules.
We remark that the rules of the operational semantics for blockchains (as well as those for statements presented above) define a {\em deterministic} transition relation. 

Note that, unlike in the original formulation of \TINYSOL{}, we do not include a rule like [Tx2] in~\cite{bartoletti2019tinysol} for rolling back a transaction in case it is non-terminating or it aborts via a \code{throw} command. 
Such a rule would require a premise that cannot be checked effectively for a Turing-complete language like \TINYSOL{} and therefore we omit it, since it is immaterial for the main contributions we give in this paper.\footnote{%
For instance, rule [Tx2] in~\cite{bartoletti2019tinysol} has an undecidable premise that checks whether the execution of the body of a contract does \emph{not} yield a final state. 
It is debatable whether such rules should appear in an operational semantics.}
In practice, termination of Ethereum smart contracts is ensured via a `gas mechanism' and is assumed by techniques for the formal analysis of smart contracts (see, e.g., \cite{AGLM/2024/reocas/outofgas}). 
However, as observed in, for instance,~\cite{GenetJS20}, proof of termination for smart contracts is non-trivial even in the presence of a `gas mechanism.' 
In the aforementioned paper, the authors present the first mechanised proof of termination of contracts written in EVM bytecode using minimal assumptions on the gas cost of operations (see the study~\cite{YangMRP19} for an empirical analysis of the effectiveness of the `gas mechanism' in estimating the computational cost of executing real-life transactions). 
We leave for future work the adaption of the results we present in this paper to the setting of \TINYSOL{} with the `gas mechanism' presented in \cite{AGLM/2024/reocas/outofgas}.


\section{Call integrity and noninterference in \TINYSOL}\label{sec:call_integrity_tinysol}
Grishchenko et al.\@ \cite{grishchenko2018} formulate the property of \emph{call integrity} for smart contracts written in the language EVM, which is the `low-level' bytecode of the Ethereum platform, and the target language to which e.g.\@ Solidity compiles.
They then prove \cite[Theorem 1]{grishchenko2018} that this property suffices to rule out reentrancy phenomena, as those described in the example in Figure~\ref{fig:reentrancy}.
We first formulate a similar property for \TINYSOL{}; this requires a few preliminary definitions.

\begin{definition}[Trace semantics]
\label{def:traces}
A \emph{trace} of method invocations is given by
\begin{center}
\begin{syntax}[h]
  \pi \IS \epsilon \OR \TRANSACT{X}{Y}{f}{\VEC{v}}{n}, \pi 
\end{syntax}
\end{center}
where $X$ is the address of the calling contract, $Y$ is the address of the called contract, $f$ is the method name, and $\VEC{v}$ and $n$ are the actual parameters.
We annotate the big-step semantics with a trace containing information on the invoked methods to yield labelled transitions of the form $\trans[\pi]_S$. 
To do this, we modify the rules in Table~\ref{fig:tinysol_semantics_statements_bss} as follows:
\begin{itemize}
\item in rules \nameref{ts_bss_skip}, \nameref{ts_bss_whilefalse}, \nameref{ts_bss_assv} and \nameref{ts_bss_assp}, every occurrence of $\ \trans_S\ $ becomes $\ \trans[\epsilon]_S\ $;
\item in rules \nameref{ts_bss_if} and \nameref{ts_bss_decv}, every occurrence of $\ \trans_S\ $ becomes $\ \trans[\pi]_S\ $;
\item rules \nameref{ts_bss_seq}, \nameref{ts_bss_whiletrue}, \nameref{ts_bss_call} and \nameref{ts_bss_dcall} respectively become:
\begin{center}\normalfont
\begin{math}
\begin{array}{l}
  \dfrac
    {
      \ENV{T}  \vdash \CONF{S_1, \ENV{SV}} \trans[\pi_1]_S \ENV{SV}'' \qquad
      \ENV{T}  \vdash \CONF{S_2, \ENV{SV}''} \trans[\pi_2]_S \ENV{SV}'
    }
    { \ENV{T} \vdash \CONF{\code{$S_1$;$S_2$}, \ENV{SV}} \trans[\pi_1,\pi_2]_S \ENV{SV}' }
\vspace*{.2cm}
\\
  \dfrac
    {
      \ENV{SV}  \vdash e \trans_e \TRUE \qquad
      \ENV{T}   \vdash \CONF{S, \ENV{SV}} \trans[\pi_1]_S \ENV{SV}'' \qquad
      \ENV{T}   \vdash \CONF{\code{while $e$ do $S$}, \ENV{SV}''} \trans[\pi_2]_S \ENV{SV}'
    }
    { \ENV{T} \vdash \CONF{\code{while $e$ do $S$}, \ENV{SV}} \trans[\pi_1,\pi_2]_S \ENV{SV}' }
\vspace*{.2cm}
\\
  \dfrac
    { \ldots \qquad \ENV{T} \vdash \CONF{S, \ENV{SV}''} \trans[\pi]_S \ENV{SV}' }
    { \ENV{T} \vdash \CONF{\CALL{e_1}{f}{\VEC{e}}{e_2}, \ENV{SV}} \trans[\TRANSACT{X}{Y}{f}{\VEC{v}}{n},\pi]_S \ENV{S}', \ENV{V} }
\vspace*{.2cm}
\\
  \dfrac
    { \ldots \qquad \ENV V(\code{this}) = X \qquad  \ENV{T} \vdash \CONF{S, \ENV{S}, \ENV{V}''} \trans[\pi]_S \ENV{SV}' }
    { \ENV{T} \vdash \CONF{\DCALL{e}{f}{\VEC{e}}, \ENV{SV}} \trans[\TRANSACT{X}{Y}{f}{\VEC{v}}{0},\pi]_S \ENV{S}', \ENV{V} }
\end{array}
\end{math}
\end{center}
\end{itemize}

\noindent
We extend this annotation to the semantics for blockchains and write $\trans[\pi]_B$ for this annotated relation.
\end{definition}

\begin{definition}[Projection]
The \emph{projection} of a trace to a specific contract $X$, written $\pi \BARBSYM_X$, is the trace of calls with $X$ as the calling address. 
Formally:
\begin{equation*}
\epsilon \BARBSYM_X \ =\ \epsilon
\qquad\qquad
  (\TRANSACT{Z}{Y}{f}{\VEC{v}}{n}, \pi) \BARBSYM_{X} \ =\ %
  \begin{cases}
    \TRANSACT{X}{Y}{f}{\VEC{v}}{n}, (\pi \BARBSYM_{X}) & \text{if $Z = X$} \tabularnewline
    \pi \BARBSYM_{X}                                   & \text{otherwise} 
  \end{cases} 
\vspace*{-.6cm}
\end{equation*}
\end{definition}

\smallskip
Notationally, given a (partial) function $f$, we write $f|_{\SETNAME X}$ for denoting the restriction of $f$ to the subset $\SETNAME X$ of its domain.

\begin{definition}[Call integrity]\label{def:call_integrity}
Let \SETNAME{A} denote the set of all contracts (addresses),
$\SETNAME{X} \subseteq \SETNAME{A}$ denote a set of trusted contracts, 
$\SETNAME{Y} \DEFSYM \SETNAME{A} \SETMINUS \SETNAME{X}$ denote all other contracts, and
$\ENV{ST}^{\SETNAME{X}}$ have domain $\SETNAME{X}$.
A contract $C \in \SETNAME{X}$ has \emph{call integrity} for $\SETNAME{Y}$ if,
for every transaction $T$
and environments $\ENV{ST}^1$ and $\ENV{ST}^2$ such that
$\ENV{ST}^1|_{\SETNAME{X}} = \ENV{ST}^2|_{\SETNAME{X}} = \ENV{ST}^{\SETNAME{X}}$,
it holds that
\begin{equation*}
           \CONF{T, \ENV{ST}^1} \trans[{\pi_1}]_B \ENV{ST}^{1'} 
  \land    \CONF{T, \ENV{ST}^2} \trans[{\pi_2}]_B \ENV{ST}^{2'} 
  \implies \pi_1\! \BARBSYM_C\ = \pi_2 \!\BARBSYM_C
  \qedhere
\end{equation*}
\end{definition}

The definition is quite complicated and contains a number of elements:
\begin{itemize}
  \item $C$ is the contract of interest.
  \item $\SETNAME{X}$ is a set of \emph{trusted} contracts, which we assume are allowed to influence the behaviour of $C$.
  This set must obviously contain $C$, since $C$ must at least be assumed to be trusted.
  Thus, a contract $C$ can have call integrity for all contracts if $\SETNAME{X} = \SET{C}$.

  \item Conversely, the set $\SETNAME{Y} = \SETNAME{A} \SETMINUS \SETNAME{X}$ is the set of addresses of all contracts that are \emph{untrusted}.\footnote{%
Note that this is formulated inversely by Grishchenko et al., who instead formulate the property for a set of \emph{untrusted} contracts $\SETNAME{A}_{\SETNAME{C}}$, corresponding to \SETNAME{Y} in the present formulation. However, using the set of \emph{trusted} addresses \SETNAME{X} seems more straightforward.}

  \item $\ENV{ST}^1$ and $\ENV{ST}^2$ are any two pairs of method/field environments that coincide (both in the code and in the values) for all the trusted contracts.\footnote{This too is inversely formulated by Grishchenko et al.}
  The point is that the contracts in \SETNAME{X} are assumed to be known, and hence invariant, whereas any contract in \SETNAME{Y} is assumed to be unknown and may be controlled by an attacker.
  Thus, we are actually quantifying over all possible contexts where the contracts in \SETNAME{X} can be run.

  \item $T$ is any transaction;
  it may be issued from any account and to any contract.
  Thus we also quantify over all possible transactions, since an attacker may request an arbitrary transaction, that is thus part of the execution context as well.
\end{itemize}

The call integrity property intuitively requires that, if we run the trusted part of the code in any execution context, then the behavior of $C$ remains the same, i.e. $C$ must make exactly the same method calls (and in exactly the same order).
Thus, to \emph{disprove} that $C$ has call integrity, it suffices to find two environments and a transaction that will induce a difference in the call trace of $C$.

The idea underlying call integrity is that the behaviour of $C$ should not depend on any untrusted code (i.e.\@ contracts in \SETNAME{Y}), even if control is transferred to a contract in \SETNAME{Y}.
The latter could, for example, happen if $C$ calls a method on $B \in \SETNAME{X}$, and $B$ then calls a method on a contract in \SETNAME{Y}.
This also means that $C$ cannot directly call any contract in \SETNAME{Y}, since that can only happen if $C$ calls a method on a contract, where the address is received as a parameter, or if it calls a method on a `hard-coded' contract address.
In both cases, we can easily pick up two environments that are able to induce different behaviors, for example by choosing a non-existing address for one context (in the first case) or by ensuring that no contract exists on the hard-coded address in one context (in the second case).
The latter possibility can seem somewhat contrived, especially if we assume that all contracts are created at the genesis event, and it might therefore be reasonable to require also that $\DOM{\ENV{ST}^1} = \DOM{\ENV{ST}^2}$, so that we at least assume that contracts exist at the same addresses.
However, on an actual blockchain, new contracts can be deployed (and in some cases also deleted) at any time, and if such a degree of realism is desired, this extra constraint should not be imposed.

\begin{remark}
\label{rem:dcall}
We want to remark here that a delegate call is recorded in the trace exactly as if it was a normal call (see the cases for \nameref{ts_bss_call} and \nameref{ts_bss_dcall}  in Definition~\ref{def:traces}).
This is what is prescribed in \cite{grishchenko2018} for defining call integrity.
Since delegate calls do not modify the balance fields (but only execute the code in the context of the caller), we believe that we could avoid including delegate calls in the traces and still obtaining a notion of ‘call-integrity' that would prevent reentrancy, since the control flow does not leave the caller contract.
However, we prefer to remain faithful to the original definition of \cite{grishchenko2018} and to the associated proof that call-integrity implies non-reentrancy.
\end{remark}

The main problem with the definition of call integrity is that it relies on a universal quantification over all possible execution contexts. 
This makes it difficult to check in practice.
However, our previous discussion indicates that call integrity may intuitively be viewed as a form of \emph{noninterference} between the trusted and the untrusted contracts.
We now see to what extent this intuition is true and formally compare the two notions.

First of all, we consider a basic lattice of security levels, made up by just two levels, namely $H$ (for \emph{high}) and $L$ (for \emph{low}), with $L < H$. 
We label every contract as high or low through a contracts-to-levels mapping $\lambda: \SETNAME{A} \rightarrow \{L,H\}$; this induces a bipartition of the contract names $\SETNAME{A}$ into the following sets:
\begin{center}
\begin{math}
  \SETNAME{L} = \SET{X \in \SETNAME{A} | \lambda(X) = L} \qquad\qquad
  \SETNAME{H} = \SET{X \in \SETNAME{A} | \lambda(X) = H} 
\end{math}
\end{center}

In this way, we create a bipartition of the state into low and high, corresponding to the fields of the low and high contracts, respectively. 
Then, we define \emph{low-equivalence} $=_L$ to be the equivalence on states such that $\ENV{S}^1 =_L \ENV{S}^2$ if and only if $\ENV{S}^1(X) = \ENV{S}^2(X)$, for every $X \in \SETNAME{L}$.

We can now adapt the notion of noninterference for multi-threaded programs by Smith and Volpano \cite{SV98} to the setting of \TINYSOL{}.

\begin{definition}[Noninterference]\label{def:noninterference}
Given a contracts-to-levels mapping $\lambda: \SETNAME{A} \rightarrow \{L,H\}$ and a contract environment $\ENV{T}$, the contracts satisfy \emph{noninterference} if, for every $\ENV{S}^1$ and $\ENV{S}^2$ and for every transaction $T$ such that
$$
  \ENV{S}^{1} =_L \ENV{S}^{2}
  \qquad
  \CONF{T, \ENV{S}^1, \ENV{T}} \trans_B \ENV{S}^{1'},\ENV{T} 
  \qquad
  \CONF{T, \ENV{S}^2, \ENV{T}} \trans_B \ENV{S}^{2'},\ENV{T}
$$

\noindent it holds that $\ENV{S}^{1'} =_L \ENV{S}^{2'}$.
\end{definition}

\begin{remark}[Incomparability]\label{rem:incomparability}
Call integrity and noninterference seem strongly related, in the sense that the first requires that the behaviour of a contract is not influenced by the (bad) execution context, whereas the second one requires that a part of the computation (the `low' one) is not influenced by the remainder context (the `high' one). 
So, one may try to prove a statement like:
``$C \in \SETNAME{X}$ has call integrity for $\SETNAME{Y} \DEFSYM \SETNAME{A} \SETMINUS \SETNAME{X}$ if and only if it satisfies noninterference w.r.t.\@ $\lambda$ such that $\SETNAME{L} = \SETNAME{X}$ and $\SETNAME{H} = \SETNAME{Y}$.''
However, both directions are false.

For the direction from right to left, consider:
\begin{lstlisting}[style=tinysol]
contract X {                contract Y {
  field balance = 0;           field balance = v;
  go() { skip }                go() { X.go()$@\kern-0.5em@ this.balance }
}                           }
\end{lstlisting}

\noindent where \code{X} is trusted and \code{Y} untrusted. 
Since \code{X} cannot invoke any method, this example satisfies call integrity. 
However, it does not satisfy noninterference. To see this, consider two environments, one assigning 1 to $\code Y$'s balance and the other one assigning 0, and the transaction $\TRANSACT{Y}{Y}{go}{}{0}$.

For the direction from left to right, consider the following:
\begin{lstlisting}[style=tinysol]
contract X {                   contract Y {                 contract Z {
  go() {                         field balance = v;           a() { skip }    
      if Y.balance = 0         }                              b() { skip }
      then @\CALL{\code Z}{\code a}{}{0}@                                           }
      else @\CALL{\code Z}{\code b}{}{0}@
  }
}
\end{lstlisting}

\noindent Assuming that both \code{X} and \code{Z} are low, the example satisfies noninterference: there is no way for \code{Y} to influence the low memory. 
By contrast, the code does not satisfy call integrity. 
Indeed, let \code{v} be 0 in one environment and 1 in the other, and consider $T$ to be $\TRANSACT{X}{X}{go}{}{0}$: in the first environment, it generates $\TRANSACT{X}{Z}{a}{}{0}$, whereas in the second one it generates $\TRANSACT{X}{Z}{b}{}{0}$.
\end{remark}

\section{A type system for noninterference and call integrity}\label{sec:typesystem_noninterference}
As demonstrated in Remark~\ref{rem:incomparability}, call integrity and noninterference are incomparable properties.
This is so because noninterference is a 2-property on the pair of \emph{stores} ($\ENV{S}^{1'}, \ENV{S}^{2'}$) resulting from two different executions, whereas call integrity is a 2-property on the pair of \emph{call traces} ($\pi_1, \pi_2$) generated during two executions.
However, the two properties have an interesting overlap, because an outgoing currency flow (i.e.\@ a method call) may also result, at least potentially, in a change of the stored values of the \code{balance} fields of the sender and of the recipient.
Therefore, every method call is \emph{also} an information flow between the two contracts involved, even when no amount of currency is transferred.
In \cite{volpano2000secure_flow_typesystem}, Volpano et al.\@ devise a type system for checking information flows, which, as they show, yields a sound approximation to noninterference.
In the following, we create an adaptation of this type system for \TINYSOL{} and show that it may \emph{also}  be used to soundly approximate call integrity.

\subsection{Types,  type environments, and typed syntax}
We begin by assuming a finite lattice $(\SECLEVELS, \ordleq)$ consisting of a set of \emph{security levels} $\SECLEVELS$, ranged over by $s$, and equipped with a partial order $\ordleq$ with $\SBOT$ and $\STOP$ as least and largest elements.
In the simplest setting, we can let $\SECLEVELS \DEFSYM \SET{L, H}$ (for `low' and `high') and define $L \ordleq L$, $L \ordleq H$, and $H \ordleq H$.
This is sufficient for ensuring bipartite noninterference, but the type system can also handle more fine-grained security control.

Besides the security-level aspect, we shall also need a conventional data-type aspect of our types, so that we can assign types to addresses representing the contract residing on that address.
For this, we shall borrow from conventional object-oriented type systems, similar to the type system for Featherweight Solidity given by Crafa, Di Pirro and Zucca in \cite{crafa2019featherweight_solidity}.
Specifically, we let the type of an address be an \emph{interface} $I$, consisting of the signatures of the fields and methods of the contract residing on that address.
To give our types a uniform format, we also include $\TINT$ and $\TBOOL$ as base types, and therefore let a data type be a pair $(B, s)$ consisting of a base type $B$ and a security level $s$.
We shall use the notations $(B, s)$ and $B_s$ interchangeably, preferring the latter except when we need indices on types.
The types are as follows:

\begin{definition}[Syntax of types and environments]
We use the following types and environments:
\begin{syntax}[h]
  B \in \BASETYPES             \IS \TINT \OR \TBOOL \OR I                                              \\
  T \in \TYPES                 \IS B_s \OR \TVAR{B_s} \OR \TCMD{s} \OR \TSPROC{\BSVEC}{s}              \\
  n \in \NAMES                 \IS \VNAMES \UNION \FNAMES \UNION \MNAMES \UNION \ANAMES \UNION \TNAMES \\
  \Gamma, \Delta \in \TYPEENVS \IS \NAMES \PARTIAL \TYPES \UNION \TYPEENVS                             \\
  \Sigma                       \IS \TNAMES \PARTIAL \TNAMES 
\end{syntax}
where $\VNAMES$ are variable names $x$, $\FNAMES$ are field names $p$, $\MNAMES$ are method names $f$, $\ANAMES$ are contract names $X$, and $\TNAMES$ are interface names $I$.
We write $\VEC{T}$ for a tuple of types $(T_1, \ldots, T_n)$.
\end{definition}

Note that for the purpose of the type system, unless otherwise noted, we shall assume that the four `magic names' $\MVAR$ are contained in the respective sets of field and variable names; i.e.\@ $\code{balance} \in \FNAMES$ and $\code{this}, \code{sender}, \code{value} \in \VNAMES$.

The meaning of the types is as follows:
\begin{itemize}
  \item The type $B_s$ is given to expressions $e$.
    It denotes that all \emph{reads} within $e$ are from containers (i.e.\@ variables and fields) of level $s$ \emph{or lower}, and the resulting value is of type $B$.

  \item The type $\TVAR{B_s}$ is given to containers.
    It denotes that a field $p$ or variable $x$ is capable of storing a value of type $B$ of level $s$ \emph{or lower}.
    
  \item $\TCMD{s}$ is a \emph{phrase type} given to \emph{code}, i.e.\@ commands $S$.
    It denotes that all \emph{assignments} in the code are made to containers whose security level is $s$ \emph{or higher}.

  \item The type $\TSPROC{\BSVEC}{s}$ is given to methods $f$. 
    It denotes that the body $S$ can be typed as $\TCMD{s}$, under the assumption that the formal parameters $\VEC{x}$ have types $\VEC{\TVAR{B_s}}$ (that is, a sequence of types $\TVAR{B_1, s_1}, \ldots, \TVAR{B_h, s_h}$).
    Note that every method declaration contains an implicit write to the \code{balance} field of the containing contract: hence, given the meaning of $\TCMD{s}$, this also means that the security level of \code{balance} must always be $s$ \emph{or higher} than the level of any method declared in an interface.
\end{itemize}

The meaning of the base types $B$ is mostly straightforward:
for integer and boolean values, we assume that the base type can be determined directly by observing the value itself; i.e.\@ $5$ is an integer and $\FALSE$ is a boolean.
However, with addresses $X \in \ANAMES$, the situation is different, as there is no inherent relationship between the address itself and its type, since addresses are \emph{declared} to have a type (and a security level), according to the contract located at the address.
The type of an address $X$ is therefore an \emph{interface} $I$, 
consisting of the signatures of the methods and fields of that contract.
The interface could in principle be extracted from the contract definition itself; however, here we prefer to let interfaces be defined separately, since it will allow multiple contracts to implement the same interface.
Interfaces are used both in the type system and to declare a tree-like hierarchy of contracts (similar to the class hierarchy of Java), mirroring the syntax of contract declarations. 
They are defined as follows:

\begin{definition}[Interface definition language]
\label{def:interfaces}
Interface declarations $ID$ are given by the language

\begin{fixqed}
\begin{syntax}[h]
  ID \IS \epsilon \OR \code{interface $I_1$ : $I_2$ \{ $IF$ $IM$ \}}\ \ ID \\
  IF \IS \epsilon \OR p : \TVAR{B_s}\ \ IF                                 \\
  IM \IS \epsilon \OR f : \TSPROC{\BSVEC}{s}\ \ IM
\end{syntax}
\end{fixqed} \qedhere
\end{definition}

In the declaration \code{interface $I_1$ : $I_2$ \{ $IF$ $IM$ \}}, we have that $I_1$ is the name of the interface and $I_2$ is the name of its parent in the hierarchy.
For the sake of simplicity, we assume that the members common to both interfaces also appear in both the parent and the child, i.e.\@ members are not automatically inherited.
We also assume that all contracts implement only a single interface.\footnote{We could also allow contracts to implement multiple interfaces; this would require the type of an address to be a \emph{set} of interfaces, rather than just a single interface. However, we shall forego that here to avoid complicating the presentation.}
Furthermore, we assume the existence of an interface $\ITOP$, corresponding to the basic implementation of a contract, which acts as the single root of the inheritance tree, corresponding to the following definition:

\begin{lstlisting}[style = tinysol]
interface @$\ITOP$ : $\ITOP$@ {
  balance : @$\TVAR{\TINT, \STOP}$@
  send    : @$\TSPROC{}{\SBOT}$@
}
\end{lstlisting}

\noindent 
This definition may look strange, since it specifies a self-inheritance relation; however, we admit it since $\ITOP$ is the root of the inheritance tree, and it alone will not inherit from anything. Moreover, the inclusion of a contract `supertype' $\ITOP_{\STOP}$ is similar to what is done in the type system developed for Featherweight Solidity by Crafa et al.\@ in \cite{crafa2019featherweight_solidity}.
This is necessary to enable us to give a type to the `magic variable' \code{sender}, which is available within the body of every method, since this variable can be bound to the address of any contract or account.
In the following section, we shall give a definition of a subtyping relation that will ensure that, for any valid interface definition $I$ (containing at least \code{balance} and \code{send}) and any security level annotation $s$, it must hold that $I_s$ is a subtype of $\ITOP_{\STOP}$, thus always allowing us to type $I_s$ up to $\ITOP_{\STOP}$.

We shall also use a \emph{typed} syntax of \TINYSOL{}, where local variables are now declared as
\begin{center}
  \code{$\TVAR{B_s}$ $x$ := $e$}
\end{center}
where $B$ is the type of the value of the expression $e$.
Likewise, we add annotated type names $I_s$ to contract declarations:
\begin{center}
  \code{contract $X$ : $I_s$ \{ $DF$ $DM$ \}}
\end{center}

\noindent where $I$ is a declared type name.
Note that the security level is given on the \emph{contract}, rather than on the interface declaration. 
This is intentional since multiple contracts may implement the same interface, but nevertheless be categorised into different security levels.

Finally, type environments are used to record the interface hierarchy and other type assignments. 
In particular, we use:
\begin{itemize}
  \item $\Sigma : \TNAMES \PARTIAL \TNAMES$ records the immediate parent of an interface;
    thus,  $\Sigma(I_1) = I_2$ whenever we have the declaration \code{interface $I_1$ : $I_2$ \{ $IF\ IM$ \}}.

  \item $\Gamma  \in \TENV \DCLSYM ((\FNAMES \UNION \MNAMES \UNION \ANAMES) \PARTIAL \TYPES) \UNION (\TNAMES \PARTIAL \TENV)$ records the statically declared definitions (for fields, methods, and contracts) and the types of interfaces.
    We store interfaces in $\Gamma$ by letting interface names $I$ return another type environment $\Gamma_I$, which records the signatures of fields and methods declared in interface $I$.
    Thus, if a contract with address $X$ implements an interface $I$, we will have that $\Gamma(X) = I_s$, for some security level $s$, and $\Gamma(I) = \Gamma_I$, where $\Gamma_I$ records the type annotations prescribed by $IF$ and $IM$, provided that $I$ is declared as \code{interface $I$ : $I'$ \{ $IF\ IM$ \}}.

  \item $\Delta : \VNAMES \PARTIAL \TYPES$ records the types of local variables.
    We record these in a separate environment since local variables are declared at runtime.
    Thus, this environment may change according to the runtime evolution of the program, whereas the other environments are static.
\end{itemize}

For type environments of kind $\Gamma$, we further assume the following well-formedness condition:

\begin{definition}[Well-formedness of $\Gamma$]\label{def:wellformedness_gamma}
We say that a type environment $\Gamma$ is \emph{well-formed}, if the following hold:
\begin{itemize}
  \item for each contract name $X$, if $X \in \DOM{\Gamma}$, then there exists $I \in \TNAMES$ such that $\Gamma(X) = I_s$ and $I \in \DOM{\Gamma}$;
  \item for each interface name $I$, if $I \in \DOM{\Gamma}$, then there exists $\Gamma_I \in \TENV$ such that $\Gamma(I) = \Gamma_I$.
  \qedhere
\end{itemize}
\end{definition}

We can now use a type environment $\Gamma$ to assign types to values, as follows:

\begin{definition}[\textsc{TypeOf}]\label{def:typeof}
The partial function $\TYPEOF{v}$ from values to types is defined as
\begin{equation*}
  \TYPEOF{v} = 
  \begin{cases}
    (\TINT, \SBOT)  & \text{if $v \in \INTEGERS$} \\
    (\TBOOL, \SBOT) & \text{if $v \in \BOOLEANS$} \\
    \Gamma(v)       & \text{if $v \in \ANAMES$ and $v \in \DOM{\Gamma}$}
  \end{cases}
\end{equation*}

\noindent and is undefined, otherwise.
\end{definition}

$\TYPEOF{\cdot}$ assigns basic types to values in an obvious way, based on the set they belong to.
However, since the type of a value has both a \emph{data} aspect and a \emph{security} aspect, we also need to assign a security level. 
Thus, we choose the lowest possible level ($\SBOT$) for the `pure' data values (integers and booleans), since they have no inherent security level attached to them. 
Hence, it is always safe to give them the lowest possible level and then coerce it up to a higher level via subtyping (see later on), when we write the value into a variable/field. 
Thence, the level will be determined by the level of the variable/field. 
In contrast, the interface and the security level of address names are provided directly by the specified type environment $\Gamma$.



\subsection{Subtyping}
$\Sigma$ gives a \emph{static} representation of the subtype-supertype relationship for interfaces; so, we need a way to decide whether a given $\Sigma$ is \emph{consistent} w.r.t.\@ this relationship.
This will be done through a set of rules for deciding whether a given interface $I_1$, declaring that it inherits from another interface $I_2$, indeed also is a \emph{subtype} of $I_2$.
The rules are given in Figure~\ref{fig:consistency_rules_sigma}; the judgment is of the form $\Sigma; \Gamma \vdash \Sigma'$, which expresses that the inheritance tree recorded in $\Sigma'$ is \emph{consistent} with the actual declarations recorded in $\Sigma$ and $\Gamma$.

\begin{figure}\centering
\begin{minipage}{0.25\textwidth}
\begin{semantics}
  \RULE[$\Sigma$-top][cons_sigma_top]
    { \begin{array}{l} \\ \ \\ \end{array}}
    { \Sigma; \Gamma \vdash (\ITOP, \ITOP) }
\end{semantics}
\end{minipage}
\begin{minipage}{0.7\textwidth}
\center
\begin{semantics}
  \RULE[$\Sigma$-rec][cons_sigma_rec]
    {\begin{array}{l} 
    I_1 \neq I_2 \qquad\qquad
    \not\exists I' . \Sigma'(I') = I_1 \qquad\qquad\quad
    \Gamma(I_1)  = \Gamma_1 \qquad\qquad
    \Gamma(I_2)  = \Gamma_2 
    \tabularnewline
     \DOM{\Gamma_2} \subseteq \DOM{\Gamma_1} \qquad
     \forall n \in \DOM{\Gamma_2} . \Sigma \vdash \Gamma_1(n) \SUBS \Gamma_2(n) \qquad
    \Sigma; \Gamma \vdash \Sigma'
    \end{array}
    }
    { \Sigma; \Gamma \vdash (I_1, I_2), \Sigma' }
\end{semantics}
\end{minipage}
\caption{Consistency rules for $\Sigma$.}
\label{fig:consistency_rules_sigma}
\end{figure}

The rules are quite simple, since they just iterate through the inheritance environment, examining each entry in turn.
$\Sigma$ only records the inheritance hierarchy, while the actual structures of the interfaces are recorded in $\Gamma$; therefore, consistency of any subpart of $\Sigma$ must be judged relative to $\Gamma$, as well as relative to the full $\Sigma$.
The base case is the rule \nameref{cons_sigma_top}, since we assume that the interface $\ITOP$ always exists and inherits from nothing besides itself; furthermore, the form of the conclusion ensures that this rule can only be applied to an inheritance tree consisting of the root node itself.
The recursive case is the rule \nameref{cons_sigma_rec}, which examines the entry $(I_1, I_2)$, for $I_1 \neq I_2$ (since no interface except $\ITOP$ may inherit from itself).
We also require that there must not exist another interface $I'$ such that $I'$ inherits from $I_1$ according to the remainder of the inheritance environment $\Sigma'$; this ensures that  the environment indeed describes a \emph{tree} and that \nameref{cons_sigma_rec} is only applicable when $I_1$ is a  \emph{leaf node} of the current inheritance tree, which is then trimmed away as we recur in the premise.
Moreover, the remaining premises of \nameref{cons_sigma_rec} ensure that all members of the supertype ($I_2$) also exist in the subtype ($I_1$) and, for each of them, that the subtyping judgment $\Sigma \vdash \Gamma_1(n) \SUBS \Gamma_2(n)$ holds.

\begin{figure}\centering
\begin{minipage}{0.45\textwidth}
\begin{semantics}
  \RULE[sub-field][typerules_sub_field]
    { \Sigma \vdash (B_1, s_1) \SUBS (B_2, s_2) }
    { \Sigma \vdash \TVAR{B_1, s_1} \SUBS \TVAR{B_2, s_2} }
\end{semantics}
\end{minipage}
\begin{minipage}{0.5\textwidth}
\begin{semantics}
  \RULE[sub-proc][typerules_sub_proc]( s_2 \ordleq s_1 )
    { \Sigma \vdash \VEC{(B_2, s_2)} \SUBS \VEC{(B_1, s_1)} }
    { \Sigma \vdash \TSPROC{\VEC{B_1, s_1}}{s_1} \SUBS \TSPROC{\VEC{B_2, s_2}}{s_2} }
\end{semantics}
\end{minipage}
\caption{Subtyping rules for interface members.}
\label{fig:subtyping_interface_members}
\end{figure}

The rules defining the subtyping judgments for interface members are given in Figure~\ref{fig:subtyping_interface_members}.
Both rules invoke a subtyping judgment for types of the form $\Sigma \vdash B_1 \SUBS B_2$ in their premise: this, finally, is the proper subtyping relation, which is defined by the rules in Figure~\ref{fig:subtyping_expressions}.
These rules are very simple:
they consist only of the reflexive and transitive rules, and of the rule \nameref{typerules_sub_name}, which reflects the interface declarations.
For this reason, the subtyping relation is parametrised with $\Sigma$ and in \nameref{typerules_sub_name} we simply look up the supertype in $\Sigma$, rather than structurally comparing their respective definitions for $I_1$ and $I_2$ in $\Gamma$ (notice that the relationship has already been verified, if $\Sigma$ is consistent with $\Gamma$).

%

\begin{figure}\centering
\begin{semantics}
  \RULE[sub-refl][typerules_sub_refl]
    { }
    { \Sigma \vdash B \SUBS B }

  \RULE[sub-name][typerules_sub_name]({\Sigma(I_1) = I_2})
    { }
    { \Sigma \vdash I_1 \SUBS I_2 }
\end{semantics} 
\begin{semantics}
  \RULE[sub-trans][typerules_sub_trans]
    { \Sigma \vdash B_1 \SUBS B_2 \AND \Sigma \vdash B_2 \SUBS B_3 }
    { \Sigma \vdash B_1 \SUBS B_3 }

  \RULE[sub-type][typerules_sub_type](s_1 \ordleq s_2)
    { \Sigma \vdash B_1 \SUBS B_2 }
    { \Sigma \vdash (B_1, s_1) \SUBS (B_2, s_2) }
\end{semantics}
\caption{The subtyping relation for base types ($B$) and expression types ($B_s$)}
\label{fig:subtyping_expressions}
\end{figure}

\begin{remark}[On the subtyping of methods and fields]
Subtyping of interface members (fields and methods) is complicated by their variances, where reading is covariant and writing is contravariant.
Methods are comparable to `write-only' fields, since subtyping should not be invoked when we are typing the \emph{declaration} of a method. Hence, the method type constructor is contravariant in its arguments, as can be seen in the rule \nameref{typerules_sub_proc}.
However, due to this contravariance, we get the rather bizarre situation that an interface supertype must be less specific (i.e.\@ have fewer members) as expected, but its methods must have \emph{more} specific arguments than in the subtype.
This hardly seems useful, but we nevertheless still allow it to give the rules a uniform shape.

With fields, the situation is more complicated.
Usually in class-based/object-oriented languages, fields are \emph{both} readable \emph{and} writable.
Hence, subtyping for fields should therefore be both covariant \emph{and} contravariant, depending on whether we read from the field or write to it.
In other words, it must be \emph{invariant}, since allowing either form of variance would lead to an unsound subtyping relation, which, in the present setup, would mean that we could create an illegal information flow according to the security levels.
However, as can be seen in the rule \nameref{typerules_sub_field}, we actually allow subtyping of the field type constructor $\TVAR{B_s}$ to be \emph{covariant} in its argument.
This is possible because of a peculiar feature of \TINYSOL:  fields in this language are \emph{not} writable outside of the contract in which they are declared. 
This is ensured directly in the syntax, since a field is only allowed to appear as the left-hand side of an assignment of the form \code{this.$p$ := $e$}.
Thus, fields are actually \emph{read-only} in all contexts in which subtyping may be invoked; therefore, we can allow covariant subtyping for fields.
This is important since we would otherwise be unable to coerce an arbitrary contract up to $\ITOP$, which is necessary to allow us to give a type to the `magic variable' \code{sender}, which is available within every method body.

This is also the reason why we separate the subtyping rules for interface members (Figure~\ref{fig:subtyping_interface_members}) from the definition of the proper subtyping relation  (Figure~\ref{fig:subtyping_expressions}), since only the latter may be used in the actual type judgments.
In the following presentation, we shall therefore only allow subtyping to be used when judging the type of an expression $e$, which limits the usage of subtyping to the right-hand side of assignments and the arguments of a method call.
This automatically ensures that the rules in Figure~\ref{fig:subtyping_interface_members} cannot be invoked, since an expression cannot return a field name or a method name (i.e.\@ something of type $\TVAR{B_s}$ or $\TSPROC{\BSVEC}{s}$); moreover, even though an expression can yield an address $X$ of type $I$, we simply look it up in $\Sigma$ by rule \nameref{typerules_sub_name}.
\end{remark}

\subsection{Safety requirement for operations}
We have left the syntactic category of operations $\op$ undefined and simply assumed that they can all be evaluated using some total deterministic relation $\trans_{\op}$, such that $\op(\VEC{v}) \trans_{\op} v$.
Furthermore, we assume that, for each operation, we can determine its signature, written $\vdash \op : \VEC{B} \to B$.
To tie these two together, we need to make one further assumption; namely that, if the signature of an operation is $\VEC{B} \to B$ and the actual parameters $\VEC{v}$ are of basic types $\VEC{B}$ (with $|\VEC{v}| = |\VEC{B}|$), then the operation can actually be performed and the resulting value $v$ is of type $B$.
As neither the operations, their semantics, or their type rules are specified, all we can do w.r.t.\@ safety is to require that this must hold.
Formally, we can specify the requirement as follows:

\begin{definition}[Safety requirement for operations]\label{def:safety_operations}
We say that $\Gamma$ satisfies the safety requirement for operations w.r.t. $\trans_{\op}$
if, for all operations $\op$ and for any argument list $\VEC{v}$, it holds that
\begin{itemize}
  \item $\vdash \op : \VEC{B} \to B$, and
  \item $|\VEC{B}| = |\VEC{v}|$, and
  \item $\forall v_i \in \VEC{v} \SUCHTHAT \TYPEOF{v_i} = (B_i, s_i)$
\end{itemize}
imply that  
\begin{itemize}
  \item for all $X \in \ANAMES$ such that $X \in \VEC{v}$, it holds that $X \in \DOM\Gamma$, and
  \item $\TYPEOF{v} = (B, \SBOT)$, where $\op(\VEC{v}) \trans_{\op} v$. \qedhere
\end{itemize}
\end{definition}

Note that, with this definition, we allow addresses to appear as arguments for operations.
However, since we assume that an operation can never yield an address, we also know that the resulting security level will be $\SBOT$, and that the resulting base type $B$ will be one of $\TINT$ or $\TBOOL$.
Moreover, operations are only concerned with the actual data and not with their security levels, since all arguments must be actual values, i.e.\@ not variables.
Therefore, in Definition \ref{def:safety_operations} we disregard the security levels; the security level will be checked in the type rule for operations given in the following section.
This simply ensures that we can give syntax-directed type rules for expressions.


\subsection{Type judgments}

\begin{figure}\centering
\begin{minipage}{0.5\textwidth}
\begin{semantics}
  \RULE[t-val][exp_t_val]({ \TYPEOF{v} = B_s }) 
    { }
    { \Sigma; \Gamma; \Delta \vdash v : B_s }

  \RULE[t-var][exp_t_var]({ \Delta(x) = \TVAR{B_s} })
    { }
    { \Sigma; \Gamma; \Delta \vdash x : B_s }

  \RULE[t-esub][exp_t_sub]
    { \Sigma; \Gamma; \Delta \vdash e : (B_1, s_1) \AND \Sigma \vdash (B_1, s_1) \SUBS (B_2, s_2) }
    { \Sigma; \Gamma; \Delta \vdash e : (B_2, s_2) }
\end{semantics}
\end{minipage}
\begin{minipage}{0.45\textwidth}
\begin{semantics}
  \RULE[t-field][exp_t_field]({ \Gamma(I)(p) = \TVAR{B_s} })
    { \Sigma; \Gamma; \Delta \vdash e : I_s }
    { \Sigma; \Gamma; \Delta \vdash e.p : B_s }

  \RULE[t-op][exp_t_op]
    { \vdash \op : \VEC{B} \to B \AND \Sigma; \Gamma; \Delta \vdash \VEC{e} : \VEC{B}_{s, \ldots, s} }
    { \Sigma; \Gamma; \Delta \vdash \op(\VEC{e}) : B_s }
\end{semantics}
\end{minipage}
\caption{Type rules for expressions.}
\label{fig:type_rules_expr}
\end{figure}

\paragraph{Typing expressions} We start with the type rules for expressions, given in Figure~\ref{fig:type_rules_expr}.
Here, judgments are of the form $\Sigma; \Gamma; \Delta  \vdash e : B_s$.
There are a few things to note:
\begin{itemize}
  \item In rule \nameref{exp_t_val}, the type of a value $v$ is determined by the $\TYPEOF{\cdot}$ function, which, in the case of integers and booleans, will set the security level to $\SBOT$, which is the lowest level.
    This is a consequence of the fact that there is no simple relationship between the datatype of a value and its security level.
    The level can then be raised safely as required by subtyping, and therefore the actual security level will be determined by the type of the variable (resp.\@ field) to which it is assigned.

  \item The rules \nameref{exp_t_var} and \nameref{exp_t_field} simply unwrap the type of the contained value from the box type of the container.
    Remember that here we assume that $x$ also covers the `magic variable' names \code{this}, \code{sender} and \code{value}, and that $p$ also covers the field name \code{balance}.

  \item Finally, in rule \nameref{exp_t_op}, we require that all arguments and the return value must be typable to the same security level $s$.
    Note in particular that we assume that no operation has an address as return type (hence, we do not allow any form of pointer arithmetic).
    Operations may be defined on addresses for their \emph{arguments} (e.g.,\@ for equality testing), but the return type must be one of the other base types, which can be given a security level.
    Thus, in rule \nameref{exp_t_op}, we also need to extract the security level $s$ from the types of the argument expressions.
\end{itemize}

\begin{figure}
\begin{minipage}{0.49\textwidth}
\begin{semantics}
  \RULE[t-skip][stm_t_skip]
    { }
    { \Sigma; \Gamma; \Delta \vdash \code{skip} : \TCMD{s} }

  \RULE[t-decv][stm_t_decv]
    { \begin{array}{l}
      \Sigma; \Gamma; \Delta                   \vdash e : B_{s'} \\
      \Sigma; \Gamma; \Delta, x : \TVAR{B_{s'}} \vdash S : \TCMD{s} 
      \end{array}
    }
    { \Sigma; \Gamma; \Delta \vdash \code{$\TVAR{B_{s'}}$ $x$ := $e$ in $S$} : \TCMD{s} }

  \RULE[t-while][stm_t_while]
    {\Sigma; \Gamma; \Delta \vdash e : \TBOOL_s \AND \Sigma; \Gamma; \Delta \vdash S : \TCMD{s}}
    { \Sigma; \Gamma; \Delta \vdash \code{while $e$ do $S$} : \TCMD{s} }

  \RULE[t-assv][stm_t_assv]
    { \Sigma; \Gamma; \Delta \vdash e : B_s }
    { \Sigma; \Gamma; \Delta \vdash \code{$x$ := $e$} : \TCMD{s} }
\WHERE{}{ \Delta(x) = \TVAR{B_s} }

  \RULE[t-call][stm_t_call](\!\!\!{
  \begin{array}{r @{~} l}
    s_1 & \ordleq s        \\
    s   & \ordleq s_2, s_3 
  \end{array}
  }\!\!\!)
    { 
    \begin{array}{r @{~} l}
      \Sigma; \Gamma; \Delta & \vdash e_1 : (I^Y,s)    \\
      \Sigma; \Gamma; \Delta & \vdash e_2 : \TINT_s \\
      \Sigma; \Gamma; \Delta & \vdash \VEC{e} : \BSVEC
    \end{array}
    }
    { \Sigma; \Gamma; \Delta \vdash \CALL{e_1}{f}{\VEC{e}}{e_2} : \TCMD{s} }
    
    \WHERE{ }{\Delta(\code{this})  = \TVAR{I^X, s_1} }
    \WHERE{ }{ \hspace*{-1.04cm}\Gamma(I^X)(\code{balance})  = \TVAR{\TINT, s_2} }
    \WHERE{ }{ \hspace*{-1.02cm}\Gamma(I^Y)(\code{balance}) = \TVAR{\TINT, s_3} }
    \WHERE{ }{ \hspace*{-.1cm}\Gamma(I^Y)(f) = \TSPROC{\BSVEC}{s} }    
\vspace*{.4cm}
        
\end{semantics}
\end{minipage}
\begin{minipage}{0.49\textwidth}
\begin{semantics}
  \RULE[t-throw][stm_t_throw]
    { }
    { \Sigma; \Gamma; \Delta \vdash \code{throw} : \TCMD{s} }

  \RULE[t-if][stm_t_if]
    { 
    \begin{array}{r @{~} l @{\qquad} r @{~} l}
                             &                     & \Sigma; \Gamma; \Delta & \vdash S_{\TRUE} : \TCMD{s}  \\
      \Sigma; \Gamma; \Delta & \vdash e : \TBOOL_s & \Sigma; \Gamma; \Delta & \vdash S_{\FALSE} : \TCMD{s} 
    \end{array}
    }
    { \Sigma; \Gamma; \Delta \vdash \code{if $e$ then $S_{\TRUE}$ else $S_{\FALSE}$} : \TCMD{s} }

  \RULE[t-seq][stm_t_seq]
    { \Sigma; \Gamma; \Delta \vdash S_1 : \TCMD{s} \AND \Sigma; \Gamma; \Delta \vdash S_2 : \TCMD{s} }
    { \Sigma; \Gamma; \Delta \vdash S_1; S_2 : \TCMD{s}}

  \RULE[t-assf][stm_t_assf]({s' \ordleq s})
    { \Sigma; \Gamma; \Delta \vdash e : B_s }
    { \Sigma; \Gamma; \Delta \vdash \code{this.$p$ := $e$} : \TCMD{s} }
\WHERE{}{\Delta(\code{this}) = I_{s'} \text{ and }  \Gamma(I)(p)  = \TVAR{B_s}}

  \RULE[t-dcall][stm_t_dcall]
    { 
    \begin{array}{l}
      \Sigma; \Gamma; \Delta \vdash e : (I^Y, s)   \\
      \Sigma \vdash (I^X, s_1) \SUBS (I^Y, s)      \\
      \Sigma; \Gamma; \Delta \vdash \VEC{e} : \BSVEC
    \end{array}
    }
    { \Sigma; \Gamma; \Delta \vdash \DCALL{e}{f}{\VEC{e}} : \TCMD{s} }
    
    \WHERE{ }{ \Delta(\code{this}) = \TVAR{I^X, s_1} }
    \WHERE{ }{ \hspace*{-.1cm}\Gamma(I^Y)(f) = \TSPROC{\BSVEC}{s} }
  \RULE[t-ssub][stm_t_ssub]({ s_2 \ordleq s_1 })
    { \Sigma; \Gamma; \Delta \vdash S : \TCMD{s_1} }
    { \Sigma; \Gamma; \Delta \vdash S : \TCMD{s_2} }
\end{semantics}
\end{minipage}
\caption{Type rules for statements}
\label{fig:type_rules_stm}
\end{figure}

\paragraph{Typing statements} Next, we consider the type rules for statements appearing in the body of method declarations; they are given in Figure~\ref{fig:type_rules_stm}.
Here, judgments are of the form $\Sigma; \Gamma; \Delta \vdash S : \TCMD{s}$, indicating that all variables \emph{written to} within $S$ are of level $s$ or \emph{higher}.
All rules are straightforward, except those for method invocations, which we now illustrate.

According to the semantics for call (cf.\@ rule \nameref{ts_bss_call}), every call includes an implicit read and write of the \code{balance} field of the caller contracts, since the call will only be performed if the value of $e_2$ is less than or equal to the value of \code{balance} (to ensure that the subtraction will not yield a negative number).
There is thus an implicit flow from \code{this.balance} to the body $S$ of the method call, similar to the case for the guard expression $e$ in an if-statement.

Furthermore, there is an implicit write to the \code{balance} field of the callee contract, and thus a flow of information from one field to the other.
This might initially seem like it would require both the caller and the callee to have the same security level for their field \code{balance}.
However, the levels \emph{can} differ, since by subtyping we can coerce one up to match the level of the other.
For this reason, we have the side condition $s \ordleq s_2, s_3$, where $s_2$ and $s_3$ are the levels of the \code{balance} fields of the caller and callee, respectively.
This enables calls from a lower security level into a higher security level, but not the other way around.
The reason behind this condition is as follows.
Suppose that the \emph{actual} level of $e_2$ were some level $s'$.
According to the semantics (Figure~\ref{fig:tinysol_semantics_statements_bss}), every method call implicitly performs a write to the \code{balance} fields of both the caller and the callee, corresponding to the following two lines of code (where $Y$ is the address of the callee):
\begin{align*}
  \code{this.balance} & \code{ := this.balance - $e_2$;} \\
  \code{$Y$.balance}  & \code{ := $Y$.balance + $e_2$}
\end{align*}
Typing these statements as in \nameref{stm_t_assf} would give us that $s' \ordleq s_2$ and $s' \ordleq s_3$; hence $s' \ordleq s_2, s_3$.
In both cases, the level of the left-hand side must of course be above-or-equal to the level of the right-hand side, and since \code{this.balance} (resp.\@ \code{$Y$.balance}) appears on both sides, we have that $s'$ must be less-than or equal-to both, otherwise the level of the right-hand side would be above the level of the left-hand side.
Finally, since both \code{balance} fields are being modified in the method body, albeit implicitly, it must also be the case that $s \ordleq s_2, s_3$, where $s$ is the level of the method body.
We then simplify this by letting $s' = s$ and requiring that $e_2$ can also be evaluated at the level $s$, which it can by subtyping if the condition $s' \ordleq s$ holds.

The case for delegate calls is simpler since there is no currency transfer associated with the call. However, there is one important point to note in rule \nameref{stm_t_dcall}: 
we have the subtyping judgment $\Sigma \vdash (I^X, s_1) \SUBS (I^Y, s)$ to require that the interface type of the caller contract is a subtype of the contract that contains the definition of $f$.
This is necessary because the body of $f$ might access fields or call methods via \code{this}; but when the body is executed in the context of the \emph{caller}, this would lead to a runtime error if the calling contract does not contain fields or methods with the same names (and of the same types).
This situation is prevented by requiring the caller contract to be a subtype (i.e.\@ a specialisation) of the callee contract, thus ensuring that the caller contract will contain \emph{at least} the same methods and fields as the callee contract.
Note that this implies that the \emph{caller contract} also must contain a method with the same name $f$ and with the same signature (or a subtype thereof).
We could remove this limitation by introducing a more complicated type rule, but we shall forego that here, since it is consistent with the purpose of \code{dcall}, i.e.\@ code reuse.

Finally, note that in both \nameref{stm_t_call} and \nameref{stm_t_dcall}, in the \emph{list} of types $\BSVEC$, the levels $\VEC{s}$, which we shall expand as $s_1', \ldots, s_h'$, do not have to be related to any of the other levels.

\begin{figure}\centering
\begin{minipage}{0.59\textwidth}
\begin{semantics}
  \RULE[t-env-$\EMPTYSET$](J \in \SET{T, M, S, F, V})
    { }
    { \Sigma; \Gamma; \Delta \vdash \ENV{J}^\EMPTYSET }

  \RULE[t-env-t][env_t_envt]
    { \Sigma; \Gamma; \Delta \vdash_X \ENV{M} \AND \Sigma; \Gamma; \Delta \vdash \ENV{T} }
    { \Sigma; \Gamma; \Delta \vdash \joine{( X,\ENV{M})}{\ENV{T}} }

  \RULE[t-env-m][env_t_envm]
    { \Sigma; \Gamma; \Delta' \vdash S : \TCMD{s} \AND \Sigma; \Gamma; \Delta \vdash_X \ENV{M} }
    { \Sigma; \Gamma; \Delta \vdash_X \joine{(f,(x_1, \ldots, x_h, S))}{\ENV{M}} }

  \WHERE{ \Gamma(X) }{= (I, s_1)}          
  \WHERE{ \Gamma(I)(\code{balance})} { =\TVAR{\TINT, s_2  }}
  \WHERE{\Gamma(I)(f)}       { = \TSPROC{(B_1, s_1'), \ldots, (B_h, s_h')}{s} }
  \WHERE{\Delta'}             { = \code{this}:\TVAR{I, s_1}, \code{value}:\TVAR{\TINT, s_2}, }
  \WHERE{}                   { \quad \code{sender}:\TVAR{\ITOP, \STOP}, }
  \WHERE{}                   { \quad x_1:\TVAR{B_1, s_1'}, \ldots, x_h:\TVAR{B_h, s_h'} }
\end{semantics}
\end{minipage}
\begin{minipage}{0.39\textwidth}
\begin{semantics}
  \RULE[t-envs][env_t_envs]
    { \Sigma; \Gamma; \Delta \vdash_X \ENV{F} \AND \Sigma; \Gamma; \Delta \vdash \ENV{S} }
    { \Sigma; \Gamma; \Delta \vdash \joine{(X,\ENV{F})}{\ENV{S}} }

  \RULE[t-envf][env_t_envf]
    { \Sigma; \Gamma; \Delta \vdash v : B_s \AND \Sigma; \Gamma; \Delta \vdash_X \ENV{F} }
    { \Sigma; \Gamma; \Delta \vdash_X \joine{(p,v)}{\ENV{F}} }
    \WHERE{    \Gamma(X)    }{ = (I, s')   \text{ and }   \Gamma(I)(p)  = \TVAR{B_s} }

  \RULE[t-envv][env_t_envv]
    { \Sigma; \Gamma; \Delta \vdash v : B_s \AND \Sigma; \Gamma; \Delta \vdash \ENV{V} }
    { \Sigma; \Gamma; \Delta \vdash \joine{(x,v)}{\ENV{V}} }
    \WHERE {\Delta(x) }{ = \TVAR{B_s}}

  \RULE[t-envsv][env_t_envsv]
    { \Sigma; \Gamma; \Delta \vdash \ENV{S} \AND \Sigma; \Gamma; \Delta \vdash \ENV{V} }
    { \Sigma; \Gamma; \Delta \vdash \ENV{SV} }
\end{semantics}
\end{minipage}
\caption{Type rules for the method-, state-, and variable environments}
\label{fig:type_rules_env}
\end{figure}

\paragraph{Typing environments} After the initial reduction step, all declarations are stored in the two environments $\ENV{ST}$, and further reductions also use the variable environment $\ENV{V}$ for local variable declarations.
Hence, we also need to be able to conclude \emph{agreement} between these environments and the type environments $\Sigma$, $\Gamma$ and $\Delta$;
this is given in Figure~\ref{fig:type_rules_env}.
Note that the check here only ensures that every declared contract member has a type; the converse check (i.e.\@ that every declared type in an interface also has an implementation) should also be performed.
However, we shall omit this in the present treatment and simply assume that it holds.

For all environments, the type judgments are relative to a $\Sigma$, $\Gamma$ and $\Delta$, to make them uniform.
However, the $\Delta$ component is actually needed only for typing $\ENV{V}$. In particular, \nameref{env_t_envf} invokes \nameref{exp_t_val} to conclude a type for the value $v$ stored in a field $p$, but this rule does not rely on $\Delta$ since $v$ is a value and so does not contain any variable to be evaluated. 
In contrast, in \nameref{env_t_envm} we build a new $\Delta'$ from the signature of the method to be able to type the body $S$ in the premise.

\begin{figure}
\begin{minipage}{0.51\textwidth}
\begin{semantics}
  \RULE[t-dec-m][dec_t_decm]
    { \Sigma; \Gamma; \Delta' \vdash S : \TCMD{s}  \AND \Sigma; \Gamma; \Delta \vdash_X DM}
    { \Sigma; \Gamma; \Delta \vdash_X \code{$f(x_1,\ldots,x_h)$ \{ $S$ \} $DM$} }
    
  \WHERE{ \Gamma(X) } { = (I, s_1)}          
  \WHERE{ \Gamma(I)(\code{balance}) }{ = \TVAR{\TINT, s_2} } 
  \WHERE{\Gamma(I)(f)}       { = \TSPROC{(B_1, s_1'), \ldots, (B_h, s_h')}{s} }
  \WHERE{\Delta'}             { = \code{this}:\TVAR{I, s_1}, \code{value}:\TVAR{\TINT, s_2}, }
  \WHERE{}                   { \quad \code{sender}:\TVAR{\ITOP, \STOP}, }
  \WHERE{}                   { \quad x_1:\TVAR{B_1, s_1'}, \ldots, x_h:\TVAR{B_h, s_h'} }
\end{semantics}
\end{minipage}
\begin{minipage}{0.47\textwidth}
\begin{semantics}
  \RULE[t-dec-f][dec_t_decf]
    { \Sigma; \Gamma; \Delta \vdash v : B_s \AND \Sigma; \Gamma; \Delta \vdash_X DF }
    { \Sigma; \Gamma; \Delta \vdash_X \code{field $p$ := $v$; $DF$} }
\WHERE {\Gamma(X) }{ = I_{s'}   \text{ and }  \Gamma(I)(p)  = \TVAR{B_s} }

  \RULE[t-dec-c][dec_t_decc]
    { \begin{array}{l} \Sigma; \Gamma; \Delta \vdash_X DF \\ \Sigma; \Gamma; \Delta \vdash_X DM \\ \Sigma; \Gamma; \Delta \vdash DC 
    \end{array}}
    { \Sigma; \Gamma; \Delta\vdash \code{contract $X$ : $I_s$ \{ $DF$ $DM$ \} $DC$} }
\end{semantics}
\end{minipage}
\caption{Type rules for declarations.}
\label{fig:type_rules_decl}
\end{figure}

\paragraph{Typing declarations} Finally, in Figure~\ref{fig:type_rules_decl}, we give the type judgments for contract declarations, that closely follow the rules for environment agreement from Figure~\ref{fig:type_rules_env} (indeed, declarations and environments $\ENV{TS}$ are two equivalent representations of the given code).
Judgments are of the form $\Sigma; \Gamma;\Delta \vdash DC$, stating that the declarations $DC$ are \emph{well-typed} w.r.t.\@ the environments $\Sigma; \Gamma; \Delta$ (here $\Delta$ is never used, so it can be easily considered as $\emptyset$; we keep it general for the sake of uniformity):
this holds if the declarations are consistent with the type information recorded in the environments, i.e.\@ every field and method must have a type, and the body of each method must be typable according to the assumptions of the type.
Note that also here we omit the rules for ensuring that all declared types in an interface also have an implementation in any contract claiming to implement that interface.

\subsection{Safety and soundness}
As is the case for the type system proposed in \cite{volpano2000secure_flow_typesystem}, our type system does not have a now-safety predicate in the usual sense, since (invariant) safety in simple type systems is a 1-property, whereas noninterference is a hyper-property (specifically, a 2-property) \cite{hyperproperties}.
Instead, the meaning of `safety' is expressed directly in the meaning of the types:
\begin{itemize}
  \item If an expression $e$ has type $B_s$, then all variables \emph{read} in the evaluation of $e$ are of level $s$ or \emph{lower}, i.e.\@ no read-up.
  \item If a statement $S$ has type $\TCMD{s}$, then all variables \emph{written} in the execution of $S$ are of level $s$ or \emph{higher}, i.e.\@ no write-down.
\end{itemize}

Intuitively, the meaning of these two types together implies that information from higher-level variables cannot flow into lower-level variables.
For a statement such as \code{$x$ := $e$} to be well-typed, it must therefore be the case that, if $\Sigma;\Gamma;\Delta \vdash x : \TVAR{B_s}$ and $\Sigma;\Gamma;\Delta \vdash e : B_{s'}$, then $s' \ordleq s$.
Since $s'$ can be coerced up to $s$ through subtyping to match the level of the variable, the statement itself can then be typed as $\TCMD{s}$.
We shall prove that our type system indeed ensures these properties in Theorems~\ref{thm:subject_reduction}-\ref{thm:soundness} below.

\begin{figure}
\begin{minipage}{0.32\textwidth}
\begin{semantics}
\vspace*{-.1cm}

  \RULE[eq-env$_V^\EMPTYSET$]
    { }
    { \Delta \vdash \ENV{V}^\EMPTYSET =_s \ENV{V}^\EMPTYSET }
\vspace*{.2cm}

  \RULE[eq-env$_F^\EMPTYSET$]
    { }
    { \Gamma \vdash_I \ENV{F}^\EMPTYSET =_s \ENV{F}^\EMPTYSET }
\vspace*{.2cm}

  \RULE[eq-env$_S^\EMPTYSET$]
    { }
    { \Gamma \vdash \ENV{S}^\EMPTYSET =_s \ENV{S}^\EMPTYSET }
\end{semantics}
\end{minipage}
\begin{minipage}{0.67\textwidth}
\begin{semantics}
  \RULE[eq-env$_V$][env_s_eq_envv](\!\!\!{
    \begin{array}{l}
      \Delta(x) = \TVAR{B_{s'}} \\
      s' \ordleq s \implies v_1 = v_2
    \end{array}
  }\!\!\!)
    { \Delta \vdash \ENV{V}^1 =_s \ENV{V}^2 }
    { \Delta \vdash \joine{(x,v_1)}{\ENV{V}^1} =_s \joine{(x,v_2)}{\ENV{V}^2} }

  \RULE[eq-env$_F$][env_s_eq_envf](\!\!\!{
    \begin{array}{l}
      \Gamma(I)(p) = \TVAR{B_{s'}} \\
      s' \ordleq s \implies v_1 = v_2
    \end{array}
  }\!\!\!)
    { \Gamma \vdash_I \ENV{F}^1 =_s \ENV{F}^2 }
    { \Gamma \vdash_I \joine{(p,v_1)}{\ENV{F}^1} =_s \joine{(p,v_2)}{\ENV{F}^2} }

  \RULE[eq-env$_S$][env_s_eq_envs]({ \Gamma(X) = I_{s'} })
    { \Gamma \vdash \ENV{S}^1 =_s \ENV{S}^2 \AND \Gamma \vdash_I \ENV{F}^1 =_s \ENV{F}^2 }
    { \Gamma \vdash \joine{(X,\ENV{F}^1)}{\ENV{S}^1} =_s \joine{(X,\ENV{F}^2)}{\ENV{S}^2} }
\end{semantics}
\end{minipage}
\vspace*{.3cm}

\begin{minipage}{0.55\textwidth}
\begin{semantics}
  \RULE[eq-env$_T$][env_s_eq_envt]({\!\!
    \begin{array}{l}
      \Gamma(X) = I_{s'} \\
      s' \ordleq s \implies \ENV{M}^1 = \ENV{M}^2
    \end{array}
  }\!\!)
    { \Gamma \vdash \ENV{T}^1 =_s \ENV{T}^2 }
    { \Gamma \vdash \joine{(X,\ENV{M}^1)}{\ENV{T}^1} =_s \joine{(X,\ENV{M}^2)}{\ENV{T}^2}  }
    \vspace*{.2cm}

  \RULE[eq-env$_{SV}$]
    { \Gamma \vdash \ENV{S}^1 =_s \ENV{S}^2 \AND \Delta \vdash \ENV{V}^1 =_s \ENV{V}^2 }
    { \Gamma; \Delta \vdash \ENV{SV}^1 =_s \ENV{SV}^2 }
\end{semantics}
\end{minipage}
\begin{minipage}{0.44\textwidth}
\begin{semantics}
\vspace*{-1.7cm}

  \RULE[eq-env$_{ST}$]
    { \Gamma \vdash \ENV{S}^1 =_s \ENV{S}^2 \AND \Gamma \vdash \ENV{T}^1 =_s \ENV{T}^2 }
    { \Gamma \vdash \ENV{ST}^1 =_s \ENV{ST}^2 }
\end{semantics}
\end{minipage}
\caption{Rules for the $s$-parameterised equivalence relation.}
\label{fig:s_equal}
\end{figure}

Before proceeding, we need to define a way to express that two \emph{states}, i.e.\@ two collections of variable and field environments $\ENV{SV}$, are equal up to a certain security level $s$.
This relation, written $\Gamma;\Delta \vdash \ENV{SV}^1 =_s \ENV{SV}^2$, is given by the rules in Figure~\ref{fig:s_equal}.
Note in particular that the definition implies that $\ENV{SV}^1$ and $\ENV{SV}^2$ must have the same domain, and this carries over to the inner environments $\ENV{F}$ inside $\ENV{S}$.
Given our annotation of security levels on interfaces as well, we also extend the $=_s$ relation to method tables $\ENV{T}$, and finally to the combined representation of state and code, i.e.\@ $\ENV{ST}$.
The following result follows directly from the definition of the $s$-parameterised equivalence relation and can be shown by a straightforward induction on the rules of $=_s$:

\begin{lemma}[Restriction]\label{lemma:restriction_s}
If $\Gamma;\Delta \vdash \ENV{SV}^1 =_s \ENV{SV}^2$ and $s' \ordleq s$, then $\Gamma;\Delta \vdash \ENV{SV}^1 =_{s'} \ENV{SV}^2$.
\end{lemma}

Next, we need the standard lemmas for strengthening and weakening of the variable environment, that can again be shown by induction on the rules of $=_s$. 

\begin{lemma}[Strengthening]\label{lemma:strengthening_env_v}
If $\Gamma;\Delta, x : \TVAR{B_{s'}} \vdash \joine{(x,v_1)}{\ENV{V}^1} =_s \joine{(x,v_2)}{\ENV{V}^2}$ then also $\Gamma;\Delta \vdash \ENV{V}^1 =_s \ENV{V}^2$.
\end{lemma}

\begin{lemma}[Weakening]\label{lemma:weakening_env_v}
If $\Gamma;\Delta \vdash \ENV{V}^1 =_s \ENV{V}^2$ and $x \notin \DOM{\ENV{V}^1} \cup \DOM{\ENV{V}^2}$, then, for any $B$, $s'$ and $v$ of type $B$, it holds that  $\Gamma;\Delta, x : \TVAR{B_{s'}} \vdash \joine{(x,v)}{\ENV{V}^1} =_s \joine{(x,v)}{\ENV{V}^2}$. 
\end{lemma}

Both lemmas can be extended directly to $\Gamma; \Delta \vdash \ENV{SV}^1 =_s \ENV{SV}^2$.
With this, we can now state the first of our main theorems, whose proof is by induction on the derivation of $\ENV{T} \vdash \CONF{S, \ENV{SV}} \trans \ENV{SV}'$ and can be found in Appendix~\ref{app:proof_subject_reduction}.

\begin{theorem}[Preservation]\label{thm:subject_reduction}
Assume that
$\Sigma;\Gamma;\Delta \vdash S : \TCMD{s}$, 
$\Sigma;\Gamma;\Delta \vdash \ENV{T}$,
$\Sigma;\Gamma;\Delta \vdash \ENV{SV}$, and
$\ENV{T} \vdash \CONF{S, \ENV{SV}} \trans \ENV{SV}'$.
Then, $\Gamma;\Delta \vdash \ENV{SV} =_{s'} \ENV{SV}'$ for any $s'$ such that $s \not\ordleq s'$.
\end{theorem}

The Preservation theorem assures us that the promise made by the type $\TCMD{s}$ is actually fulfilled:
if $S$ is typed as $\TCMD{s}$, then every variable or field written to in $S$ will be of level $s$ \emph{or higher}; hence every variable or field of a level that is \emph{strictly lower} than or incomparable to $s$ will be unaffected.
Thus, the pre- and post-transition states will be equal on all values stored in variables or fields of level $s'$ or lower, since they cannot have been changed during the execution of $S$.
In other words, what is shown to be `preserved' in this theorem is the \emph{values} at levels lower than or incomparable to $s$.

Note that the theorem does not show preservation of data types for the environments (as is otherwise usually required in preservation proofs for type systems); i.e.\@ we do not show that variables of type $\TINT$ still contain integer typed values after the transitions, and similarly for booleans.
As can be seen in Figure~\ref{fig:type_rules_env}, the type judgment $\Sigma;\Gamma;\Delta \vdash \ENV{SV}$ only ensures that every field and variable in $\ENV{SV}$ has \emph{some} type in $\Gamma$ and $\Delta$.
The \emph{number} of declared fields and variables cannot change between the pre- and post-states of a transition (this is ensured by the rule \nameref{ts_bss_decv});
only the stored values can change, but there is no inherent relationship between a value and its assigned security level.

Our next theorem assures us that the type of an expression is also in accordance with the intended meaning, namely: if $\Sigma;\Gamma;\Delta \vdash e : B_s$, then every variable (or field) read from in $e$ will be of level $s$ or lower (i.e.\@ no read-down of values from a higher level).
We express this requirement by considering two different states, $\ENV{SV}^1$ and $\ENV{SV}^2$, which must agree on all values of level $s$ and lower; then,
evaluating $e$ w.r.t.\@ either of these states should yield the same result. 
The proof is by induction on the derivation of $\Sigma;\Gamma;\Delta \vdash e : B_s$ and is in Appendix~\ref{app:proof_safety_expressions}.

\begin{theorem}[Safety for expressions]
\label{thm:safety_expressions}
Assume that
$\Sigma;\Gamma;\Delta \vdash e : B_s$,
$\Sigma;\Gamma;\Delta \vdash \ENV{SV}^1$,
$\Sigma;\Gamma;\Delta \vdash \ENV{SV}^2$, and
$\Gamma;\Delta \vdash \ENV{SV}^1 =_s \ENV{SV}^2$.
Then, $\ENV{SV}^1 \vdash e \trans_e v$ and $\ENV{SV}^2 \vdash e \trans_e v$.
\end{theorem}

Finally, we can use the preceding two theorems to show soundness for the type system.
The soundness theorem expresses that, if a statement $S$ is well-typed \emph{to any level} $s_1$ and we execute $S$ with any two states $\ENV{SV}^1$ and $\ENV{SV}^2$ that agree \emph{up to any level} $s_2$, then the resulting states $\ENV{SV}^{1'}$ and $\ENV{SV}^{2'}$ will still agree on all values up to level $s_2$.
This ensures noninterference, since any difference in values of a \emph{higher} level than $s_2$ cannot induce a difference in the computation of values at any lower levels.

\begin{theorem}[Soundness]
\label{thm:soundness}
Assume that
$\Sigma;\Gamma;\Delta \vdash S : \TCMD{s_1}$,
$\Sigma;\Gamma;\Delta \vdash \ENV{T}$,
$\Sigma;\Gamma;\Delta \vdash \ENV{SV}^1$,
$\Sigma;\Gamma;\Delta \vdash \ENV{SV}^2$,
$\Gamma;\Delta \vdash \ENV{SV}^1 =_{s_2} \ENV{SV}^2$,
$\ENV{T} \vdash \CONF{S, \ENV{SV}^1} \trans \ENV{SV}^{1'}$, and
$\ENV{T} \vdash \CONF{S, \ENV{SV}^2} \trans \ENV{SV}^{2'}$. 
Then, $\Gamma;\Delta \vdash \ENV{SV}^{1'} =_{s_2} \ENV{SV}^{2'}$.
\end{theorem}
\begin{proof}[Proof Sketch]
There are two cases to consider.
If $s_1 \not\ordleq s_2$ (i.e.\@ $s_2$ is either strictly below $s_1$, or they are incomparable), then,
by Theorem~\ref{thm:subject_reduction}, we can conclude that $\Gamma;\Delta \vdash \ENV{SV}^1 =_{s_2} \ENV{SV}^{1'}$ and
$\Gamma;\Delta \vdash \ENV{SV}^2 =_{s_2} \ENV{SV}^{2'}$;
as we know that $\Gamma;\Delta \vdash \ENV{SV}^1 =_{s_2} \ENV{SV}^2$, it therefore also holds that $\Gamma;\Delta \vdash \ENV{SV}^{1'} =_{s_2} \ENV{SV}^{2'}$. 
If $s_1 \ordleq s_2$, we proceed by induction on the inferences of the transitions;
all details are in Appendix~\ref{app:proof_soundness}.
\end{proof}

Theorem~\ref{thm:soundness} corresponds to the soundness theorem proved by Volpano et al.  \cite{volpano2000secure_flow_typesystem} for their While-like language.
However, given the class-based nature of \TINYSOL, we can actually take this result one step further and allow even parts of the code to vary.
Specifically, given two `method table' environments, $\ENV{T}^1$ and $\ENV{T}^2$, we just require that these two environments agree up to the same level $s_2$ to ensure agreement of the resulting two states $\ENV{SV}^{1'}$ and $\ENV{SV}^{2'}$.
We state this in the following theorem:

\begin{theorem}[Extended soundness]\label{thm:extended_soundness}
Assume that
$\Sigma;\Gamma;\Delta \vdash S : \TCMD{s_1}$,
$\Sigma;\Gamma;\Delta \vdash \ENV{T}^1$,
$\Sigma;\Gamma;\Delta \vdash \ENV{T}^2$,
$\Gamma \vdash \ENV{T}^1 =_{s_2} \ENV{T}^2$,
$\Sigma;\Gamma;\Delta \vdash \ENV{SV}^1$,
$\Sigma;\Gamma;\Delta \vdash \ENV{SV}^2$,
$\Gamma;\Delta \vdash \ENV{SV}^1 =_{s_2} \ENV{SV}^2$,
$\ENV{T}^1 \vdash \CONF{S, \ENV{SV}^1} \trans \ENV{SV}^{1'}$, and
$\ENV{T}^2 \vdash \CONF{S, \ENV{SV}^2} \trans \ENV{SV}^{2'}$. 
Then, $\Gamma;\Delta \vdash \ENV{SV}^{1'} =_{s_2} \ENV{SV}^{2'}$.
\end{theorem}
\begin{proof}
By induction on the sum of the lengths of the inferences for
$\ENV{T}^1 \vdash \CONF{S, \ENV{SV}^1} \trans_S \ENV{SV}^{1'}$, and
$\ENV{T}^2 \vdash \CONF{S, \ENV{SV}^2} \trans_S \ENV{SV}^{2'}$. 
All the cases are like the corresponding ones in the proof of Theorem~\ref{thm:soundness}, except those for method invocations (that are the only ones where $\ENV T$ plays a role).
We only consider the case in which \nameref{ts_bss_call} was the last one used in the inferences; the case for \nameref{ts_bss_dcall} is similar.
There are two cases to consider:
\begin{itemize}
  \item If $s_1 \not\ordleq s_2$, then $\ENV{T}^1$ and $\ENV{T}^2$ only agree up to level $s_2$, but may differ on other levels, including $s_1$ (which is either strictly above $s_2$, or they are incomparable).
  Thus the call may be to two different methods (albeit with the same signature), namely $\ENV T^1 (Y)(f)  = (\VEC x , S_1)$ and $\ENV T ^2 (Y)(f) = (\VEC x , S_2)$.
  However, as $S$ is typed to level $s_1$, this cannot induce a difference at level $s_1$ or lower. In particular, by the premises of rule \nameref{ts_bss_call} we have that $\ENV{T}^1 \vdash \CONF{S_1, \ENV{SV}^1} \trans_S \ENV{SV}^{1'}$ and $\ENV{T}^2 \vdash \CONF{S_2, \ENV{SV}^2} \trans_S \ENV{SV}^{2'}$. By two applications of Theorem~\ref{thm:subject_reduction}, we can then conclude that
$\Gamma;\Delta \vdash \ENV{SV}^1 =_{s_2} \ENV{SV}^{1'}$ and $\Gamma;\Delta \vdash \ENV{SV}^2 =_{s_2} \ENV{SV}^{2'}$.  As we know by assumption that $\Gamma;\Delta \vdash \ENV{SV}^1 =_{s_2} \ENV{SV}^2$, we can therefore also conclude that $\Gamma;\Delta \vdash \ENV{SV}^{1'} =_{s_2} \ENV{SV}^{2'}$.
  \item If $s_1 \ordleq s_2$, then $\ENV{T}^1$ and $\ENV{T}^2$ agree on all levels up to level $s_2$ \emph{including} level $s_1$.
By assumption, $\Gamma \vdash \ENV{T}^1 =_{s_2} \ENV{T}^2$ and so, by \nameref{env_s_eq_envt}, $\ENV{T}^1(Y) = \ENV{T}^2(Y)$, since $Y$ resides at level $s_1$;
in particular, $\ENV T^1 (Y)(f) = \ENV T ^2 (Y)(f) = (\VEC x , S')$. The rest of the proof then proceeds like in the case for \nameref{ts_bss_call} of Appendix~\ref{app:proof_soundness}.
\qedhere
\end{itemize}
\end{proof}

\subsection{Extending the type system to transactions}
A transaction is nothing but a method call with parameters that are values and \code{sender} set to an account address, which corresponds to a minimal implementation of $\ITOP$.
Thus, the theorems from the preceding section can easily be extended to transactions and blockchains.

A blockchain consists of a set of contract declarations $DC$, followed by a list of transactions $T$.
Hence, we can conclude $\Sigma;\Gamma;\Delta \vdash DC~T : \TCMD{s}$, if it holds that $\Sigma;\Gamma;\Delta \vdash DC$ and $\Sigma;\Gamma;\Delta \vdash T : \TCMD{s}$.
The latter can be simply concluded by the following rules:
\begin{center}
\begin{semantics}
  \RULE[t-empty]
    { }
    {\Sigma;\Gamma;\Delta \vdash \epsilon : \TCMD{s}}
  \tabularnewline 
  \tabularnewline
  \tabularnewline
\end{semantics}\hspace*{0.5cm}
\begin{semantics}
  \RULE[t-trans]
    { \Sigma;\Gamma;\Delta' \vdash \CALL{Y}{f}{\VEC{v}}{n} : \TCMD{s} \AND \Sigma;\Gamma;\Delta \vdash T : \TCMD{s} }
    { \Sigma;\Gamma;\Delta \vdash \TRANSACT{A}{Y}{f}{\VEC{v}}{n}, T : \TCMD{s} }

  \WHERE{\Delta'}{ = \code{this} : \TVAR{\Gamma(A)} }
\end{semantics}
\end{center}

\noindent This gives us the following two results:
\begin{lemma}\label{lemma:DC}
If $\Sigma;\Gamma;\Delta \vdash DC$ and $\CONF{DC, \ENV{ST}^{\EMPTYSET}} \trans_{DC} \ENV{ST}$, then $\Sigma; \Gamma;\Delta \vdash \ENV{ST}$.
\end{lemma}

\begin{lemma}\label{lemma:subject_reduction_transactions}
If 
$\Sigma;\Gamma;\Delta \vdash \TRANSACT{A}{Y}{f}{\VEC{v}}{n}, T : \TCMD{s}$ and 
$\Sigma; \Gamma;\Delta \vdash \ENV{ST}$ and 
$\CONF{\TRANSACT{A}{Y}{f}{\VEC{v}}{n}; T, \ENV{ST}} \trans_B \CONF{T, \ENV{S}'; \ENV{T}}$,
then also 
$\Sigma;\Gamma;\Delta \vdash \ENV{S}'$.
\end{lemma}

As the initial step (the `genesis event') does nothing except transforming the declaration $DC$ into the environment representation $\ENV{ST}$, the first result is obvious, and as the rule \nameref{ts_trans} just unwraps a transaction step into a call to the corresponding method.
The second result follows directly from the Preservation theorem;
this can then be generalised in an obvious way to the whole transaction list.
Likewise, the Safety and Soundness theorems can be extended to transactions in the same manner.

\subsection{Noninterference and call integrity}
As immediately evident from Definition~\ref{def:noninterference} and Theorem~\ref{thm:soundness}, well-typedness ensures noninterference (as usual, we denote by $\trans*_B$ the reflexive and transitive closure of $\trans_B$):

\begin{corollary}[Noninterference]\label{cor:noninterference}
Assume a set of security levels $\SECLEVELS \DEFSYM \SET{L, H}$, with $L \ordleq L \ordleq H \ordleq H$, and furthermore that 
$\Sigma;\Gamma;\Delta \vdash T : \TCMD{s}$, 
$\Sigma;\Gamma;\Delta \vdash \ENV{ST}^1$, 
$\Sigma;\Gamma;\Delta \vdash \ENV{ST}^2$,
$\Gamma \vdash \ENV{ST}^1 =_L \ENV{ST}^2$,
$\CONF{T, \ENV{ST}^1} \trans*_B \ENV{ST}^{1'}$, and
$\CONF{T, \ENV{ST}^2} \trans*_B \ENV{ST}^{2'}$.
Then, $\Gamma \vdash \ENV{ST}^{1'} =_L \ENV{ST}^{2'}$. 
\end{corollary}
\begin{proof}
By Theorem~\ref{thm:soundness} and Lemma~\ref{lemma:subject_reduction_transactions}.
\end{proof}

From Corollary~\ref{cor:noninterference}, we then obviously also have that $\Gamma \vdash \ENV{S}^{1'} =_L \ENV{S}^{2'}$, regardless of whether $s$ is $L$ or $H$.
In particular, we can assign security levels to entire contracts, as well as all their members.
Thus, our type system can be used to ensure noninterference according to Definition~\ref{def:noninterference}.

As we previously argued in Remark~\ref{rem:incomparability}, noninterference and call integrity are incomparable properties.
However, as our next theorem shows, well-typedness actually also ensures call integrity.
This is surprising, so before stating the theorem, we should give some hints as to why this is the case.

The definition of call integrity (Definition~\ref{def:call_integrity}) requires the execution of any code in a contract $C$ to be unaffected by all contracts in an `untrusted set' $\SETNAME{Y}$, regardless of whether parts of the code in $\SETNAME{Y}$ execute before, meanwhile or after the code in $C$.
This is expressed by a quantification over all possible traces resulting from a change in $\SETNAME{Y}$, i.e.\@ either in the code or in the values of the fields.
Regardless of any such change, it must hold that the sequence of method calls originating from $C$ be the same.

Noninterference, on the other hand, says nothing about execution traces, but only speaks of the correspondence between values residing in the memory before and after the execution step.
The two counter-examples used in Remark~\ref{rem:incomparability} made use of this fact:
\begin{itemize}
  \item The first counter-example had $C$ be unable to perform any method call, thus obviously satisfying call integrity, but allowed different \code{balance} values to be transferred into it from a `high' context by means of a method call, thereby violating noninterference.
  However, this situation is ruled out by well-typedness (cf.\@ rule \nameref{stm_t_call}), because well-typedness disallows \emph{any} method calls from a `high' to a `low' context, precisely because every method call may transfer the \code{value} parameter along with each call.

  \item The second counter-example had an \code{if} statement in $C$ (the `low' context) depend on a field value in a `high' context.
  The two branches then perform two different method calls, thus enabling a change of the `high' context to induce two different execution traces for $C$.
  Thus, the example satisfies noninterference, because no value stored in the memory is changed, but it obviously does not satisfy call integrity.
  However, this situation is also ruled out by well-typedness, because \nameref{stm_t_if} does not allow the boolean guard expression $e$ in a `low' context to depend on a value from a `high' context.
\end{itemize}

Thus, both of the two counter-examples would be rejected by the type system.
With a setting of $L$ for the `trusted' segment and $H$ for the `untrusted',\footnote{This counter-intuitive naming can perhaps best be thought of as indicating our level of \emph{distrust} in a contract.} no values or computations performed in the untrusted segment can affect the values in the trusted segment, nor the value of any expression in this segment, nor can it even perform a call into the trusted segment.
On the other hand, the trusted segment \emph{can} call out into the untrusted part, but such a call cannot then reenter the trusted segment: it must return before any further calls from the trusted segment can happen.

\begin{theorem}[Well-typedness implies call integrity]\label{thm:welltypedness_callintegrity}
Let $\SECLEVELS \DEFSYM \SET{L, H}$ with $L \ordleq L \ordleq H \ordleq H$.
Fix the two sets of addresses $\SETNAME{X}$ and $\SETNAME{Y}$ as in Definition~\ref{def:call_integrity}, such that $\SETNAME{A} = \SETNAME{X} \UNION \SETNAME{Y}$ and $\SETNAME{A} = \DOM{\ENV{T}}$.
Fix a type assignment $\Gamma$ such that 
\begin{itemize}
  \item $\forall X \in \SETNAME{X} \SUCHTHAT \Gamma(X) = I_L$ for some $I$ where 
  \begin{itemize}
    \item $\forall p \in \Gamma(I) \SUCHTHAT \Gamma(I)(p) = \TVAR{B_L}$, and
    \item $\forall f \in \Gamma(I) \SUCHTHAT \Gamma(I)(f) = \TSPROC{\BSVEC}{L}$, for any $\VEC{B_s}$
  \end{itemize}
\item and with the level $H$ given to all other interfaces, fields and methods.
\end{itemize}
Also assume that
$\Sigma;\Gamma;\Delta \vdash T : \TCMD{s}$,
$\Sigma;\Gamma;\Delta \vdash \ENV{ST}^1$,
$\Sigma;\Gamma;\Delta \vdash \ENV{ST}^2$,
$\Gamma \vdash \ENV{ST}^1 =_L \ENV{ST}^2$,
$\CONF{T, \ENV{ST}^1} \trans[\pi_1]_B \ENV{ST}^{1'}$, and
$\CONF{T, \ENV{ST}^2} \trans[\pi_2]_B \ENV{ST}^{2'}$.
Then, $\pi_1 \BARBSYM_X = \pi_2 \BARBSYM_X$, for any $X \in \SETNAME{X}$. 
\end{theorem}
\begin{proof}
By induction on the length of the transaction $T$.
The base case (viz., $T = \epsilon$) is trivial, since $\pi_1 = \pi_2 = \epsilon$.
For the inductive case, let $T = \TRANSACT{A}{Y}{f}{\VEC{v}}{n}, T'$.
We consider two possible cases.

If $Y \in \SETNAME{Y}$, then the result is immediate by \nameref{stm_t_call}, because no call into \SETNAME{X} is allowed due to the implicit write to \code{balance} in every method call.
Thus, neither of the traces $\pi_1$ or $\pi_2$ would contain any method call from any $X$ in $\SETNAME{X}$, and therefore $\pi_1 \BARBSYM_X = \pi_2 \BARBSYM_X$ obviously holds since $\pi_1\BARBSYM_X = \epsilon = \pi_2\BARBSYM_X$.

If $Y \in \SETNAME{X}$, we know
by assumption that $\Gamma(Y) = I_L$ for some $I$; hence $\ENV{T}^1(Y) = \ENV{T}^2(Y) = \ENV{M}^Y$ and the method called is the same in both runs, say $\ENV{M}^Y(f) = (\VEC{x}, S)$.
Let $\ENV{V} = \makeenv{(\VEC{x} , \VEC{v}) , (\code{this} , Y) , (\code{sender} , A) , (\code{value} , n)}$; then,
\[
\pi_1 = \TRANSACT{A}{Y}{f}{\VEC{v}}{n}, \pi_1', \pi_1''
\qquad\qquad
\pi_2 = \TRANSACT{A}{Y}{f}{\VEC{v}}{n}, \pi_2', \pi_2''
\]
where
\begin{equation}
\label{eq:trans}
  \ENV{T}^1 \vdash \CONF{S, \ENV{S}^1, \ENV{V}} \trans[\pi_1']_S \ENV{S}^{1''}, \ENV{V}^{1'} 
  \qquad\qquad
  \ENV{T}^2 \vdash \CONF{S, \ENV{S}^1, \ENV{V}} \trans[\pi_2']_S \ENV{S}^{2''}, \ENV{V}^{2'}
\end{equation}
and
\[
\CONF{T', \ENV{S}^{1''},\ENV{T}^1} \trans[\pi_1'']_B \ENV{ST}^{1'}
\qquad\qquad
\CONF{T', \ENV{S}^{2''},\ENV{T}^2} \trans[\pi_2'']_B \ENV{ST}^{2'}
\]
By the induction hypothesis, we have that $\pi_1'' \BARBSYM_X = \pi_2'' \BARBSYM_X$, for every $X \in \SETNAME X$.
To prove that $\pi_1' \BARBSYM_X = \pi_2' \BARBSYM_X$, we
then proceed by a second induction, on the inference of the transitions in \eqref{eq:trans}. 
First of all, by Theorem~\ref{thm:extended_soundness} we know that $\Gamma \vdash \ENV{S}^{1'}, \ENV{V}^{1'} =_L \ENV{S}^{2'}, \ENV{V}^{2'}$.
Hence, the last rule used in both inferences is the same: in all cases, this is syntax-driven; moreover, whenever there is a guard that may change the evolution, this guard is evaluated at the same value, because of Theorem~\ref{thm:safety_expressions}.
There are 4 base cases: \nameref{ts_bss_skip}, \nameref{ts_bss_assv}, \nameref{ts_bss_assp} and \nameref{ts_bss_whilefalse}. All cases are trivial, since no call is performed and so $\pi_1' = \epsilon = \pi_2'$.
There are 6 inductive cases, according to the last rule used in the inferences:
\begin{description}
  \item[\nameref{ts_bss_decv}:] Then $S = \code{$\TVAR{B_{s'}}$ $x$ := $e$ in $S'$}$. 
    By Theorem~\ref{thm:safety_expressions}, $\ENV{S}^1, \ENV{V} \vdash e \trans_e v$ and $\ENV{S}^2, \ENV{V} \vdash e \trans_e v$. Then,
    \begin{align*}
       \ENV{T}^1 \vdash \CONF{S', \ENV{S}^1, \joine{(x,v)}{\ENV{V}}} \trans[\pi_1']_S \ENV{S}^{1'}, \joine{(x,v_1)}{\ENV{V}^{1'}} \\
      \ENV{T}^2 \vdash \CONF{S', \ENV{S}^2, \joine{(x,v)}{\ENV{V}}} \trans[\pi_2']_S \ENV{S}^{1'}, \joine{(x,v_2)}{\ENV{V}^{2'}}
    \end{align*}
    and, by the second induction hypothesis, we conclude that $\pi_1' \BARBSYM_X = \pi_2' \BARBSYM_X$, for all $X \in \SETNAME X$.

  \item[\nameref{ts_bss_seq}:] Then $S = S_1; S_2$ and so $\pi_1' = \pi_1^1 , \pi_1^2$ and $\pi_2' = \pi_2^1 , \pi_2^2$.
    By two applications of the second induction hypothesis, we obtain that $\pi_1^1 \BARBSYM_X = \pi_2^1 \BARBSYM_X$ and $\pi_1^2 \BARBSYM_X = \pi_2^2 \BARBSYM_X$, that allow us to easily conclude.

  \item [\nameref{ts_bss_if}:] Then, $S = \code{if $e$ then $S_{\TRUE}$ else $S_{\FALSE}$}$.
    By Theorem~\ref{thm:safety_expressions}, $\ENV{S}^1, \ENV{V} \vdash e \trans_e b$ and $\ENV{S}^2, \ENV{V} \vdash e \trans_e b$; so, the same branch is chosen in both contexts and the same command $S_b$ is executed.
    The case then holds for $S_b$ by the second induction hypothesis.

  \item[\nameref{ts_bss_whiletrue}:]
    The argument is the same as for \nameref{ts_bss_if} above.

\item[\nameref{ts_bss_call}:] Then $S = \CALL{e_1}{g}{\VEC{e}}{e_2}$. By Theorem~\ref{thm:safety_expressions}, $\ENV{S}^1, \ENV{V} \vdash e_1 \trans_e Z$ and $\ENV{S}^2, \ENV{V} \vdash e_1 \trans_e Z$, for some address $Z$ such that $\Gamma(Z) = I_s$ for some interface $I$ and security level $s$.
Moreover, $\ENV{T}^1(Z)(g) = (\VEC{x_1}, S_1)$ and $\ENV{T}^2(Z)(g) = (\VEC{x_2}, S_2)$, for $|\VEC{x_1}| = |\VEC{x_2}| = |\VEC e|$.
Moreover, $\VEC{e}$ and $e_2$ are both typed as $L$ and, by Theorem \ref{thm:safety_expressions}, are evaluated at the same values $\VEC{v'}$ and $n'$.
There are two cases, depending on the level $s$:
\begin{itemize}
  \item If $s=L$, then $S_1 = S_2 = S'$ and $\VEC{x_1} = \VEC{x_2} = \VEC{x'}$.
Now, let $\ENV{V}' = \makeenv{(\VEC{x'} , \VEC{v'}) , (\code{this} , Z) , (\code{sender} , Y) , (\code{value} , n')}$; then,
  \begin{equation*}
    \qquad\qquad \ENV{T}^1 \vdash \CONF{S', \ENV{S}^1, \ENV{V}'} \trans[\hat\pi_1]_S \ENV{S}^{1'}, \ENV{V}^{1'} \qquad
    \ENV{T}^2 \vdash \CONF{S', \ENV{S}^2, \ENV{V}'} \trans[\hat\pi_2]_S \ENV{S}^{1'}, \ENV{V}^{2'}
   \end{equation*}
  By the second induction hypothesis, $\hat\pi_1 \BARBSYM_X\ =\ \hat\pi_2 \BARBSYM_X$, for any $X \in \SETNAME X$; so,
  \begin{equation*}
    \pi_1' \BARBSYM_X{} = \left(\TRANSACT{Y}{Z}{g}{\VEC{v'}}{n'} , \hat\pi_1\right) \BARBSYM_X{} =\ \left(\TRANSACT{Y}{Z}{g}{\VEC{v'}}{n'} , \hat\pi_2\right) \BARBSYM_X = \pi_2' \BARBSYM_X{}
  \end{equation*}

  \item If $s = H$, then $S_1$ and $S_2$ may differ.
    However, the first scenario depicted in the proof of this theorem applies, since $Z \in \SETNAME{Y}$.
    So, by letting $\hat\pi_1$ and $\hat\pi_2$ be the traces generated by $S_1$ and $S_2$ respectively,
   we know that $\hat\pi_1 \BARBSYM_X\ = \epsilon =\ \hat\pi_2 \BARBSYM_X$, for all $X \in \SETNAME X$, because the call can never \emph{reenter} any contract in the `low' segment $\SETNAME{X}$; in particular, no further call from $X$ can appear in the traces.
Hence,
    \begin{equation*}
         \pi_1' \BARBSYM_X{}  =\
         \left(\TRANSACT{Y}{Z}{g}{\VEC{v'}}{n'}\right) \BARBSYM_X{} =\ \pi_2' \BARBSYM_X{}
    \end{equation*}
\end{itemize}
    
\item[\nameref{ts_bss_dcall}:] Then $S = \DCALL{e}{g}{\VEC{e}}$. Like in the previous case, 
$e$ and $\VEC{e}$ are both typed as $L$ and, by Theorem \ref{thm:safety_expressions}, are evaluated at the same values $Z$ and $\VEC{v'}$; moreover, $\Gamma(Z) = I_s$, $\ENV{T}^1(Z)(g) = (\VEC{x_1}, S_1)$ and $\ENV{T}^2(Z)(g) = (\VEC{x_2}, S_2)$, for $|\VEC{x_1}| = |\VEC{x_2}| = |\VEC e|$.
The proof then proceeds like in the previous case, with $n' = 0$.
 \qedhere  
\end{description}
\end{proof}

Theorem~\ref{thm:welltypedness_callintegrity} tells us that \emph{every} contract $X$ in the trusted segment $\SETNAME{X}$ has call integrity w.r.t.\@ the untrusted segment $\SETNAME{Y}$.
This is thus a stronger condition than that of Definition~\ref{def:call_integrity}, which only defines call integrity for a \emph{single} contract $C \in \SETNAME{X}$.
This means that our type system will reject cases where e.g.\@ $C$ calls another contract $Z \in \SETNAME{X}$ and $Z$ calls \code{send()} methods of different contracts, depending on a `high' value.
As \code{send()} is always ensured to do nothing, such calls could never lead to $C$ being reentered, so this would actually still be safe, even though $Z$ itself would not satisfy call integrity.
Thus, this is an example of what resides in the `slack' of our type system.
However, this situation seems rather contrived, since it depends specifically on the \code{send()} method, which is always ensured to do nothing except returning.
For practical purposes, it would be strange to imagine a contract $C \in \SETNAME{X}$ having call integrity w.r.t.\@ $\SETNAME{Y}$, but \emph{without} the other contracts in $\SETNAME{X}$ also satisfying call integrity w.r.t.\@ $\SETNAME{Y}$.
Thus, our type system seems to yield a reasonable approximation to the property of call integrity.

\begin{remark}
As we have previously remarked, the side condition $s \ordleq s_2, s_3$ for method calls is instrumental in ensuring that well-typedness also ensures call-integrity, yet in the case of delegate calls this side condition is not needed.
The absence of this side condition means that an untrusted (i.e.\@ High) contract, say $Y$, \emph{can} perform a \emph{delegate} call into the trusted (i.e.\@ Low) segment in a well-typed program, say, to a contract $X$, unlike the case for ordinary method calls.
However, this does not violate the call-integrity property because,  in the case of delegate calls, the code is executed in the context of the \emph{caller} contract.
The call trace would register a call from $Y$ to $X$, e.g.
\begin{equation*}
  \ENV{T} \vdash \CONF{\DCALL{Y}{f}{\VEC{e}}, \ENV{SV}} \trans[\TRANSACT{Y}{X}{f}{\VEC{v}}{0},\pi]_S \ENV{S}', \ENV{V}
\end{equation*}
but the magic variables \code{this} and \code{sender} are \emph{not} rebound in a delegate call, so if the body of $f$ performs any further calls, say to a method $g$ in a contract $Z$, then that call would \emph{also} be registered in the call trace with $Y$ as the sender.\footnote{Of course, for such a call to be well-typed, it would also be necessary that both the delegate call to \code{$X$.$f$} and the method call to \code{$Z$.$g$} were typable as $\TCMD{H}$, to permit the delegate call from High to Low.}
The control flow does not actually reenter the trusted segment, so well-typedness still implies call-integrity.
\end{remark}


\section{Examples and limitations}\label{sec:examples}
Let us see a few examples of the application of the type system.
To begin with, consider the first counter-example in Remark~\ref{rem:incomparability}, which should be ill-typed by the type system.
In the counter-example we say that \code{X} is Low and \code{Y} is High, so we let them respectively implement the interfaces $\code{IL}$ and $\code {IH}$ defined as follows:

\begin{lstlisting}[style = tinysol]
interface IL : I@$^\top$@ {         interface IH : I@$^\top$@ {
  balance : @$\TVAR{\TINT,L}$@           balance : @$\TVAR{\TINT,H}$@            contract X : IL@$_L$@ { ... }
  send    : @$\TSPROC{}{\SBOT}$@           send    : @$\TSPROC{}{\SBOT}$@
  go      : @$\TSPROC{}{L}$@            go      : @$\TSPROC{}{H}$@             contract Y : IH@$_H$@ { ... }
}                           }
\end{lstlisting}
A part of the failing typing derivation for the body of the method \code{Y.go()} in the declaration of contract \code{Y} is:
\begin{equation}\label{eq:typing_example1}
  \dfrac
  { \dfrac
    { \dfrac
      { \Delta(\code{this}) = \TVAR{\code{IH}_H} }
      { \Sigma;\Gamma;\Delta \vdash \code{this} : \code{IH}_H } \qquad
      \begin{array}{l}
      \Gamma(\code{IH})(\code{balance}) \\
      \qquad\qquad = \TVAR{\TINT_H} 
      \end{array}}
    { \Sigma;\Gamma;\Delta \vdash \code{this.balance} : \TINT_H }
    \qquad 
     \dfrac
      { \TYPEOF{\code{X}} = \Gamma(\code X) = \code {IL}_L }
      { \Sigma;\Gamma;\Delta \vdash \code{X} : \code {IL}_L } \qquad 
      \begin{array}{l}
      \Gamma(\code {IL})(\code{go}) \\
      \qquad \neq \TSPROC{}{H} 
      \end{array}}
  { \Sigma;\Gamma;\Delta \not\vdash \CALL{\code X}{\code{go}}{}{\code{this.balance}} : \TCMD{H} }
\end{equation}

We have that $\Sigma;\Gamma;\Delta \vdash \code{this.balance} : {\TINT_H}$ in contract \code{Y}, so in order for the declaration of method \code{go()} in \code{Y} to be well-typed, the body of the method must be typable as $\TSPROC{}{H}$ by rule \nameref{dec_t_decm}.
However, as the derivation in \eqref{eq:typing_example1} illustrates, this constraint cannot be satisfied, because the lookup $\Gamma(\code {IL})(\code{go})$ yields $\TSPROC{}{L}$, that cannot be promoted to $\TSPROC{}{H}$ through subtyping, because the $\TSPROC{\VEC{B_s}}{s}$ type constructor is contravariant in $s$.

The above example is simple, since the name \code{X} is `hard-coded' directly in the body of \code{Y.go()}, and therefore the type check fails already while checking the contract definition.
However, suppose \code{X} were instead received as a parameter.
Then the signature of the method \code{Y.go} would have instead to be 
\begin{center}
\code{go~:~$\TSPROC{\code {IH}_H}{H}$} 
\end{center}
and the type check would then fail at the call-site, if a Low address were passed.
The following shows a part of the failing typing derivation for the call \code{Y.go(X):this.balance}, where the parameter \code{X} is assumed to implement the interface $\code {IL}$ as before (the typing of \code{this.balance} is like in the leftmost part of \eqref{eq:typing_example1}):
\begin{equation}\label{eq:typing_example2}
  \dfrac
  { \cdots\quad
    \dfrac
      { \TYPEOF{\code Y} = \Gamma(\code{Y}) = \code {IH}_H }
      { \Sigma;\Gamma;\Delta \vdash \code{Y} : \code {IH}_H } \qquad 
      \begin{array}{l}
      \Gamma(\code {IH})(\code{go}) \\
      \quad = \TSPROC{\code {IH}_H}{H} 
      \end{array}
  \quad
    \dfrac
    { \dfrac
      { \TYPEOF{\code X} = \Gamma(\code{X}) = \code {IL}_L }
      { \Sigma;\Gamma;\Delta \vdash \code{X} : \code {IL}_L } 
      \quad
      \dfrac
      { \Sigma(\code {IL}) = \code{I}^\top\!\! \neq \code {IH}
      }
      { \Sigma \not\vdash \code {IL} \subs \code {IH} }
    }
    { \Sigma;\Gamma;\Delta \not\vdash \code{X} : \code {IH}_H } 
  }
  { \Sigma;\Gamma;\Delta \not\vdash \CALL{\code Y}{\code{go}}{\code X}{\code{this.balance}} : \TCMD{H} }
\end{equation}
The method call expects a parameter of type $\code {IH}_H$, but $\code {IL}_L$ cannot be coerced up to $\code {IH}_H$ through subtyping, because of the interfaces definition ($\code {IH}$ and $\code {IL}$ are unrelated in the interface inheritance tree, since they are both children of \code{I}$^\top$).

Thus, the type system  prevents calls from High to Low, regardless of whether the Low address is `hard-coded' or passed as a parameter to a High method.
However, the aforementioned examples also illustrate a limitation of our type system approach to ensuring call integrity: the entire blockchain must be checked, i.e.\@ both the contracts \emph{and} the transactions.
This is necessary since the type check can fail at the call-site of a method, as in the example shown in \eqref{eq:typing_example2}, and the call-site of any method can be a transaction.

Next, we shall briefly consider two examples, reported by Grishchenko et al.\@ in~\cite{grishchenko2018}, of Solidity contracts that are misclassified w.r.t.\@ reentrancy by the static analyser Oyente \cite{oyente}; a false positive and a false negative example.

The false negative example relies on a misplaced update of a field value, just as in the example in Figure~\ref{fig:reentrancy} (page~\pageref{fig:reentrancy}).%
\footnote{It also involves the presence of a `fallback function', which is a special feature of Solidity.
It is a parameterless function that is implicitly invoked by \code{send()}, thus allowing the recipient to execute code upon reception of a currency transfer.
This feature is not present in \TINYSOL{}, yet we can achieve a similar effect by simply allowing the mandatory \code{send()} method to have an arbitrary method body, rather than just \code{skip}.
This has no effect on the type system and associated proofs, since the \code{send()} method is treated as any other method therein.
Hence, this situation is in principle the same as if the sender had invoked some other method than \code{send()}, similarly to the example in Figure~\ref{fig:reentrancy}.
}
In this example, suppose \code{X} were assigned the level $L$ and \code{Y} the level $H$.
With a transaction $\TRANSACT{A}{\code{X}}{\code{transfer}}{\code{Y}}{n}$ (for any address $A$ and any amount of currency $n$), the type system would then correctly reject this blockchain because of the inherent flow from High to Low that is implicit in the call \code{X.transfer(this)} issued by \code{Y}.
The typing derivation would fail in a similar manner as the situation depicted in \eqref{eq:typing_example2}.

The false positive example of Grishchenko et al.\@ from~\cite{grishchenko2018} is also similar to the example in Figure~\ref{fig:reentrancy}, but this time just with the assignment to the guard variable correctly placed \emph{before} the method call (i.e.\@ with lines 6 and 7 switched in Figure~\ref{fig:reentrancy}).
This too would be rejected by our type system, since it does not take the ordering of statements in sequential composition into account (i.e.\@ rule \nameref{stm_t_seq}).
Thus, this example constitutes a false positive for our type system as well, which is hardly surprising.

\begin{figure}[t]
%
%
%
\begin{lstlisting}[style=tinysol]
contract X : IBank@$_L$@ {                           contract Y : IBank@$_H$@ {
  field owner = A;                                field credit = 0;
  transfer(recipient, amount) {                   deposit(owner) {
    if this.sender = this.owner then                this.credit = this.value;
       recipient.deposit(this.sender)$amount        this.owner = owner
    else skip                                     }
  }                                               ...
  ...                                           }
}
\end{lstlisting}
\caption{A two-bank setup.}
\label{fig:twobank}
\end{figure}

Let us now consider a true positive example.
Figure~\ref{fig:twobank} illustrates a part of the code for two banks, which would allow users to store some of their assets and also to transfer assets between them.\footnote{\TINYSOL{} does not have a `mapping' type such as in Solidity, so the setup here is limited to a single user.}
We assume both banks implement the same interface \code{IBank} (so, they must have the same fields and methods -- even though we only explicitly write the relevant ones and hide the reminder in the dots `\code{...}'), but with different security settings: \code{X} is $L$ and \code{Y} is $H$, meaning the latter is untrusted.
There is no callback from \code{Y}, so in this setup a blockchain with a transaction $\TRANSACT{A}{\code{X}}{\code{transfer}}{\code{Y},1}{0}$ would actually be accepted by the type system, because the Low values from \code{X} can safely be coerced up (via subtyping) to match the setting of High on \code{Y}.

To conclude, we now provide some typing insights on delegate calls. Consider the pointer to implementation pattern provided in Figure~\ref{fig:pimpl}.
Assume that contract $\code C$ has been assigned some interface $\code I$ and security level $s_1$; hence, we have to infer that
\begin{equation*}
  \Sigma;\Gamma;\Delta \vdash \code{contract C : I$_{s_1}$ \{} DF\ DM \code{\}}
\end{equation*}

\noindent where $DF$ and $DM$ are the field and method declarations for $\code C$. In particular, this requires to infer, by rule by \nameref{dec_t_decc}, that
\begin{equation*}
  \Sigma;\Gamma;\Delta \vdash_{\code C} \code{f}_i\code{(}\VEC x\code{) \{ dcall impl.f}_i\code{(}\VEC x\code{) \}}
\end{equation*}

Now, \nameref{dec_t_decm} must be used and its premises give us that
\[
\begin{array}{ll}
  \Gamma(\code C)  = (\code I, s_1)              
  &
  \Gamma(\code I)(\code{balance}) = \TVAR{\TINT,s_2}            
  \\
  \Gamma(\code I)(\code{f}_i)  = \TSPROC{\VEC{B_i, s'_i}}{s}\qquad\qquad
  &
  \Sigma;\Gamma;\Delta'  \vdash \code{dcall impl.f}_i\code ( \VEC x \code ) : \TCMD s
\end{array}
\]
where $\Delta' = \code{this}:\TVAR{\code I, s_1}, \code{value}:\TVAR{\TINT, s_2}, \code{sender}:\TVAR{\ITOP, \STOP}, \VEC x : (\VEC{B_i, s'_i})$.
Then, \nameref{stm_t_dcall} is needed to infer the last judgement; from its premises, we have that
\begin{equation*}
\Sigma; \Gamma; \Delta' \vdash \code{impl} : (\code I', s)
\qquad\qquad
\Sigma \vdash (\code I, s_1) \SUBS (\code I', s)
\vspace*{-.5cm}
\end{equation*}

\noindent To infer the last judgement, it must be that $s_1 \ordleq s$ and that there must exist a $j \geq 0$ such that
$\overbrace{\Sigma(\ldots\Sigma(}^j \code I)\ldots) = \code I'$.
Finally, 
\begin{equation*}
\Gamma(\code I')(\code f) = \TSPROC{\VEC{B_i, s'_i}}{s}
\qquad\qquad
\Sigma; \Gamma; \Delta' \vdash \VEC{x} : (\VEC{B_i, s'_i})
\end{equation*}

With this in mind, we can complete the typing of the pointer to implementation pattern with the definitions provided in Figure~\ref{fig:pimpl-ex}. 
In particular, we can set $s_1 = \SBOT = L$ and $s = \STOP = H$, assuming a two-level security lattice. 
If this were an ordinary call, the side condition $s \ordleq s_2, s_3$ on the rule \nameref{stm_t_call} would force both \code{balance} fields to be at level $H$, even though there is no credit movement between the caller and the callee. 
However, this side condition is not present in the type rule for delegate calls (rule \nameref{stm_t_dcall}), so we can also accept e.g.\@ $s_2 = s_3 = L$, because there is no information flow to and from the \code{balance} fields in the case of a delegate call.


\begin{figure}
\begin{minipage}{0.4\textwidth}
\begin{lstlisting}[style = tinysol]
contract C : I@$_{L}$@ {
  field balance := 10;
  field owner := Bob;
  field impl := X;
  send() { skip }
  Update(addr) { 
    if sender = owner 
      then impl := addr 
      else skip 
  }
  @$f_1$@(@$\VEC{x}$@) { dcall impl.@$f_1$@(@$\VEC{x}$@) }
  @$\vdots$@
  @$f_n$@(@$\VEC{x}$@) { dcall impl.@$f_n$@(@$\VEC{x}$@) }
}  
\end{lstlisting}
\end{minipage}
\qquad\qquad
\begin{minipage}{0.4\textwidth}
\begin{lstlisting}[style = tinysol]
interface I : I' {        
  balance : @$\TVAR{\TINT, s_2}$@
  impl : @$\TVAR{\code I'_H}$@
  send : @$\TSPROC{}{L}$@
  Update : @$\TSPROC{\code I'_H}{H}$@
  @$f_1$@ : @$\TSPROC{\VEC{B_1, s'_1}}{H}$@
  @$\vdots$@           
  @$f_n$@ : @$\TSPROC{\VEC{B_n, s'_n}}{H}$@
}  
\end{lstlisting}
\end{minipage}

\begin{minipage}{0.4\textwidth}
\begin{lstlisting}[style = tinysol]
contract X : I@$'_H$@ {
  field balance := 20;
  send() { skip }
  @$f_1$@(@$\VEC{x}$@) { @$S_1$@ }        
  @$\vdots$@           
  @$f_n$@(@$\VEC{x}$@) { @$S_n$@ }
}  
\end{lstlisting}
\end{minipage}
\qquad\qquad
\begin{minipage}{0.4\textwidth}
\begin{lstlisting}[style = tinysol]
interface I' : @$\ITOP$@ {                
  balance : @$\TVAR{\TINT, s_3}$@
  send : @$\TSPROC{}{L}$@
  @$f_1$@ : @$\TSPROC{\VEC{B_1, s'_1}}{H}$@
  @$\vdots$@           
  @$f_n$@ : @$\TSPROC{\VEC{B_n, s'_n}}{H}$@
}  
\end{lstlisting}
\end{minipage}
\caption{A possible realization for the \emph{pointer to implementation} pattern.}
\label{fig:pimpl-ex}
\end{figure}

\section{Related work}
\label{sec:relwork}
In light of the visibility and immutability of smart contracts, which makes it hard to correct errors once they are deployed in the wild, it is not surprising that there has been a substantial research effort within the formal methods community on developing formal techniques to prove safety properties of those programs---see, for instance,~\cite{tolmach2020survey_smart_contracts} for a survey.
The literature on this topic is already huge and the whole gamut of techniques from the field of verification and validation has been adapted to the smart-contract setting. 
For example, this includes contributions employing frameworks based on finite-state machines to design and synthesise Ethereum smart contracts~\cite{MavridouL18}, a variety of static analysis techniques and accompanying tools, such as those presented in~\cite{FeistGG19,KalraGDS18,eThor,TsankovDDGBV18}, and deductive verification~\cite{CassezFA22,CassezFQ22,PZR20}, amongst others. 
The Dafny-based approach reported in~\cite{CassezFQ22} is able to model  arbitrary reentrancy in a setting with the `gas mechanism', whereas~\cite{BramEMSS21} presents a way to analyse safety properties of smart contracts exhibiting reentrancy in a gas-free setting.

The study in~\cite{HuZLZKC21} is close in spirit to ours in that it proposes a type system to ensure the absence of information flows that violate integrity policies in Solidity smart contracts.
That work also presents a type verifier and its prototype implementation within the K-framework~\cite{RosuS10}, which is then applied to analyse more than one hundred smart contracts. 
However, their technique has not been related to call integrity, which, by contrast, is the focus of our work.
Thus, our contribution in the present paper complements this work and serves to further highlight the utility and applicability of secure-flow types in the smart-contract setting.
However, there are also clear differences between this aforementioned work and the present one.
Most notably, our type system uses a more refined subtyping relation, which also handles subtyping of method and address types, whereas subtyping is not defined for the former in \cite{HuZLZKC21}, and the latter is not given a type altogether.
This gives us a more fine-grained control over the information flow, since it allows us to assign different security levels to a contract and its members.
For example, a High contract might have certain Low methods, which hence would not be callable from another High contract, whereas High methods would.
This is in line with standard object-oriented principles, e.g.\@ Java-style visibility modifiers.

Another approach to using a type system to ensure smart-contract safety in a Solidity-like language is presented by Crafa et al.~in~\cite{crafa2019featherweight_solidity}.
This work is indeed related to ours in that both are based on well-known typing principles from object-oriented languages, especially subtyping for contract/address types and the inclusion of a `default' supertype for all contracts, similar to our $\ITOP$.
However, the aim of~\cite{crafa2019featherweight_solidity} is rather different from ours, in that the type system offered in that paper seeks to prevent \emph{runtime errors} that do not stem from a negative account balance, e.g.\@ those resulting from attempts to access nonexistent members of a contract.
Incidentally, such runtime errors would \emph{also} be prevented by our type system (rules \nameref{stm_t_call} and \nameref{stm_t_assf} in particular), due to our use of `interfaces' as address types, if the converse check (ensuring every declared type in an interface has an implementation) were also performed.
However, our focus has been on checking the currency flow, rather than preventing runtime errors of this kind.

The aforementioned paper~\cite{crafa2019featherweight_solidity} introduced Featherweight Solidity (FS). 
Like \TINYSOL{}, FS is a calculus that formalises the core features of Solidity and, as mentioned above, it supports the static analysis of safety properties of smart contracts via type systems. 
Therefore, the developments in the present paper might conceivably have been carried out in FS instead of \TINYSOL{}. 
Our rationale for using \TINYSOL{} is that it provided a very simple language that was sufficient to express the property of call integrity, thus allowing us to focus on the core of this property. 
Of course, `simplicity' is a subjective criterion and the choice of one language instead of another is often a matter of
preference and convenience. 
To our mind, \TINYSOL{} is slightly simpler than FS, which includes functionalities such as fallback functions and different kinds of exceptions. 
Moreover, the big-step semantics of \TINYSOL{} provided was more convenient for the development of our type system than the small-step semantics given for FS. 
Furthermore, unlike FS, \TINYSOL{} also formalises the semantics of blockchains. 
Having said so, \TINYSOL{} and FS are quite similar and it would be interesting to study their similarities in more detail. 
To this end, in future work, we intend to carry out a formal comparison of these two core languages and to see which adaptations to our type system are needed when formulated for FS. 
In particular, we note that FS handles the possibility of an explicit type conversion (type cast) of \code{address} to \code{address payable} by augmenting the \code{address} type with type information about the contract to which it refers. 
This distinction is not present in our version of \TINYSOL{}, as we require all contracts and accounts to have a default \code{send()} function, so all addresses are in this sense `payable'. 
However, our type system does not depend on the presence of a \code{send()} function, so this difference is not important here.

\section{Conclusion and future work}
\label{sec:concl}
In this paper we studied two security properties, namely call integrity and noninterference, in the setting of \TINYSOL{}, a minimal calculus for Solidity smart contracts. 
To this end, we rephrased the syntax of \TINYSOL{} to emphasise its object-oriented flavour, gave a new big-step operational semantics for that language and used it to define call integrity and noninterference. 
Those two properties have some similarities in their definition, in that they both require that some part of a program is not influenced by the other part. 
However, we showed that the two properties are actually incomparable. 
Nevertheless, we provided a type system for noninterference and showed that well-typed programs \emph{also} satisfy call integrity. 
Hence, programs that are accepted by our type systems lie at the intersection of call integrity and noninterference.
A challenging development of our work would be to prove whether the type system exactly characterises the intersection of these two properties, or to find another characterisation of this set of programs. 

There is a publicly available implementation of an interpreter for \TINYSOL, together with a type checker based on the type system described in the present paper.
The code for this implementation is available in \cite{tinysol_interpreter}. 
The repository also contains a small sample of typable and untypable example contracts drawn from the conference version of the present paper \cite{AGL/2024/ecoop/flowtypes}.
A direction for future work is to apply our static analysis methodology (and the associated implementation) to concrete case studies, to better understand the benefits of using a completely static proof technique for call integrity. 
This may also include expanding the collection of examples with \TINYSOL{} versions of actual contracts, to be able to carry out a larger-scale testing of the type checker, similar to the experiments in \cite{HuZLZKC21}. To do so, it would be useful to consider the extension of \TINYSOL{} with a `gas mechanism' \cite{AGLM/2024/reocas/outofgas}, allowing one to prove the termination of transactions and to compute their computational cost.

However, many actual contracts also rely on a feature that we have not included in the present formulation of \TINYSOL,
namely the \emph{fallback function}.
This construct is a peculiar feature of the Solidity smart-contract language, which was the source of a spectacular security breach (the so-called \emph{Parity Wallet Hack}, described in e.g.\@ \cite{HuZLZKC21}).
A fallback function is a special, parameterless method, which is executed in case a method call \emph{fails}, because the callee contract does not contain a method by the called name.
With fallback functions,  in such a case, control would \emph{still} be transferred to the callee contract, and the callee can then use the fallback function to perform some error handling, or even to forward the call to some other method via a delegate call.
This feature is also used to implement a  form of dynamic dispatch of method calls in Solidity.
Unfortunately, the fallback function construct is untypable by ordinary syntactic means, such as the type system presented in the present paper.
It admits a form of introspection into the language, because the body of a fallback function has access to information \emph{about} the failed method call, such as the name of the non-existing method, and the list of actual parameters.
This information depends on how the method was called (i.e.\@ on the form of the failed method call), and it is therefore only available at runtime.
Hence, it cannot be given a static  type, which is necessary for a static type check.
On the other hand, the present type system already prohibits  calls to non-existing methods because of our interface types for contracts, thereby rendering fallback functions useless in well-typed programs.
This is doubly problematic from a practical perspective, since the fallback function \emph{is} part of the Solidity language.
Work is currently underway to develop a method based on a \emph{semantic approach} to type soundness (cf.\@ \cite{caires2007, timany/2024/acm/semantic_typing, ahmed2004semantic_types_phd}) to allow type safety to be recovered for \TINYSOL{} contracts containing fallback functions, without disallowing usage of the construct altogether.


A potential limitation of the approach presented in this paper is that the entire blockchain must be checked to show call integrity of a contract. 
Indeed, since a typing derivation can fail at the call-site and the call-site of a method can be a transaction, transactions must be well-typed too. 
In passing, we note that this kind of problem is also present in~\cite{SV98,volpano2000secure_flow_typesystem} (and, in general, in many works on type systems for security), where the whole code needs to be typed in order to obtain the desired guarantees.
We think that an important avenue for future work, and one we intend to pursue, is to explore whether, and to what extent, other typing disciplines (like the semantic approach mentioned above) can be employed to mitigate this problem. 

\appendix

\section{Proof of Theorem~\ref{thm:subject_reduction}}\label{app:proof_subject_reduction}
\begin{proof}
By induction on the derivation of $\ENV{T} \vdash \CONF{S, \ENV{SV}} \trans_S \ENV{SV}'$.
There are 4 base cases (corresponding to the rules of Fig. \ref{fig:tinysol_semantics_statements_bss} that have no occurrence of $\trans_S$ in their premise):
\begin{itemize}
  \item If \nameref{ts_bss_skip} was used, then $S$ = \code{skip} and
    the result is immediate, since \code{skip} does not affect $\ENV{SV}$.

  
    \item If \nameref{ts_bss_whilefalse} was used, then $S$ = \code{while $e$ do $S'$}, with $e$ that evaluates to $\FALSE$.
        The result is immediate, since $\ENV{SV}$ is not modified in the transition.

  \item If \nameref{ts_bss_assv} was used, then $S$ = \code{$x$ := $e$} and
    from its premise we know that  $\ENV{SV} \vdash e \trans_e v$ and $\ENV{V}' = \ENV{V}\EXTEND{x}{v}$.
    By the premise of \nameref{stm_t_assv},     $\Delta(x) = \TVAR{B_s}$ and $\Sigma;\Gamma;\Delta \vdash e : B_s$.
    As $s'$ is strictly lower than $s$, we therefore conclude $\Gamma \vdash \ENV{SV} =_{s'} \ENV{S}, \ENV{V}\EXTEND{x}{v}$.

  \item If \nameref{ts_bss_assp} was used, then $S$ = \code{this.$p$ := $e$}.
    The argument is the same as above, except that the update affects $\ENV{S}$, rather than $\ENV{V}$.
\end{itemize}
For the inductive step, we distinguish the last rule used in the inference; we have the following 6 cases to consider:
\begin{itemize}
    \item If \nameref{ts_bss_seq} was used, then $S$ = \code{$S_1$;$S_2$} and
    from its premise we know that 
    \begin{align*} 
      \ENV{T} \vdash \CONF{S_1, \ENV{SV}} \trans_S \ENV{SV}''  \qquad\qquad
      \ENV{T} \vdash \CONF{S_2, \ENV{SV}''} \trans_S \ENV{SV}'
    \end{align*}
   Moreover, the last rule used to type $S$ is \nameref{stm_t_seq}, whose premise ensures that
    $\Sigma;\Gamma;\Delta \vdash S_1 : \TCMD{s}$ and $\Sigma;\Gamma;\Delta \vdash S_2 : \TCMD{s}$.
    By two applications of the induction hypothesis, the statement holds for both premises; 
    so, we obtain that $\Gamma;\Delta \vdash \ENV{SV} =_{s'} \ENV{SV}''$ and
    $\Gamma;\Delta \vdash \ENV{SV}'' =_{s'} \ENV{SV}'$, and we conclude by transitivity.

  \item If \nameref{ts_bss_if} was used, then $S$ = \code{if $e$ then $S_\TRUE$ else $S_\FALSE$} and
    from its premise we know that  $\ENV{T} \vdash \CONF{S_b, \ENV{SV}} \trans_S \ENV{SV}'$, 
    where $b \in \SET{\TRUE, \FALSE}$ depending on the evaluation value of $e$.
    By \nameref{stm_t_if}, we know that     
    $\Sigma;\Gamma;\Delta \vdash S_\TRUE : \TCMD{s}$ and $\Sigma;\Gamma;\Delta \vdash S_\FALSE : \TCMD{s}$;
    regardless of which branch was chosen, the statement then holds by the induction hypothesis.

  \item If \nameref{ts_bss_whiletrue} was used, then $S$ = \code{while $e$ do $S'$}, with $e$ that evaluates to $\TRUE$; from its premise, we know that 
  $\ENV{T}  \vdash \CONF{S', \ENV{SV}} \trans_S \ENV{SV}'' $ and $\ENV{T}  \vdash \CONF{\code{while $e$ do $S'$}, \ENV{SV}''} \trans_S \ENV{SV}'$.
  By the premise of rule \nameref{stm_t_while}, we know that $\Sigma;\Gamma;\Delta \vdash S' : \TCMD{s}$.
  The statement then holds by two applications of the induction hypothesis.

  \item If \nameref{ts_bss_decv} was used, then $S$ = \code{$\TVAR{B_{s''}}$ $x$ := $e$ in $S'$}, where $x \notin \DOM{\ENV{V}'}$;
  from its premise we know that 
    \begin{align*}
      \ENV{SV} \vdash e \trans_e v \qquad\qquad
      \ENV{T}  \vdash \CONF{S', \ENV{S}\ , \joine{(x,v)}{\ENV{V}}} \trans_S \ENV{S}'\ , \joine{(x,v')}{\ENV{V}'}
    \end{align*}
    By the premises of \nameref{stm_t_decv}, we have that 
    $\Sigma; \Gamma; \Delta \vdash e : B_{s''}$ and $\Sigma; \Gamma; \Delta, x : \TVAR{B_{s''}} \vdash S' : \TCMD{s}$.
    We can then apply the induction hypothesis to conclude that $\Gamma;\Delta, x : \TVAR{B_{s''}} \vdash \ENV{S}\ , \joine{(x,v)}{\ENV{V}} =_{s'} \ENV{S}'\ , \joine{(x,v')}{\ENV{V}'}$.
    By Lemma~\ref{lemma:strengthening_env_v}, we can then conclude $\Gamma;\Delta \vdash \ENV{SV} =_{s'} \ENV{SV}'$, as required.

  \item If \nameref{ts_bss_call} was used, then $S = \CALL{e_1}{f}{\VEC{e}}{e_2}$ and it was typed by using \nameref{stm_t_call}.
  From the premise of \nameref{ts_bss_call}, we know that $\ENV{V}(\code{this}) = X$ (the caller), 
  $\ENV{SV} \vdash e_1 \trans_e Y$ (the callee), and there are two writes to the \code{balance} fields of caller and callee (which are typed as $\TVAR{\TINT,s_2}$ and $\TVAR{\TINT,s_3}$, for $s \ordleq s_2,s_3$), that generate the new environment $\ENV{S}''$ (that only differs from $\ENV{S}$ for the values assigned to the balances); so,  $\Gamma \vdash \ENV{S} =_{s'} \ENV{S}''$.
  Moreover, $\ENV{SV} \vdash e_2 \trans_e n$, $\ENV{SV} \vdash \VEC{e} \trans_e \VEC{v}$ and $\ENV{T} \vdash \CONF{S', \ENV{SV}''} \trans_S \ENV{SV}'$, 
  where $S'$ is the body of the method and
  $\ENV{V}'' = \makeenv{(\code{this}, Y)), (\code{sender}, X), (\code{value}, n),(\VEC{x}, \VEC{v}})$. 
  
  By the premises of \nameref{stm_t_call}, we know that
    \[
    \qquad\Sigma;\Gamma;\Delta \vdash e_1 : I^Y_s \quad \Gamma(I^Y)(f) = \TSPROC{\VEC{B_s}}{s}  \quad 
          \Sigma;\Gamma;\Delta \vdash \VEC{e} : \VEC{B_s}  \quad  \Sigma;\Gamma;\Delta \vdash e_2 : \TINT_s \quad
    \Delta(\code{this}) = \TVAR{I^X,s_1} 
    \]
  Moreover, from \nameref{dec_t_decm} we also know that $\Sigma;\Gamma;\Delta' \vdash S' : \TCMD{s}$, where $\Delta' = \code{this} : \TVAR{I^Y_s},\code{sender} : \TVAR{I^\top,s_\top},\code{value} : \TVAR{\TINT_s},\VEC{x} : \VEC{\TVAR{{B_s}}}$, where $\Sigma \vdash (I^X,s_1) \SUBS (I^\top,s_\top)$.
  Hence, $\Sigma;\Gamma;\Delta' \vdash \ENV{V}''$ holds by \nameref{env_t_envv}.
By the induction hypothesis, $\Gamma;\Delta' \vdash \ENV{SV}'' =_{s'} \ENV{SV}'$; so, $\Gamma \vdash \ENV{S} =_{s'} \ENV{S}'$
that trivially implies $\Gamma;\Delta \vdash \ENV{SV} =_{s'} \ENV{S}', \ENV{V}$, as desired.

  \item If \nameref{ts_bss_dcall} was used, then $S = \DCALL{e}{f}{\VEC{e}}$; from its premise, we know that 
  $\ENV{SV} \vdash e \trans_e Y$, $\ENV{SV} \vdash \VEC{e} \trans_e \VEC{v}$ and $\ENV{T} \vdash \CONF{S', \ENV{S},\ENV{V}''} \trans_S \ENV{SV}'$, 
  where $S'$ is the body of the method and
  $\ENV{V}'' = \makeenv{(\code{this}, \ENV{V}(\code{this})), \linebreak (\code{sender}, \ENV{V}(\code{sender})),  (\code{value}, \ENV{V}(\code{value})), (\VEC{x}, \VEC{v})}$. 

  By the premises of \nameref{stm_t_dcall}, we know that
    \[
    \qquad\quad \Sigma;\Gamma;\Delta \vdash e : I^Y_s \quad \Gamma(I^Y)(f) = \TSPROC{\VEC{B_s}}{s}  \quad 
          \Sigma;\Gamma;\Delta \vdash \VEC{e} : \VEC{B_s}  \quad
    \Delta(\code{this}) = \TVAR{I^X,s_1} 
    \quad  \Sigma \vdash  : (I^X,s_1) \SUBS (I^Y,s) 
    \]
  Moreover, from \nameref{dec_t_decm} we also know that $\Sigma;\Gamma;\Delta' \vdash S' : \TCMD{s}$, where $\Delta' = \code{this} : \TVAR{I^Y_s},\code{sender} : \TVAR{I^\top,s_\top},\code{value} : \TVAR{\TINT_s},\VEC{x} : \VEC{\TVAR{{B_s}}}$, where $\Sigma \vdash (I^X,s_1) \SUBS (I^\top,s_\top)$.
  Hence, $\Sigma;\Gamma;\Delta' \vdash \ENV{V}''$ holds by \nameref{env_t_envv}.
By the induction hypothesis, $\Gamma;\Delta' \vdash \ENV{S},\ENV{V}'' =_{s'} \ENV{SV}'$
that trivially implies $\Gamma;\Delta \vdash \ENV{SV} =_{s'} \ENV{S}', \ENV{V}$, as desired.
 \qedhere
\end{itemize}
\end{proof}

\section{Proof of Theorem~\ref{thm:safety_expressions}}\label{app:proof_safety_expressions}
\begin{proof}
By induction on the derivation of $\Sigma;\Gamma;\Delta \vdash e : B_s$.
There are 2 base cases:
\begin{itemize}
  \item If \nameref{exp_t_val} was used to conclude $\Gamma \vdash e : B_s$, then $e$ = $v$. 
    The result is immediate, since the value does not depend on $\ENV{SV}$.

  \item If \nameref{exp_t_var} was used, then $e$ = $x$ and, from the side condition of the rule, we have that $\Delta(x) = \TVAR{B_s}$.
    By assumption, $\Gamma;\Delta \vdash \ENV{SV}^1 =_s \ENV{SV}^2$; so, by \nameref{env_s_eq_envv}, $\ENV{V}^1(x) = \ENV{V}^2(x)$.
    Thus, we can conclude by \nameref{Exp-Var}.
\end{itemize}
For the inductive case, we distinguish the last rule used in the inference; we have the following 3 cases:
\begin{itemize}
  \item If \nameref{exp_t_field} was used, then $e$ = $e'.p$; moreover, 
    from the premise and side condition of that rule, we know that $\Sigma;\Gamma;\Delta \vdash e' : I_s$ and $\Gamma(I)(p) = \TVAR{B_s}$.
    By assumption, $\Gamma;\Delta \vdash \ENV{SV}^1 =_s \ENV{SV}^2$;
    therefore, by the induction hypothesis, $\ENV{SV}^1 \vdash e' \trans_e X$ and $\ENV{SV}^2 \vdash e' \trans_e X$.
    Then, $\ENV{S}^1(X)(p) = \ENV{S}^2(X)(p)$ and we can conclude by \nameref{Exp-Field}.

  \item If \nameref{exp_t_op} was used, then $e$ = $\op(e_1, \ldots, e_n)$.
    From the premise, we know that each of the arguments $e_i$ are typable as $\Sigma;\Gamma;\Delta \vdash e_i : (B_i,s)$; so, by $n$ applications of the induction hypothesis, we get that $\ENV{SV}^1 \vdash e_i \trans_e v_i$ and $\ENV{SV}^2 \vdash e_i \trans_e v_i$. We can conclude by rule \nameref{Exp-Op}.

  \item If \nameref{exp_t_sub} was used, then we know from the premise that $\Sigma;\Gamma;\Delta \vdash e : B'_{s'}$ and $\Sigma \vdash (B',s') \subs (B,s)$.
    By rule \nameref{typerules_sub_type}, we know that $s' \ordleq s$;
    hence, by Lemma~\ref{lemma:restriction_s}, $\Gamma;\Delta \vdash \ENV{SV}^1 =_{s'} \ENV{SV}^2$ also holds.
    The claim holds because of the induction hypothesis.
\qedhere
\end{itemize}
\end{proof}

\section{Details of the Proof of Theorem~\ref{thm:soundness}}\label{app:proof_soundness}
\begin{proof}
%
%
We consider here the case when $s_1 \ordleq s_2$;
we proceed by induction on the sum of the lengths of the inferences for
$\ENV{T} \vdash \CONF{S, \ENV{SV}^1} \trans_S \ENV{SV}^{1'}$, and
$\ENV{T} \vdash \CONF{S, \ENV{SV}^2} \trans_S \ENV{SV}^{2'}$. 
The first observation is that the choice of the last rule in both inferences is syntax-driven (according to the outmost operator in $S$), and so both inferences terminate with the very same rule of Fig. \ref{fig:tinysol_semantics_statements_bss}. The only case when the syntax does not univocally identifies the rule to use is when  $S$ is a \code{while}. So, we preliminarily prove that also in this case no uncertainty is possible. Indeed, let $\ENV{SV}^1 \vdash e \trans_e b_1$ and $\ENV{SV}^2 \vdash e \trans_e b_2$,
where $b_1,b_2 \in \SET{\TRUE, \FALSE}$. Since we are assuming that $s_1 \ordleq s_2$ and by hypothesis $\Gamma \vdash \ENV{SV}^1 =_{s_2} \ENV{SV}^2$, we can conclude that $b_1 = b_2$ by Theorem~\ref{thm:safety_expressions}. Hence, both inferences terminate either with \nameref{ts_bss_whilefalse} or \nameref{ts_bss_whiletrue}, according to the evaluation of the guard.

There are 4 base cases (corresponding to the rules of Fig. \ref{fig:tinysol_semantics_statements_bss} that have no occurrence of $\trans_S$ in their premise):
\begin{itemize}
  \item If \nameref{ts_bss_skip} was used to infer both transitions, then $S$ = \code{skip} and the result is immediate, since \code{skip} does not affect $\ENV{SV}^1$ and $\ENV{SV}^2$.

 
  \item If \nameref{ts_bss_whilefalse} is used to conclude both transitions, then $S$ = \code{while $e$ do $S'$} and
    the result is immediate, since $\ENV{SV}^{1'} = \ENV{SV}^1$ and $\ENV{SV}^{2'} = \ENV{SV}^2$.

   \item If \nameref{ts_bss_assv} was used to infer both transitions, then $S$ = \code{$x$ := $e$}. From the premises of that rule, we know that
    $\ENV{SV}^1 \vdash e \trans_e v_1$ and $\ENV{SV}^2 \vdash e \trans_e v_2$; thus,
    $\ENV{SV}^{1'} = \ENV{S}^1, \ENV{V}^1\EXTEND{x}{v_1}$ and $\ENV{SV}^{2'} = \ENV{S}^2, \ENV{V}^2\EXTEND{x}{v_2}$.
   By the premise of \nameref{stm_t_assv}, $\Delta(x) = \TVAR{B,s_1}$ and $\Sigma;\Gamma;\Delta \vdash e : (B,s_1)$.
   Since by assumption $s_1 \ordleq s_2$ and $\Gamma;\Delta \vdash \ENV{SV}^1 =_{s_2} \ENV{SV}^2$,
  by Theorem~\ref{thm:safety_expressions} we can conclude that $v_1 = v_2 = v$.
  Thus, $\Gamma;\Delta \vdash \ENV{S}^1, \ENV{V}^1\EXTEND{x}{v} =_{s_2} \ENV{S}^2, \ENV{V}^2\EXTEND{x}{v}$ also holds.

  \item If \nameref{ts_bss_assp} was used, then $S$ = \code{this.$p$ := $e$}.
  The argument is then the same as above, except that it is $\ENV{S}$, rather than $\ENV{V}$, that is updated.

\end{itemize}
For the inductive step, we distinguish the last rule used in both inferences (as we argued above, the two transitions have been inferred by using the same last rule). We have the following 6 cases to consider:
\begin{itemize}
  \item If \nameref{ts_bss_seq} was used to conclude both transitions, then $S$ = \code{$S_1$;$S_2$}. From the premises of that rule, we know that
  \[
    \ENV{T} \vdash \CONF{S_1, \ENV{SV}^1} \trans_S \ENV{SV}^{1''}  \qquad
    \ENV{T} \vdash \CONF{S_2, \ENV{SV}^{1''}} \trans_S \ENV{SV}^{1'}
  \]
and 
  \[
    \ENV{T} \vdash \CONF{S_1, \ENV{SV}^2} \trans_S \ENV{SV}^{2''}  \qquad
    \ENV{T} \vdash \CONF{S_2, \ENV{SV}^{2''}} \trans_S \ENV{SV}^{2'}
  \]
Moreover, from the premises of \nameref{stm_t_seq}, we know that $\Sigma;\Gamma;\Delta \vdash S_1 : \TCMD{s_1}$ and $\Sigma;\Gamma;\Delta \vdash S_2 : \TCMD{s_1}$. By two applications of the induction hypothesis, we conclude that $\Gamma;\Delta \vdash \ENV{SV}^{1''} =_{s_2} \ENV{SV}^{2''}$ and
$\Gamma;\Delta \vdash \ENV{SV}^{1'}  =_{s_2} \ENV{SV}^{2'}$, as required.

  \item If \nameref{ts_bss_if} was the last rule used in both inferences, then $S$ = \code{if $e$ then $S_\TRUE$ else $S_\FALSE$}. Like in the case of the \code{while} discussed at the beginning of this proof, we have that $b_1 = b_2 = b \in \SET{\TRUE, \FALSE}$, where
$\ENV{SV}^1 \vdash e \trans_e b_1$ and $\ENV{SV}^2 \vdash e \trans_e b_2$; thus, the same branch $S_b$ is chosen in both inferences, that is
$\ENV{T} \vdash \CONF{S_{b}, \ENV{SV}^1} \trans_S \ENV{SV}^{1'}$ and $\ENV{T} \vdash \CONF{S_{b}, \ENV{SV}^2} \trans_S \ENV{SV}^{2'}$.
Regardless of the value of $b$, from the premise of \nameref{stm_t_if} we know that $\Sigma;\Gamma;\Delta \vdash e : s_1$ and $\Gamma \vdash S_b :  \TCMD{s_1}$. The statement then holds by the induction hypothesis.

  \item If \nameref{ts_bss_whiletrue} is used to conclude both the inferences, then $S$ = \code{while $e$ do $S'$} and
    from the premises of that rule we know that
    \begin{align*}
      \ENV{T} \vdash \CONF{S', \ENV{SV}^1} \trans_S \ENV{SV}^{1''} \qquad
      \ENV{T} \vdash \CONF{\code{while $e$ do $S'$}, \ENV{SV}^{1''}} \trans_S \ENV{SV}^{1'} 
    \end{align*}
and
    \begin{align*}
      \ENV{T} \vdash \CONF{S', \ENV{SV}^2} \trans_S \ENV{SV}^{2''} \qquad
      \ENV{T} \vdash \CONF{\code{while $e$ do $S'$}, \ENV{SV}^{2''}} \trans_S \ENV{SV}^{2'}
    \end{align*}
  Moreover, by the premise of \nameref{stm_t_while}, we know that $\Sigma;\Gamma;\Delta \vdash e : s_1$ and $\Sigma;\Gamma;\Delta \vdash S' : \TCMD{s_1}$.
  The statement then holds by two applications of the induction hypothesis, like in the case for \nameref{ts_bss_seq} above.

  \item If \nameref{ts_bss_decv} was the last rule in both inferences, then $S$ = \code{$\TVAR{B}$ $x$ := $e$ in $S'$}; from the premises of that rule, we have that
  \[
    \ENV{SV}^1 \vdash e \trans_e v_1 \qquad
    \ENV{T} \vdash \CONF{S', \ENV{S}^1, \joine{(x,v_1)}{\ENV{V}^1}} \trans_S \ENV{S}^{1'}, \joine{(x,v_1')}{\ENV{V}^{1'}}
  \]
and
    \[
    \ENV{SV}^2 \vdash e \trans_e v_2 \qquad
    \ENV{T}    \vdash \CONF{S', \ENV{S}^2, \joine{(x,v_2)}{\ENV{V}^2}} \trans_S \ENV{S}^{2'}, \joine{(x,v_2')}{\ENV{V}^{2'}}
  \]
  where $x \notin \DOM{\ENV{V}^{1}} \cup \DOM{\ENV{V}^{2}}$.
   
  From the premise of \nameref{stm_t_decv}, we know that $\Sigma;\Gamma;\Delta \vdash e : B_s$ and $\Sigma;\Gamma;\Delta, x : \TVAR{B_s} \vdash S' : \TCMD{s_1}$.
  In order to apply the induction hypothesis, we must show that
  \begin{equation}
  \label{eq:equality}
    \Gamma; \Delta, x : \TVAR{B_s} \vdash \ENV{S}^1, \joine{(x, v_1)}{\ENV{V}^1} = _{s_2} \ENV{S}^2, \joine{(x,v_2)}{\ENV{V}^2}
  \end{equation}
 If $s \ordleq s_2$, then \eqref{eq:equality} holds by Theorem~\ref{thm:safety_expressions}: indeed, $v_1 = v_2 = v$, since all variables read within $e$ are of a lower level than $s_2$, and by assumption, $\ENV{SV}^1$ and $\ENV{SV}^2$ agree on all values up to, and including, $s_2$.
 If $s \not\ordleq s_2$, then 
\eqref{eq:equality} holds by definition of the $\Gamma;\Delta \vdash \cdot =_s \cdot$ relation.

Hence, by the induction hypothesis we have that
$\Gamma;\Delta, x : \TVAR{B_s} \vdash \ENV{S}^{1'}, \joine{(x,v_1')}{\ENV{V}^{1'}} =_{s_2} \ENV{S}^{2'}, \joine{(x,v_2')}{\ENV{V}^{2'}}$;
by Lemma~\ref{lemma:strengthening_env_v}, we can then conclude $\Gamma;\Delta \vdash \ENV{SV}^{1'} =_{s_2} \ENV{SV}^{2'}$, as required.

  \item If \nameref{ts_bss_call} was the last rule, then $S = \CALL{e_1}{f}{\VEC{e}}{e_2}$ and so it has been typed by using \nameref{stm_t_call}.
  
First, from the premises of \nameref{ts_bss_call} we know that $\ENV{V}^1(\code{this}) = X_1$ and $\ENV{V}^2(\code{this}) = X_2$. 
From the premise of \nameref{stm_t_call}, we have that $\Delta(\code{this}) = \TVAR{I^X,s_1'}$ with $s_1' \ordleq s_1$.
Since $\Delta \vdash \ENV V^1 =_{s_2} \ENV V^2$ and $s_1 \ordleq s_2$, we can conclude that $X_1 = X_2 = X$, by \nameref{env_s_eq_envv}.

Second, from the premises of \nameref{ts_bss_call} and \nameref{stm_t_call}, we have that 
    \begin{align*}
    \ENV{SV}^1 \vdash e_1 \trans_e Y_1 \qquad
    \ENV{SV}^2 \vdash e_1 \trans_e Y_2 \qquad
    \Sigma;\Gamma;\Delta \vdash e_1 : (I^Y,s_1)
\\
    \ENV{SV}^1 \vdash e_2 \trans_e n_1 \qquad
    \ENV{SV}^2 \vdash e_2 \trans_e n_2 \qquad
    \Sigma;\Gamma;\Delta \vdash e_2 : (\TINT,s_1)
  \end{align*}
By Theorem~\ref{thm:safety_expressions}, we conclude that $Y_1 = Y_2 = Y$ and $n_1 = n_2 = n$, since we are assuming that $s_1 \ordleq s_2$.

Third, for the list of actual parameters, fix an arbitrary expression $e_i \in \VEC e$.
By the premises of \nameref{ts_bss_call} and \nameref{stm_t_call}, we have that
  \[
    \ENV{SV}^1 \vdash e_i \trans_e v_i^1 \qquad
    \ENV{SV}^2 \vdash e_i \trans_e v_i^2 \qquad
    \Sigma;\Gamma;\Delta \vdash e_i : (B_i,s_i)
  \]
  If $s_i \ordleq s_2$, then by Theorem~\ref{thm:safety_expressions} we have that $v_i^1 = v_i^2$.
Otherwise, it may be the case that $v_i^1 \neq v_i^2$, but since $s_i \not\ordleq s_2$, then this assignment will still satisfy the condition that the variable-environments will agree up to level $s_2$.

Being the callee and the caller the same in both transitions, by \nameref{env_s_eq_envf} we have that $\Gamma \vdash_{I^X} \ENV F^{1X} =_{s_2} \ENV F^{2X}$ and $\Gamma \vdash_{I^Y} \ENV F^{1Y} =_{s_2} \ENV F^{2Y}$, where $\ENV F^{iZ} = \ENV S^i(Z)$ for $i \in \{1,2\}$ and $Z \in \{X,Y\}$.
Since \nameref{stm_t_call} entails that $\Gamma(I^X)(\code{balance}) = \TVAR{\TINT,s_2'}$ and $\Gamma(I^Y)(\code{balance}) = \TVAR{\TINT,s_3'}$, with $s_1 \ordleq s_2',s_3'$, we have that
\begin{equation}
\label{eq:indHypOne}
\Gamma \vdash \ENV{S}^{1''} =_{s_2} \ENV{S}^{2''}
\end{equation}
where 
    \begin{align*}
\ENV{S}^{1''} = \ENV{S}^1
    \EXTEND{X}{\ENV{F}^{1X} [\code{balance -= } n]}
    \EXTEND{Y}{\ENV{F}^{1Y} [\code{balance += } n]}
\\
\ENV{S}^{2''} = \ENV{S}^2
    \EXTEND{X}{\ENV{F}^{2X} [\code{balance -= } n]}
    \EXTEND{Y}{\ENV{F}^{2Y} [\code{balance += } n]}
    \end{align*}
From the premises of \nameref{ts_bss_call} we have that
\begin{equation}
\label{eq:indTrans}
    \ENV{T} \vdash \CONF{S', \ENV{SV}^{1''}} \trans_S \ENV{SV}^{1'} \qquad
    \ENV{T} \vdash \CONF{S', \ENV{SV}^{2''}} \trans_S \ENV{SV}^{2'}
\end{equation}
where $\ENV T (Y)(f) = (\VEC x , S')$ and
    \begin{align*}
\ENV{V}^{1''} = \{(\code{this}, Y) , (\code{sender}, X) , (\code{value}, n) , (x_1, v_1^1) , \ldots , (x_k, v_k^1)\}
\\
\ENV{V}^{2''} = \{(\code{this}, Y) , (\code{sender}, X) , (\code{value}, n) , (x_1, v_1^2) , \ldots , (x_k, v_k^2)\}
    \end{align*}
From what we proved above, by \nameref{env_s_eq_envv} it easily follows that
\begin{equation}
\label{eq:indHypTwo}
\Delta' \vdash \ENV{V}^{1''} =_{s_2} \ENV{V}^{2''}
\end{equation}
where $\Delta' = \code{this} : \TVAR{I^Y,s_1},\code{sender} : \TVAR{I^\top,s_\top},\code{value} : \TVAR{\TINT,s_1},\VEC{x} : \VEC{\TVAR{{B_s}}}$
and $\Sigma \vdash (I^X,s_1') \SUBS (I^\top,s_\top)$.

Finally, from the premise of \nameref{dec_t_decm}, we have that $\Sigma;\Gamma;\Delta' \vdash S' : \TCMD{s_1}$.
This, together with 
$\Gamma \vdash \ENV{T}^{1} =_{s_2} \ENV{T}^{2}$ (that holds by hypothesis),
$\Gamma;\Delta' \vdash \ENV{SV}^{1''} =_{s_2} \ENV{SV}^{2''}$ (that holds because of \eqref{eq:indHypOne} and \eqref{eq:indHypTwo}), and
\eqref{eq:indTrans}, allows us to apply the inductive hypothesis and obtain 
 that $\Gamma;\Delta' \vdash \ENV{SV}^{1'} =_{s_2} \ENV{SV}^{2'}$.
This allows us to conclude that $\Gamma;\Delta \vdash \ENV{S}^{1'}, \ENV{V}^1 =_{s_2} \ENV{S}^{2'}, \ENV{V}^2$, as required.

  \item If \nameref{ts_bss_dcall} was used, then $S = \DCALL{e}{f}{\VEC{e}}$ and so it has been typed by using \nameref{stm_t_dcall}.
  By the premises of \nameref{stm_t_dcall}, we know that
    \[
    \qquad\quad \Sigma;\Gamma;\Delta \vdash e : (I^Y,s_1) \quad \Gamma(I^Y)(f) = \TSPROC{\VEC{B_s}}{s_1}  \quad 
          \Sigma;\Gamma;\Delta \vdash \VEC{e} : \VEC{B_s}  \quad
    \Delta(\code{this}) = \TVAR{I^X,s_1'} 
    \quad  \Sigma \vdash  : (I^X,s_1') \SUBS (I^Y,s_1) 
    \]

From $\Delta \vdash \ENV V^1 =_{s_2} \ENV V^2$, we obtain that $\ENV V^1(\code{this})=\ENV V^2(\code{this})$, being $s_1' \ordleq s_1$.
Moreover, like in the case for \nameref{ts_bss_call}, we can prove that the callee (given by evaluating $e$ and yielding $Y$) is the same in the two transitions; the same happens for the actual parameters (obtained by evaluating $\VEC e$ and yielding $\VEC{v_1}$ and $\VEC{v_2}$) whose level is smaller than or equal to $s_2$. So, by \nameref{env_t_envv} and \nameref{env_s_eq_envv}
\begin{equation}
\label{eq:indHyp}
\Sigma;\Gamma;\Delta' \vdash \ENV{V}^{1''}
\qquad
\Sigma;\Gamma;\Delta' \vdash \ENV{V}^{2''}
\qquad
\Delta' \vdash \ENV{V}^{1''} =_{s_2} \ENV{V}^{2''}
\end{equation}
where
    \begin{align*}
\Delta' & = \code{this} : \TVAR{I^Y,s_1},\code{sender} : \TVAR{I^\top,s_\top},\code{value} : \TVAR{\TINT,s_1},\VEC{x} : \VEC{\TVAR{{B_s}}}
\\
  \ENV{V}^{1''} & = \makeenv{(\code{this}, \ENV{V}^1(\code{this})), (\code{sender}, \ENV{V}^1(\code{sender})),  (\code{value}, \ENV{V}^1(\code{value})), (\VEC{x}, \VEC{v_1})} 
\\
  \ENV{V}^{2''} & = \makeenv{(\code{this}, \ENV{V}^2(\code{this})), (\code{sender}, \ENV{V}^2(\code{sender})),  (\code{value}, \ENV{V}^2(\code{value})), (\VEC{x}, \VEC{v_2})} 
    \end{align*}
and $\ENV T (Y)(f) = (\VEC x , S')$. 
 From the premises of \nameref{ts_bss_dcall} we have that 
\[
\ENV{T} \vdash \CONF{S', \ENV{S}^1,\ENV{V}^{1''}} \trans_S \ENV{SV}^{1'}
\qquad
\ENV{T} \vdash \CONF{S', \ENV{S}^2,\ENV{V}^{2''}} \trans_S \ENV{SV}^{2'}
\]
  Moreover, from \nameref{dec_t_decm} we also know that $\Sigma;\Gamma;\Delta' \vdash S' : \TCMD{s_1}$; so we can apply the induction hypothesis (thanks to \eqref{eq:indHyp}) to obtain $\Gamma;\Delta' \vdash \ENV{S},\ENV{V}'' =_{s_2} \ENV{SV}'$
that trivially implies $\Gamma;\Delta \vdash \ENV{SV} =_{s_2} \ENV{S}', \ENV{V}$, as desired.
 \qedhere
\end{itemize}
\end{proof}

\newpage
\bibliographystyle{plainnat}
\bibliography{literature}

\begin{thebibliography}{37}
\providecommand{\natexlab}[1]{#1}
\providecommand{\url}[1]{\texttt{#1}}
\expandafter\ifx\csname urlstyle\endcsname\relax
  \providecommand{\doi}[1]{doi: #1}\else
  \providecommand{\doi}{doi: \begingroup \urlstyle{rm}\Url}\fi

\bibitem[Aceto et~al.(2024)Aceto, Gorla, and Lybech]{AGL/2024/ecoop/flowtypes}
Luca Aceto, Daniele Gorla, and Stian Lybech.
\newblock {A Sound Type System for Secure Currency Flow}.
\newblock In Jonathan Aldrich and Guido Salvaneschi, editors, \emph{38th
  European Conference on Object-Oriented Programming (ECOOP 2024)}, volume 313
  of \emph{Leibniz International Proceedings in Informatics (LIPIcs)}, pages
  1:1--1:27, Dagstuhl, Germany, 2024. Schloss Dagstuhl -- Leibniz-Zentrum
  f{\"u}r Informatik.
\newblock ISBN 978-3-95977-341-6.
\newblock \doi{10.4230/LIPIcs.ECOOP.2024.1}.
\newblock URL
  \url{https://drops.dagstuhl.de/entities/document/10.4230/LIPIcs.ECOOP.2024.1}.

\bibitem[Aceto et~al.(2025)Aceto, Gorla, Lybech, and
  Hamdaqa]{AGLM/2024/reocas/outofgas}
Luca Aceto, Daniele Gorla, Stian Lybech, and Mohammad Hamdaqa.
\newblock Preventing out-of-gas exceptions by typing.
\newblock In Tiziana Margaria and Bernhard Steffen, editors, \emph{Leveraging
  Applications of Formal Methods, Verification and Validation. REoCAS
  Colloquium in Honor of Rocco De Nicola}, pages 409--426, Cham, 2025. Springer
  Nature Switzerland.
\newblock ISBN 978-3-031-73709-1.

\bibitem[Ahmed(2004)]{ahmed2004semantic_types_phd}
Amal~Jamil Ahmed.
\newblock \emph{Semantics of Types for Mutable State}.
\newblock PhD thesis, Princeton University, 2004.
\newblock URL \url{http://www.ccs.neu.edu/home/amal/ahmedsthesis.pdf}.

\bibitem[Atzei et~al.(2017)Atzei, Bartoletti, and Cimoli]{ABC17}
Nicola Atzei, Massimo Bartoletti, and Tiziana Cimoli.
\newblock A survey of attacks on ethereum smart contracts (sok).
\newblock In \emph{Proc. of {POST}}, volume 10204 of \emph{LNCS}, pages
  164--186. Springer, 2017.
\newblock \doi{10.1007/978-3-662-54455-6\_8}.

\bibitem[Bartoletti et~al.(2019)Bartoletti, Galletta, and
  Murgia]{bartoletti2019tinysol}
Massimo Bartoletti, Letterio Galletta, and Maurizio Murgia.
\newblock A minimal core calculus for solidity contracts.
\newblock In Cristina P{\'e}rez-Sol{\`a}, Guillermo Navarro-Arribas, Alex
  Biryukov, and Joaquin Garcia-Alfaro, editors, \emph{Data Privacy Management,
  Cryptocurrencies and Blockchain Technology}, pages 233--243, Cham, 2019.
  Springer International Publishing.
\newblock ISBN 978-3-030-31500-9.
\newblock \doi{10.1007/978-3-030-31500-9\_15}.

\bibitem[Br{\"{a}}m et~al.(2021)Br{\"{a}}m, Eilers, M{\"{u}}ller, Sierra, and
  Summers]{BramEMSS21}
Christian Br{\"{a}}m, Marco Eilers, Peter M{\"{u}}ller, Robin Sierra, and
  Alexander~J. Summers.
\newblock Rich specifications for {Ethereum} smart contract verification.
\newblock \emph{Proc. {ACM} Program. Lang.}, 5\penalty0 ({OOPSLA}):\penalty0
  1--30, 2021.
\newblock URL \url{https://doi.org/10.1145/3485523}.

\bibitem[Caires(2007)]{caires2007}
Luís Caires.
\newblock Logical semantics of types for concurrency.
\newblock In Till Mossakowski, Ugo Montanari, and Magne Haveraaen, editors,
  \emph{Algebra and Coalgebra in Computer Science}, pages 16--35, Berlin,
  Heidelberg, 08 2007. Springer Berlin Heidelberg.
\newblock ISBN 978-3-540-73857-2.
\newblock \doi{10.1007/978-3-540-73859-6_2}.

\bibitem[Cassez et~al.(2022{\natexlab{a}})Cassez, Fuller, and
  Asgaonkar]{CassezFA22}
Franck Cassez, Joanne Fuller, and Aditya Asgaonkar.
\newblock Formal verification of the {Ethereum 2.0 Beacon Chain}.
\newblock In \emph{28th International Conference on Tools and Algorithms for
  the Construction and Analysis of Systems}, volume 13243 of \emph{LNCS}, pages
  167--182. Springer, 2022{\natexlab{a}}.
\newblock URL \url{https://doi.org/10.1007/978-3-030-99524-9\_9}.

\bibitem[Cassez et~al.(2022{\natexlab{b}})Cassez, Fuller, and
  Quiles]{CassezFQ22}
Franck Cassez, Joanne Fuller, and Horacio Mijail~Anton Quiles.
\newblock Deductive verification of smart contracts with {Dafny}.
\newblock In \emph{27th International Conference on Formal Methods for
  Industrial Critical Systems}, volume 13487 of \emph{LNCS}, pages 50--66.
  Springer, 2022{\natexlab{b}}.
\newblock URL \url{https://doi.org/10.1007/978-3-031-15008-1\_5}.

\bibitem[Clarkson and Schneider(2010)]{hyperproperties}
Michael~R. Clarkson and Fred~B. Schneider.
\newblock Hyperproperties.
\newblock \emph{J. Comput. Secur.}, 18\penalty0 (6):\penalty0 1157--1210, 2010.
\newblock \doi{10.3233/JCS-2009-0393}.

\bibitem[Crafa et~al.(2019)Crafa, Pirro, and
  Zucca]{crafa2019featherweight_solidity}
Silvia Crafa, Matteo~Di Pirro, and Elena Zucca.
\newblock Is solidity solid enough?
\newblock In \emph{Financial Cryptography Workshops}, 2019.

\bibitem[dao()]{dao2016}
dao.
\newblock The dao smart contract.
\newblock
  \url{http://etherscan.io/address/0xbb9bc244d798123fde783fcc1c72d3bb8c189413#code},
  2016.

\bibitem[Dickerson et~al.(2018)Dickerson, Gazzillo, Herlihy, Saraph, and
  Koskinen]{dickerson2018proof_carrying_smart_contracts}
Thomas~D. Dickerson, Paul Gazzillo, Maurice Herlihy, Vikram Saraph, and Eric
  Koskinen.
\newblock Proof-carrying smart contracts.
\newblock In \emph{Financial Cryptography Workshops}, 2018.

\bibitem[Feist et~al.(2019)Feist, Grieco, and Groce]{FeistGG19}
Josselin Feist, Gustavo Grieco, and Alex Groce.
\newblock Slither: a static analysis framework for smart contracts.
\newblock In \emph{Proceedings of the 2nd International Workshop on Emerging
  Trends in Software Engineering for Blockchain}, pages 8--15. {IEEE} / {ACM},
  2019.
\newblock URL \url{https://doi.org/10.1109/WETSEB.2019.00008}.

\bibitem[Foundation(2022)]{solidity2022}
Ethereum Foundation.
\newblock Solidity documentation.
\newblock \url{https://docs.soliditylang.org/}, 2022.
\newblock Accessed: 2024-01-15.

\bibitem[Friðriksson(2024)]{tinysol_interpreter}
Arnór Friðriksson.
\newblock Interpreter and type checker for {\TINYSOL}.
\newblock \url{https://github.com/Zepeacedust/TinySol-Type-checker}, 2024.
\newblock Accessed: 2025-03-03.

\bibitem[Genet et~al.(2020)Genet, Jensen, and Sauvage]{GenetJS20}
Thomas Genet, Thomas~P. Jensen, and Justine Sauvage.
\newblock Termination of {Ethereum}'s smart contracts.
\newblock In \emph{Proc. of the 17th International Joint Conference on
  e-Business and Telecommunications - Volume 2: SECRYPT}, pages 39--51.
  ScitePress, 2020.
\newblock URL \url{https://doi.org/10.5220/0009564100390051}.

\bibitem[Goguen and Meseguer(1982)]{goguen_mesegier1982security_policies}
J.~A. Goguen and J.~Meseguer.
\newblock Security policies and security models.
\newblock In \emph{1982 IEEE Symposium on Security and Privacy}, pages 11--20,
  1982.
\newblock \doi{10.1109/SP.1982.10014}.

\bibitem[Grishchenko et~al.(2018)Grishchenko, Maffei, and
  Schneidewind]{grishchenko2018}
Ilya Grishchenko, Matteo Maffei, and Clara Schneidewind.
\newblock A semantic framework for the security analysis of {Ethereum} smart
  contracts.
\newblock In Lujo Bauer and Ralf K{\"u}sters, editors, \emph{Principles of
  Security and Trust}, pages 243--269, Cham, 2018. Springer International
  Publishing.
\newblock ISBN 978-3-319-89722-6.

\bibitem[Hu et~al.(2021)Hu, Zhuang, Lin, Zhang, Kan, and Cao]{HuZLZKC21}
Xinwen Hu, Yi~Zhuang, Shangwei Lin, Fuyuan Zhang, Shuanglong Kan, and Zining
  Cao.
\newblock A security type verifier for smart contracts.
\newblock \emph{Comput. Secur.}, 108:\penalty0 102343, 2021.
\newblock URL \url{https://doi.org/10.1016/j.cose.2021.102343}.

\bibitem[Kalra et~al.(2018)Kalra, Goel, Dhawan, and Sharma]{KalraGDS18}
Sukrit Kalra, Seep Goel, Mohan Dhawan, and Subodh Sharma.
\newblock {\textsc{ZEUS}:} analyzing safety of smart contracts.
\newblock In \emph{25th Annual Network and Distributed System Security
  Symposium}. The Internet Society, 2018.
\newblock URL
  \url{https://www.ndss-symposium.org/wp-content/uploads/2018/02/ndss2018\_09-1\_Kalra\_paper.pdf}.

\bibitem[Luu et~al.(2016)Luu, Chu, Olickel, Saxena, and Hobor]{oyente}
Loi Luu, Duc-Hiep Chu, Hrishi Olickel, Prateek Saxena, and Aquinas Hobor.
\newblock Making smart contracts smarter.
\newblock In \emph{Proc. SIGSAC Conf. on Computer and Communications Security},
  page 254–269. ACM, 2016.
\newblock URL \url{https://doi.org/10.1145/2976749.2978309}.

\bibitem[Mavridou and Laszka(2018)]{MavridouL18}
Anastasia Mavridou and Aron Laszka.
\newblock Designing secure {Ethereum} smart contracts: {A} finite state machine
  based approach.
\newblock In \emph{22nd Conference on Financial Cryptography and Data
  Security}, volume 10957 of \emph{LNCS}, pages 523--540. Springer, 2018.
\newblock URL \url{https://doi.org/10.1007/978-3-662-58387-6\_28}.

\bibitem[Nielson and
  Nielson(2007)]{nielson_nielson2007semantics_with_applications}
Hanne~Riis Nielson and Flemming Nielson.
\newblock \emph{Semantics with Applications: An Appetizer}.
\newblock Springer-Verlag London, 2007.
\newblock \doi{10.1007/978-1-84628-692-6}.

\bibitem[parity a()]{wallet17a}
parity a.
\newblock The parity wallet breach.
\newblock
  \url{https://www.coindesk.com/30-million-ether-reported-stolen-parity-wallet-breach/},
  2017.

\bibitem[parity b()]{wallet17b}
parity b.
\newblock The parity wallet vulnerability.
\newblock \url{https://paritytech.io/blog/security-alert.html}, 2017.

\bibitem[Park et~al.(2020)Park, Zhang, and Rosu]{PZR20}
Daejun Park, Yi~Zhang, and Grigore Rosu.
\newblock End-to-end formal verification of {Ethereum 2.0 Deposit Smart
  Contract}.
\newblock In Shuvendu~K. Lahiri and Chao Wang, editors, \emph{Computer Aided
  Verification - 32nd International Conference, {CAV} Proceedings, Part {I}},
  volume 12224 of \emph{Lecture Notes in Computer Science}, pages 151--164.
  Springer, 2020.
\newblock URL \url{https://doi.org/10.1007/978-3-030-53288-8\_8}.

\bibitem[Rosu and Serbanuta(2010)]{RosuS10}
Grigore Rosu and Traian{-}Florin Serbanuta.
\newblock An overview of the {K} semantic framework.
\newblock \emph{J. Log. Algebraic Methods Program.}, 79\penalty0 (6):\penalty0
  397--434, 2010.
\newblock URL \url{https://doi.org/10.1016/j.jlap.2010.03.012}.

\bibitem[Schneidewind et~al.(2020)Schneidewind, Grishchenko, Scherer, and
  Maffei]{eThor}
Clara Schneidewind, Ilya Grishchenko, Markus Scherer, and Matteo Maffei.
\newblock ethor: Practical and provably sound static analysis of ethereum smart
  contracts.
\newblock In \emph{Proc. of {SIGSAC} Conf. on Computer and Communications
  Security}, pages 621--640. {ACM}, 2020.
\newblock URL \url{https://doi.org/10.1145/3372297.3417250}.

\bibitem[Seijas et~al.(2016)Seijas, Thompson, and
  McAdams]{seijas2016scripting_smart_contracts}
Pablo~Lamela Seijas, Simon~J. Thompson, and Darryl McAdams.
\newblock Scripting smart contracts for distributed ledger technology.
\newblock \emph{IACR Cryptol. ePrint Arch.}, 2016:\penalty0 1156, 2016.

\bibitem[Smith and Volpano(1998)]{SV98}
Geoffrey Smith and Dennis~M. Volpano.
\newblock Secure information flow in a multi-threaded imperative language.
\newblock In \emph{Proc. of 25th POPL}, pages 355--364. {ACM}, 1998.

\bibitem[smlxl inc.(2023)]{evmcodes}
smlxl inc.
\newblock An {Ethereum} virtual machine opcodes interactive reference, 2023.
\newblock URL \url{https://evm.codes/}.

\bibitem[Timany et~al.(2024)Timany, Krebbers, Dreyer, and
  Birkedal]{timany/2024/acm/semantic_typing}
Amin Timany, Robbert Krebbers, Derek Dreyer, and Lars Birkedal.
\newblock A logical approach to type soundness.
\newblock \emph{J. ACM}, 71\penalty0 (6), November 2024.
\newblock ISSN 0004-5411.
\newblock \doi{10.1145/3676954}.
\newblock URL \url{https://doi.org/10.1145/3676954}.

\bibitem[Tolmach et~al.(2020)Tolmach, Li, Lin, Liu, and
  Li]{tolmach2020survey_smart_contracts}
Palina Tolmach, Yi~Li, Shang-Wei Lin, Yang Liu, and Zengxiang Li.
\newblock A survey of smart contract formal specification and verification.
\newblock \emph{ACM Computing Surveys (CSUR)}, 54\penalty0 (7):\penalty0
  148:1--148:38, 2020.
\newblock URL \url{https://doi.org/10.1145/3464421}.

\bibitem[Tsankov et~al.(2018)Tsankov, Dan, Drachsler{-}Cohen, Gervais,
  B{\"{u}}nzli, and Vechev]{TsankovDDGBV18}
Petar Tsankov, Andrei~Marian Dan, Dana Drachsler{-}Cohen, Arthur Gervais,
  Florian B{\"{u}}nzli, and Martin~T. Vechev.
\newblock Securify: Practical security analysis of smart contracts.
\newblock In \emph{Proc. of {SIGSAC} Conference on Computer and Communications
  Security}, pages 67--82. {ACM}, 2018.
\newblock URL \url{https://doi.org/10.1145/3243734.3243780}.

\bibitem[Volpano et~al.(2000)Volpano, Smith, and
  Irvine]{volpano2000secure_flow_typesystem}
Dennis Volpano, Geoffrey Smith, and Cynthia Irvine.
\newblock A sound type system for secure flow analysis.
\newblock \emph{Journal of Computer Security}, 4, 08 2000.
\newblock \doi{10.3233/JCS-1996-42-304}.

\bibitem[Yang et~al.(2019)Yang, Murray, Rimba, and Parampalli]{YangMRP19}
Renlord Yang, Toby Murray, Paul Rimba, and Udaya Parampalli.
\newblock Empirically analyzing {Ethereum}'s gas mechanism.
\newblock In \emph{Proc. of {IEEE} European Symposium on Security and Privacy
  Workshops}, pages 310--319. {IEEE}, 2019.
\newblock URL \url{https://doi.org/10.1109/EuroSPW.2019.00041}.

\end{thebibliography}
\end{document}